\documentclass[aps,prl,reprint,superscriptaddress,amsmath,amssymb,floatfix,nobibnotes]{revtex4-2}
\usepackage{graphicx}
\usepackage{amsthm}
\usepackage{xcolor}
\usepackage{tikz}
\usetikzlibrary{arrows.meta}

\usepackage[lmargin=.7in,rmargin=.7in,tmargin=.6in,bmargin=1in]{geometry}
\usepackage{amsmath,amssymb,amsfonts,enumitem}
\usepackage{amsthm}
\usepackage{adjustbox}
\usepackage{braket}
\usepackage{tabularx}
\usepackage{booktabs}
\usepackage{minitoc}
\usepackage[normalem]{ulem}
\usepackage{tikz}
\usepackage{bbm}
\usepackage{graphicx}
\usepackage{textcomp}
\usepackage{mathtools}
\usepackage{bm}
\usepackage{times}
\usepackage[ruled,vlined]{algorithm2e}
\usepackage{algcompatible}

\newcommand{\norm}[1]{\left\lVert #1 \right\rVert}

\usepackage[colorlinks]{hyperref}
\hypersetup{
    linkcolor=blue,
    citecolor=blue,
    urlcolor=blue
}

\usepackage{amsmath,amssymb,amsfonts}
\usepackage{comment}
\newtheorem{theorem}{Theorem}

\newtheorem{lemma}{Lemma}

\def\be{\begin{equation}}
\def\ee{\end{equation}}

\usepackage[normalem]{ulem}
\newcommand\erase{\bgroup\markoverwith{\textcolor{red}{\rule[.5ex]{2pt}{0.4pt}}}\ULon}
\newcommand\eraseblue{\bgroup\markoverwith{\textcolor{blue}{\rule[.5ex]{2pt}{0.8pt}}}\ULon}

\newcommand{\PC}{\mathcal{P}}

\renewcommand{\geq}{\geqslant}
\renewcommand{\leq}{\leqslant}

\DeclareMathOperator*{\argmin}{arg\,min}
\renewcommand{\vec}[1]{\boldsymbol{#1}}

\newcommand{\bs}{\textsf{BS}}

\def\be{\begin{equation}}
\def\ee{\end{equation}}

\def\bs{\begin{split}}
\def\es{\end{split}}
\def\bea{\begin{eqnarray}}
\def\eea{\end{eqnarray}}
\def\w{\omega}

\def\k{^{(k)}}

\def\k{^{(k)}}

\newcommand{\EPFLQSE}{Centre for Quantum Science and Engineering, \'{E}cole Polytechnique F\'{e}d\'{e}rale de Lausanne (EPFL), Lausanne, Switzerland}
\newcommand{\EPFL}{Institute of Physics, \'{E}cole Polytechnique Fédérale de
Lausanne (EPFL), Lausanne, Switzerland}
\newcommand{\KEIO}{Quantum Computing Center, Keio University, Hiyoshi 3-14-1, Kohoku-ku, Yokohama 223-8522, Japan}

\usepackage{tocloft}

\usepackage{dsfont}

\begin{document}
\doparttoc 
\faketableofcontents
\title{Simultaneous computation of multiple excited states via double bracket flows}
\author{Reyhaneh Aghaei Saem }
\thanks{These authors contributed equally to this work.}
\affiliation{\EPFL} \affiliation{\EPFLQSE} 
\author{Mehrad Sahebi}
\thanks{These authors contributed equally to this work.}
\affiliation{\EPFL} \affiliation{\EPFLQSE} 
\author{Zo\"{e} Holmes}
\affiliation{\EPFL} \affiliation{\EPFLQSE}
\author{Yudai Suzuki}
\affiliation{\EPFL} \affiliation{\EPFLQSE} 
\affiliation{\KEIO}
\date{\today}
\begin{abstract}
    Efficient computation of excited states is essential for studying strongly correlated many-body systems and molecular electronic structure. 
    In this work, we propose a framework for simultaneously simulating multiple excited states based on dynamical flows. 
    The main idea is to exploit a common double bracket flow that drives different pure initial states toward different target excited states without such prior information.
    We establish a necessary condition for convergence and a sufficient condition for exponential convergence. We further develop two scheduling strategies: Chebyshev scheduling, which achieves optimal convergence near a fixed point but requires prior spectral information, and adaptive scheduling, which achieves comparable performance in numerical simulations without spectral information. 
    Our results enable new classical and quantum algorithms for excited-state computation without prior preparation of lower-energy eigenstates.
\end{abstract}

\maketitle

\setcounter{secnumdepth}{1}
\section{Introduction}

Computing excited states is central to understanding a broad range of physical and chemical phenomena.
For instance, excitation spectra provide insights into collective excitations and quantum phases in many-body systems~\cite{Sachdev2011}, while electronic excited states are crucial for describing photochemical reactions and molecular spectroscopy~\cite{dreuw2005,gonzalez2012,lischka2018}. 
Characterizing these phenomena often requires several excited states rather than a single excitation~\cite{lischka2018,curchod2018}.
Therefore, efficient computational methods for calculating excited states are essential for advancing research in quantum physics and chemistry.

The computation of excited states poses a challenge for both classical and quantum methods. 
A major difficulty in existing approaches lies in their dependence on prior knowledge of lower-energy eigenstates or spectral information.
In classical electronic-structure calculations, methods such as equation-of-motion coupled-cluster theory~\cite{stanton1993theequationofmotion,krylov2008equationofmotion} and linear-response approaches~\cite{runge1984density,casida1996timedependent,koch1990excitation} typically construct excited states based on a ground-state reference. 
Similar dependencies also arise in quantum algorithms: variational deflation methods~\cite{Higgott2019variationalquantum} typically require previously determined lower-energy eigenstates, while filter-based methods~\cite{kitaev1995quantum,abrams1999quantum,ge2017faster,lin2020optimal,lin2020nearoptimalground,dong2022groundstate} often rely on spectral information to isolate the desired energy region.
A review of existing eigenstate computation methods and their limitations is given in Appendix~\ref{sec:known-methods}.

\begin{figure}[t]
\centering
\definecolor{flowA}{HTML}{080D8B}\definecolor{flowB}{HTML}{C93166}\definecolor{flowC}{HTML}{EC7800}
\definecolor{flowink}{HTML}{1A1A1A}\definecolor{flowgray}{HTML}{6E6E6E}\definecolor{flowbox}{HTML}{F1F0EC}
\definecolor{flowdark}{HTML}{4A4A4A}\definecolor{flowlight}{HTML}{D6D6D6}
\definecolor{flowrest1}{HTML}{8F8F8F}\definecolor{flowrest2}{HTML}{A2A2A2}\definecolor{flowrest3}{HTML}{B5B5B5}\definecolor{flowrest4}{HTML}{C8C8C8}\definecolor{flowrest5}{HTML}{DBDBDB}
\begin{tikzpicture}[x=1cm, y=1cm, font=\footnotesize, text=flowink]
  \useasboundingbox (0,0) rectangle (8.60,5.05);
  
  \node[anchor=north west, inner sep=0pt] at (0.000,5.045) {(a)};
  
  \draw[flowink, line width=0.9pt, fill=flowbox, rounded corners=0.178cm] (3.104,2.899) rectangle (4.596,4.436);
  \node at (3.850,3.668) {\Large$U$};
  \node[anchor=base, inner sep=0pt, text=flowgray] at (3.850,4.556) {double-bracket flow};
  
  \foreach \c/\y/\w/\i in {flowA/2.942/3.289/1, flowB/3.668/3.668/2, flowC/4.393/4.046/3} {
    \draw[\c, line width=1.4pt] (1.567,\y) .. controls (2.413,\y) and (2.258,\w) .. (3.104,\w);
    \fill[\c] (1.480,\y) circle (0.100);
    \node[anchor=east, inner sep=0pt] at (1.280,\y) {$|\phi_{\i}\rangle$};
  }
  
  \foreach \c/\w/\y/\i in {flowA/3.289/2.402/1, flowB/3.668/3.270/2, flowC/4.046/3.810/3} {
    \draw[\c, line width=1.4pt] (4.596,\w) .. controls (5.291,\w) and (5.131,\y) .. (5.627,\y);
    \fill[\c] (5.753,\y) -- (5.583,\y+0.085) -- (5.583,\y-0.085) -- cycle;
    \draw[\c, line width=2.4pt] (5.836,\y) -- (6.929,\y);
    \node[anchor=west, inner sep=0pt] at (7.082,\y) {$|\lambda_{\i}\rangle$};
  }
  
  \draw[flowrest1, line width=1.0pt] (5.836,4.188) -- (6.929,4.188);
  \draw[flowrest2, line width=1.0pt] (5.836,4.379) -- (6.929,4.379);
  \draw[flowrest3, line width=1.0pt] (5.836,4.539) -- (6.929,4.539);
  \draw[flowrest4, line width=1.0pt] (5.836,4.683) -- (6.929,4.683);
  \draw[flowrest5, line width=1.0pt] (5.836,4.829) -- (6.929,4.829);
 
  \draw[flowgray, line width=0.7pt] (8.077,2.338) -- (8.077,4.794);
  \fill[flowgray] (8.077,4.914) -- (8.022,4.784) -- (8.132,4.784) -- cycle;
  \node[rotate=90, anchor=base, inner sep=0pt, text=flowgray] at (8.470,3.626) {Energy};
  
  \node[anchor=north west, inner sep=0pt] at (0.000,2.052) {(b)};
  
  \fill[flowbox, rounded corners=0.127cm] (3.000,0.515) rectangle (4.700,2.052);
  
  \def\flowlevels{1.025/1, 1.275/2, 1.474/3, 1.640/4, 1.773/5, 1.882/6}
  \foreach \y/\k in \flowlevels {
    \ifnum\k=1 \draw[flowdark, line width=1.8pt] (0.850,\y) -- (1.850,\y);
    \else      \draw[flowlight, line width=0.9pt] (0.850,\y) -- (1.850,\y); \fi
    \ifcase\k\or \draw[flowA, line width=1.8pt] (3.350,\y) -- (4.350,\y);
             \or \draw[flowB, line width=1.8pt] (3.350,\y) -- (4.350,\y);
             \or \draw[flowC, line width=1.8pt] (3.350,\y) -- (4.350,\y);
             \else \draw[flowlight, line width=0.9pt] (3.350,\y) -- (4.350,\y); \fi
    \draw[flowdark, line width=1.8pt] (5.850,\y) -- (6.850,\y);
  }
  \node[anchor=base, inner sep=0pt, text=flowgray] at (1.350,0.685) {only ground state};
  \node[anchor=base, inner sep=0pt] at (3.850,0.685) {this work};
  \node[anchor=base, inner sep=0pt, text=flowgray] at (6.350,0.685) {diagonalization};
  
  \draw[flowgray, line width=0.7pt] (1.350,0.375) -- (6.240,0.375);
  \fill[flowgray] (6.350,0.375) -- (6.220,0.430) -- (6.220,0.320) -- cycle;
  \node[anchor=base, inner sep=0pt, text=flowgray] at (1.350,0.075) {$r=1$};
  \node[anchor=base, inner sep=0pt, text=flowgray] at (3.850,0.075) {rank $r$};
  \node[anchor=base, inner sep=0pt, text=flowgray] at (6.350,0.075) {$r=d$};
\end{tikzpicture}
\caption{\textbf{Simultaneous computation of $r$ lowest eigenstates using a common double-bracket flow.}
(a) The central idea is that a common unitary $U$ generated by the double-bracket flow (DBF) enables the computation of $r$ eigenstates from $r$ orthonormal reference states $\{\ket{\phi_i}\}_{i=1}^{r}$.
(b) The DBF minimizes the cost function $\mathrm{Tr}(\rho H)$ over the manifold of rank-$r$ density matrix. 
The extreme cases $r=1$ and $r=d$ correspond to ground-state preparation and Hamiltonian diagonalization, respectively. We focus on the intermediate-rank regime to enable the simultaneous computation of multiple excited states.}
\label{fig:scheme}
\end{figure}
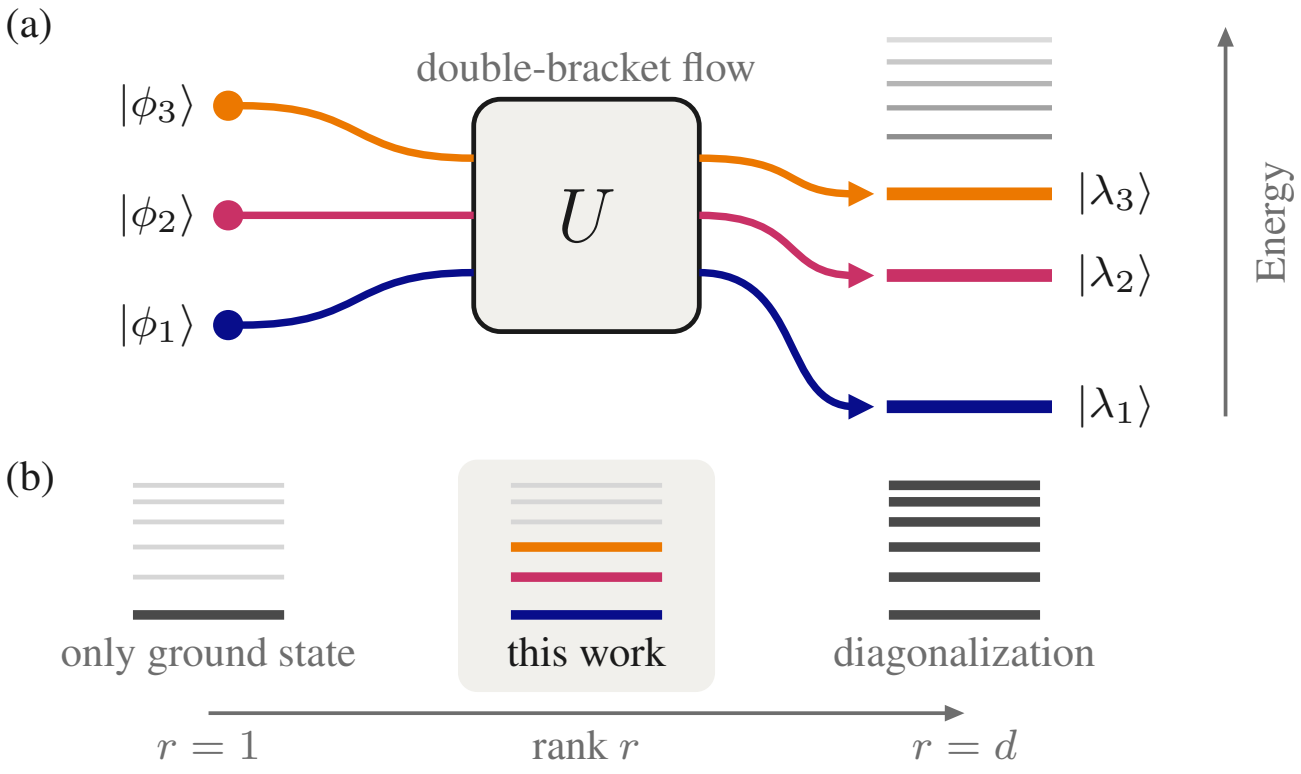

\begin{figure}[t]
\centering
\input{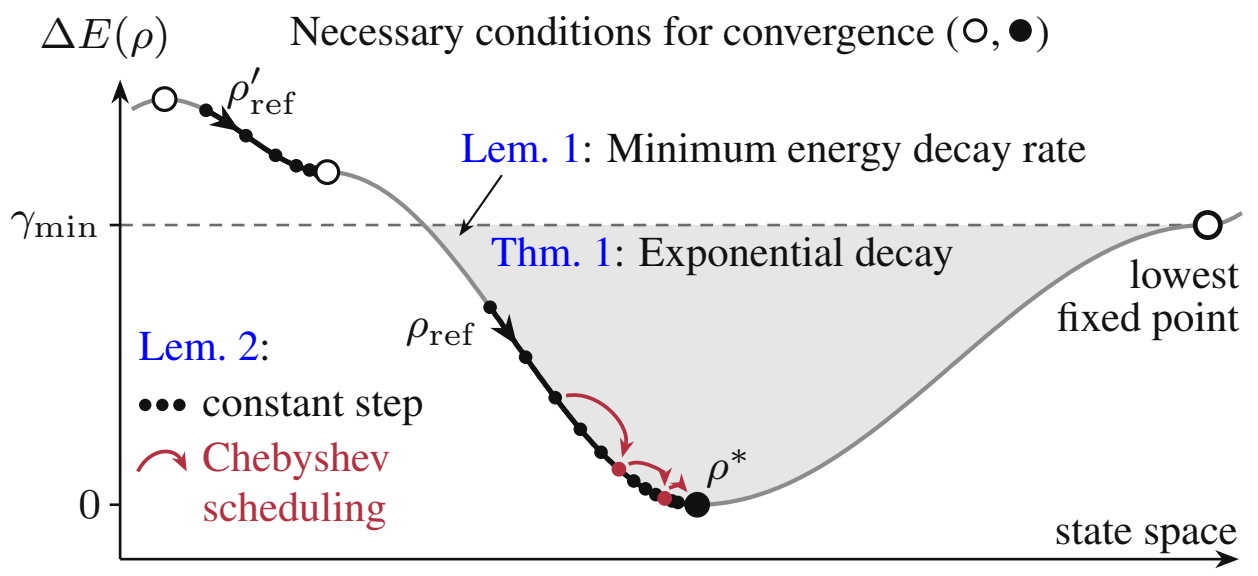}
\caption{ \textbf{Summary of this work.}
This schematic illustrates the energy landscape of the DBF in terms of the energy excess $\Delta E(\rho) = E(\rho)-E^{*}$, where $E^{*}=\mathrm{Tr}(\rho^{*} H)$ is the global energy minimum of the target state $\rho^{*}$ in Eq.~\eqref{eq:target}.
We establish a necessary condition for convergence to $\rho^{*}$ and a sufficient condition for exponential convergence (Theorem~\ref{theorem:convergence-rate}).
The latter is derived by a lower bound on the energy dissipation rate (Lemma~\ref{lemma:gradient}).
We also introduce a Chebyshev scheduling strategy that achieves optimal convergence in the regime near a fixed point~(Lemma~\ref{lemma:chebyshev}).
}
\label{fig:landscape}
\end{figure}

In this work, we introduce a framework for the simultaneous computation of multiple excited states based on a double-bracket flow (DBF)~\cite{brockett1991, helmke1994}, which can be implemented either via classical or quantum simulation methods.
The DBF is a Riemannian gradient flow that minimizes the cost function $E(\rho)=\mathrm{Tr}(\rho H)$ over the manifold of a rank-$r$ density matrix $\rho$, where $H$ denotes a target Hamiltonian acting on $d$-dimensional Hilbert space.
Since its global minimum is a mixture of the $r$ lowest-energy eigenstates, the DBF provides a natural framework for accessing them as sketched in Fig.~\ref{fig:scheme}.

The DBF has been extensively studied in two extreme cases: ground-state preparation ($r=1$)~\cite{gluza2026double,robbiati2026double,zander2026role,suzuki2026grovers,alghadeer2025doublebracket,kolotouros2026quantum,arcialopez2026quantum,wiersema2023optimizing} and Hamiltonian diagonalization ($r=d$)~\cite{glazek1993renormalization,glazek1994perturbative,wegner1994flow,kehrein2006,gluza2024doublebracket,xiaoyue2024strategies}.

Here we consider the intermediate-rank regime ($1<r<d$) and apply a common DBF to drive different initial states toward different target excited states.

Our theoretical analysis establishes rigorous convergence properties for the simultaneous computation of multiple excited states.
In particular, we derive necessary conditions for convergence to a target excited state from the corresponding pure state and sufficient conditions for exponential convergence.
Furthermore, we develop two scheduling strategies: Chebyshev scheduling for optimal convergence with spectral information, and adaptive scheduling, which achieves effective convergence without such information, as demonstrated numerically.
Figure~\ref{fig:landscape} summarizes our technical contributions.

\section{DBF for the simultaneous computation of multiple excited states}

Minimizing the energy of a pure state yields the ground state of a Hamiltonian.
We extend this principle to the simultaneous computation of the ground state and the first $r-1$ excited states. Let us write the Hamiltonian in its eigenbasis as $H=\sum_{j=1}^{d}\lambda_j |\lambda_j\rangle\langle \lambda_j|$  with eigenvalues $\lambda_1  < \dots < \lambda_d$.
The idea is to seek a common unitary $U$ that maps the $r$ orthonormal reference states $\{|\phi_i\rangle\}_{i=1}^r$ to the $r$ lowest-energy eigenstates $\{|\lambda_i\rangle\}_{i=1}^r$ up to a phase (see Fig.~\ref{fig:scheme} for a sketch of this). 
Since both sets are orthonormal, such a unitary transform exists.

To obtain the unitary $U$, we minimize the mean energy $E(\rho)=\mathrm{Tr}(\rho H)$ over a set of rank-$r$ density matrices $\rho$.
To specify which reference state should prepare each eigenstate, we assign the reference states distinct weights $p_1>\cdots>p_r>0$, with $\sum_{i=1}^r p_i=1$, and form the rank-$r$ mixture
\begin{equation}
\label{eq:rho}
    \rho_{\mathrm{ref}}=\sum_{i=1}^r p_i |\phi_i\rangle\langle\phi_i| \, .
\end{equation}
At the global minimum, the weights $p_{i}$ are assigned in descending order to the energy eigenstates in ascending order.
The unique global energy minimum  is therefore given by
\begin{equation} \label{eq:target}
    \rho^*=\sum_{i=1}^r p_i |\lambda_i\rangle\langle\lambda_i| \, .
\end{equation}
Since the weights are distinct, any minimizing unitary $U_{\mathrm{opt}}$ maps $|\phi_i\rangle$ to $|\lambda_i\rangle$ up to a phase. Thus, once we have found $U_{\mathrm{opt}}$, we can prepare any of the $r$ lowest eigenstates by applying it to the corresponding reference state.

To search for such a unitary, we use the DBF. 
Starting from $\rho(0)=\rho_{\mathrm{ref}}$, we evolve the mixture according to
\begin{equation}
    \label{eq:C}
        \partial_t \rho(t)=[[\rho(t),H],\rho(t)] = [W(t), \rho(t)] \, ,
\end{equation}
where $W(t)=[\rho(t),H]$ is anti-Hermitian and hence generates a unitary evolution $\rho(t)=U(t)\rho_{\mathrm{ref}}U^\dagger(t)$.

To see why this choice of generator lowers the mean energy, we differentiate $E(\rho(t))=\mathrm{Tr}(\rho(t)H)$ to obtain
\begin{equation}
    \partial_t E\left(\rho(t)\right) = - \| W(t) \|_{HS}^2 \le 0 \,,
        \label{eq:E-C}
\end{equation}
where $\|\cdot\|_{HS}$ denotes the Hilbert-Schmidt norm (see Appendix~\ref{app:energy} for the details).
We refer to $ \| [\rho, H] \|_{HS}^2$ as the energy dissipation rate, because the norm of the commutator determines the rate of energy decay along the DBF.
This indicates that the mean energy decreases along the flow and becomes constant only at a fixed point, characterized by $W(t)=0$.
That is, $\rho(t) = \sum_{i=1}^{r} p_i |\phi_i(t)\rangle\langle\phi_i(t)|$ and $H$ share an eigenbasis, and hence the fixed points can be expressed as
\begin{equation}
\label{eq:placements}
    \rho_\sigma=\sum_{a=1}^{d}p_{\sigma(a)} |\lambda_a\rangle\langle \lambda_a| \,,
\end{equation}
where $\sigma$ is a permutation of the $d$ energy levels, with $p_i=0$ for $i>r$. 
Among these fixed points, the unique global energy minimum~\cite{gross1988rayleigh} is given by $\rho^{*}$ in Eq.~\eqref{eq:target}.

We utilize the DBF for the simultaneous computation of multiple excited states.
Specifically, we apply the common unitary $U(t)$ to each initial pure state $\ket{\phi_i}$ in the orthonormal set to drive it to the corresponding $i$-th eigenstate $\ket{\lambda_{i}}$ for $i=1,\ldots,r$.
However, convergence to the global energy minimum $\rho^{*}$ in Eq.~\eqref{eq:target} is not guaranteed for an arbitrary reference state $\rho_{\mathrm{ref}}$, since the DBF may converge to a fixed point other than the global minimum; in this case, applying $U(t)$ to $\ket{\phi_i}$ may fail to yield the corresponding target eigenstate.

We establish in Appendix~\ref{app:thm1} that convergence to the global energy minimum $\rho^{*}$ requires the reference state $\rho_{\mathrm{ref}}$ to have nonzero overlap with every target eigenstate.
More precisely, the following necessary condition must hold:
\begin{equation} \label{eq:A_i}
    A_i := \langle \lambda_i|P_r|\lambda_i \rangle>0
\end{equation}
for all $i \le r$, where $P_r=\sum_{j=1}^{r}|\phi_j\rangle\langle\phi_j|$ is the projector onto the span of the $r$ reference states.
In other words, if the reference subspace has no overlap, convergence to the global minimum is impossible.

This condition is consistent with ground-state preparation, which corresponds to the special case $r=1$, where the initial state must have nonzero overlap with the ground state. 
It also highlights an important distinction from Hamiltonian diagonalization: at full rank ($r=d$), the overlap condition is automatically satisfied, whereas in the intermediate-rank regime, it imposes a nontrivial constraint on the initial reference subspace. 
We also numerically verify that the necessary condition; see End Matter.

\section{Our framework}

We now introduce two practical implementations for the DBF, since the continuous-time flow in Eq.~\eqref{eq:C} cannot be implemented exactly on classical or quantum computers.

\medskip
\textit{Discretized DBF.--}
    The first approach is a straightforward discretization of Eq.~\eqref{eq:C}~\cite{moore1994numerical,helmke1994}:
\begin{equation}
\label{eq:D}
    \rho_{k+1}=V_{k}\rho_k V_{k}^{\dagger}\, , \quad V_{k}=e^{s_k W_k} ,
\end{equation}
where $\rho_k = \sum_{i=1}^{r} p_i | \phi_i^{(k)} \rangle \langle \phi_i^{(k)} | $ is the state at step $k$, with $\rho_{0}=\rho_{\mathrm{ref}}$ and $W_k= [\rho_{k},H]$.
The corresponding evolution of each pure state is given by $|\phi_i^{(k+1)}\rangle =  e^{s_k W_k}|\phi_i^{(k)}\rangle$.
This derivation is analogous to solving the von Neumann equation, $\frac{d}{dt} \rho = [W(t), \rho]$, where $W(t)$ is regarded as constant over each time step; under this assumption, the solution is given by $\rho(s+t)= e^{sW(t)}\rho(t) e^{-sW(t)}$.
Although $W(t)$ is time-dependent, the update in Eq.~\eqref{eq:D} agrees with the continuous-time evolution in Eq.~\eqref{eq:C} to first order in the step size $s$.
This scheme can be implemented numerically on a classical computer using standard matrix operations.

\textit{Product-formula DBF.--}
The second approach employs a product-formula approximation to the discretized DBF in Eq.~\eqref{eq:D}~\cite{gluza2024doublebracket,gluza2026double}:

\begin{align}
\label{eq:PF}
    \rho_{k+1} = V_k \rho_k V_k^\dagger \, ,
\end{align}
where $V_{k}$ approximates $e^{s_{k}W_{k}}$, and $\rho_k$ again denotes the state at step $k$.
One example is the group-commutator approximation~\cite{chen2022efficient},
\begin{equation}
\begin{split}
    V_k  &= 
    e^{i\tau_k H}e^{i\theta_k \rho_k}e^{-i\tau_k H}e^{-i\theta_k \rho_k} \\
    & = e^{s_{k} [\rho_{k}, H]} + \mathcal{O}(s_k^{3/2})
\end{split}
\end{equation}
with positive parameters $\tau_k,\theta_k >0$, satisfying $s_{k}=\tau_k \theta_k$.
Although the product-formula approximation can also be computed classically, its structure is particularly suited to quantum implementation~\cite{gluza2024doublebracket,gluza2026double,son2025quantum}.

For both approaches, we apply the corresponding unitary $\prod_{j=0}^{k} V_{j}$ for the discretized and the product-formula DBFs to an initial pure state $\ket{\phi_{i}}$ in the orthonormal set to obtain the $i$-th target excited state. 
The corresponding algorithms are provided in Methods as Algorithms~\ref{alg:dbf} and~\ref{alg:dbf-pf} respectively.

In Appendices~\ref{app:energy-landscape} and~\ref{app:thm1-theorem-proof}, we show that the necessary condition for convergence to the global minimum in Eq.~\eqref{eq:A_i} also applies to the discretized and product-formula DBFs, provided that the step size $s$ is sufficiently small for these flows to faithfully reproduce the continuous-time one.
Specifically, we require $s_k < \bar{s}$ with 
\begin{equation}
\label{eq:step-conditions}
\begin{aligned}
    \textup{discrete:} \; \bar{s} = \frac{2}{\gamma_{\max}}, \quad
    \textup{product-formula:} \; \bar{s} = \frac{12}{7\gamma_{\max}} ,
\end{aligned}
\end{equation}
where $\gamma_{\max} = p_1 \Omega$, with $\Omega = \lambda_d - \lambda_1$ denoting the spectral width of $H$, and $p_1$ the largest weight. 
For the product-formula DBF, we further fix the ratio of the angles as $\theta_k/\tau_k=\Omega/p_1$.

\section{Convergence rates}

We now characterize the convergence speed to the global minimum in terms of the mean energy $E(\rho)$.
As shown in Eq.~\eqref{eq:E-C}, the mean energy is non-increasing along the continuous-time DBF, where its rate of decrease is governed by the norm of the generator $W$.
We further show that the mean energy is also non-increasing along the discretized and product-formula DBFs, as long as the step sizes satisfy the conditions in Eq.~\eqref{eq:step-conditions} (see Appendix~\ref{app:energy} for details):
\begin{equation}
    E(\rho_{k+1}) - E(\rho_k) \le  - s_k \, (1 - \frac{s_k}{\bar{s}})\| W_k \|_{HS}^2 \,.
        \label{eq:E-D}
\end{equation}
Namely, the right-hand side of Eq.~\eqref{eq:E-D} is non-positive under the step-size conditions in Eq.~\eqref{eq:step-conditions}.

To quantify convergence to the target state, we track the excess energy 
\begin{equation}
    \Delta E(\rho)=E(\rho)-E^* , \quad  E^*=\mathrm{Tr}(\rho^*H) \, .
\end{equation}
We first determine the smallest positive excess energy among the fixed points of the flow. Below this threshold, $\rho^*$ is the only fixed point. That is, below the lowest saddle point, the energy dissipation rate is nonzero, ensuring exponential convergence.
Before showing this, we first characterize this lowest fixed (saddle) point. 

A simple way to obtain another fixed point is to exchange the weights on two energy levels in $\rho^*$. Let $\rho_{\sigma_{i,j}}$ denote the state obtained by exchanging $p_i$ and $p_j$, where $i<j$ and $i\le r$. Since only the contributions from levels $i$ and $j$ change, its excess energy is
\begin{align}
\label{eq:gamma-ij}
    \gamma_{ij}
    := \Delta E(\rho_{\sigma_{i,j}}) \nonumber &= (p_i\lambda_j+p_j\lambda_i)-(p_i\lambda_i+p_j\lambda_j) \nonumber \\
    &= (p_i-p_j)(\lambda_j-\lambda_i) \, .
\end{align}
The lowest saddle point is one adjacent swap away from $\rho^*$,
\begin{equation}
\label{eq:gamma-min}
    \gamma_{\min}=\min_{1\leq i\leq r}\gamma_{i,i+1} \, .
\end{equation}
To see why, exchange two unequal weights on neighboring energy levels. Only this pair is now out of order. A small rotation mixing these levels shifts population toward the lower level and decreases the mean energy. For $1<r<d$, at least one other pair still has more population on its lower level. Mixing that pair raises the mean energy, so the exchanged state is a saddle.
Now consider a fixed point with the smallest positive excess energy. It must have an adjacent pair of weights out of order. Correcting that pair gives a fixed point of strictly lower energy, which by minimality must be $\rho^*$. 

The lowest saddle point therefore lies at energy $E^*+\gamma_{\min}$. We now use this separation to bound the energy dissipation rate. The proof of the bound is given in Appendix~\ref{app:lemma1}.

\begin{lemma}[Lower bound on the energy dissipation rate, informal]
\label{lemma:gradient}
    For any state $\rho$ of the form in Eq.~\eqref{eq:rho}, the following lower bound on the energy dissipation rate for the DBF holds:
    \begin{equation}
    \label{eq:gradient}
        \| [\rho, H] \|_{HS}^{2} \geq 2 \Delta E (\rho) \left(\gamma_{\min} - \Delta E(\rho)\right) \,.
    \end{equation}
\end{lemma}
Lemma~\ref{lemma:gradient} establishes that the energy dissipation rate is strictly positive whenever $0<\Delta E(\rho)<\gamma_{\min}$. 
This is consistent with the above argument: when $\Delta E \ge \gamma_{\min}$, the trajectory may approach a non-target fixed point, and the resulting lower bound is trivially zero.
Therefore, the condition $\Delta E(\rho)<\gamma_{\min}$ serves as a sufficient condition for convergence to the global energy minimum, i.e., $\rho^{*}$.
Using Lemma~\ref{lemma:gradient}, we establish the convergence rate toward the target excited states in terms of the fidelity $f_i(\ket{\phi}) := | \langle \lambda_i | \phi \rangle|^2$ under the condition $\Delta E(\rho) <\gamma_{\min}$.

\begin{theorem}[Sufficient conditions for exponential convergence, informal]
\label{theorem:convergence-rate}
    Let the state $\rho$ evolve according to one of the three flows: the continuous-time DBF, the discretized DBF, or the product-formula flow, defined in Eqs.~\eqref{eq:C},~\eqref{eq:D} and~\eqref{eq:PF}, respectively.
    For the latter two flows, assume that the step sizes $s_k < s_{\max}< \bar{s}$, with some $s_{\max}>0$ and $\bar{s}$ given by Eq.~\eqref{eq:step-conditions}.
    We define the flow time $t$ as the continuous evolution time for the continuous-time DBF and as $t=\sum_{k=0}^{K-1}s_k$ after $K$ steps for the discretized and product-formula DBFs. The fidelity of the $i$-th state with its target eigenstate at time t is defined as  $F_i(t) = f_i(\ket{\phi_i(t)})$ for the continuous-time DBF and $F_i(t) = f_i(\ket{\phi_i^{(K)}})$ for the discretized and product-formula DBFs.

    If the initial energy excess satisfies $\Delta E (\rho_\mathrm{ref})<\gamma_{\min}$, then the energy excess and the infidelities $1-F_i$ of all $r$ states decay exponentially with the flow time $t$.
    Specifically, 
    \begin{equation}
    \label{eq:rate}
        1 - F_i(t) \le \frac{\Delta E\left(\rho(t)\right)}{\gamma_{\min}} \le c_0 \, e^{-2 c \gamma_{\min} t} \, ,
    \end{equation}
    for all $i \le r$, where the prefactor $c_0 = \Delta E (\rho_\mathrm{ref})/ \left(\gamma_{\min} - \Delta E(\rho_\mathrm{ref})\right)$ depends only on the initial state.
    The decay constant is $c=1$ for the continuous-time DBF. For the discretized and product-formula DBFs, $c$ satisfies $0<c<1$, approaches 1 as the step sizes decrease, and is bounded below by $1/4$ whenever $s_k \le \bar s/2$.
\end{theorem}

Theorem~\ref{theorem:convergence-rate} shows that all three DBFs converge exponentially to the target excited states whenever $\Delta E(\rho_\mathrm{ref})<\gamma_{\min}$. 
We note that the constant $c$ captures the effect of discretization. As the step sizes decrease, the convergence rate approaches $2\gamma_{\min}$, i.e., the rate of the continuous-time flow. Moreover, step sizes up to half the upper bound $\bar{s}$ guarantee at least one quarter of this rate. Appendix~\ref{app_sec:exponential-convergence-proof} gives the complete statement, including the explicit values of $c$ for the discretized and product-formula DBFs.
In End Matter, we numerically demonstrate that the sufficient condition in Theorem~\ref{theorem:convergence-rate} is satisfied.
Moreover, numerical simulations show that initial states violating $\Delta E(\rho_\mathrm{ref})<\gamma_{\min}$ also exhibit convergence rates faster than $2\gamma_{\min}$, suggesting that the sufficient condition is conservative and that the method exhibits exponential convergence over a substantially broader range of initial conditions.

Convergence rates in Theorem~\ref{theorem:convergence-rate} further provide a quantitative estimate of the computational cost of reaching a given accuracy. Let $\varepsilon$ denote the target infidelity for each of the $r$ states, and let $\kappa=\gamma_{\max}/\gamma_{\min}$. By Eq.~\eqref{eq:rate}, the continuous-time DBF reaches $1-F_i(t)\le\varepsilon$ after a flow time of $t=\ln(c_0/\varepsilon)/(2\gamma_{\min})$. For the discretized and product-formula DBFs, the guaranteed energy decrease per step is maximized by choosing a constant step size $s_k=s=\bar{s}/2$. With this choice, $(1+\kappa)\ln(c_0/\varepsilon)$ steps suffice for the discretized DBF, while $(1+7\kappa/6)\ln(c_0/\varepsilon)$ steps suffice for the product-formula DBF, as shown in Appendix~\ref{app for steps}. 
Both step counts grow logarithmically with $1/\varepsilon$ and linearly with $\kappa$. Thus, for the discretized DBF, the step size $\bar{s}/2=1/\gamma_{\max}$ gives the best guarantee, although it need not be the fastest choice near the target.

\section{Optimal and adaptive scheduling strategies}

We can further accelerate convergence of the discretized and product-formula DBFs by appropriately choosing the step size $s$ and the weights $p_i$.
A natural choice of the weights $p_i$ and step size $s$ is given by  $
p_i-p_{i+1} \propto (\lambda_{i+1}-\lambda_{i})^{-1}
$ and 
\begin{equation}
\label{eq:s-opt}
    s^\star=\frac{2}{\gamma_{\max}+\gamma_{\min}},
\end{equation}
respectively.
Each choice is optimal near a fixed point when the other is held fixed. 
This optimality follows from minimizing the largest contraction factor $\left|1-s\gamma_{ij}\right|$ among the off-diagonal elements $\left|(\rho-\rho^*)_{ij}\right|$, where the contraction factor is the multiplicative factor by which each element changes in one iteration; see Appendix~\ref{app:optimal_step_schedules}.
Intuitively, the step size is chosen to accelerate the contraction of the off-diagonal elements, particularly those that decay most slowly.

In practice, however, the step size need not remain constant. We thus develop two scheduling strategies: Chebyshev scheduling, which uses nonconstant step sizes to achieve optimal convergence near a fixed point, and adaptive scheduling, which updates both $p_i$ and $s$ without requiring spectral information.

\textit{Optimal scheduling.--}
We first introduce a non-constant step-size schedule, referred to as \textit{Chebyshev scheduling}~\cite{young1953richardson}. The scheme uses an $m$-periodic sequence $\{s_k\}_{k=1}^{m}$ where 
 \begin{align} \label{chebyschMain}
  {s^\star_k}^{-1}:=\frac{\gamma_{\max}+\gamma_{\min}}{2}+\frac{\gamma_{\max}-\gamma_{\min}}{2}\cos\!\Big(\frac{(2k-1)\pi}{2m}\Big) \ 
\end{align}
for $k=1,\cdots,m$. The following lemma establishes the optimality of this scheme in the near-target regime. The proof is provided in Appendix~\ref{app:chebyshev}.

\begin{lemma}(Optimality of Chebyshev scheduling)
\label{lemma:chebyshev}
   The $m$-periodic Chebyshev schedule is optimal among all $m$-periodic step-size schedules near the target fixed points. Moreover, the average contraction factor per iteration is monotonically decreasing in $m$ and approaches $\frac{\sqrt{\kappa}-1}{\sqrt{\kappa}+1}$ in the limit of large $m$, where $\kappa= \frac{\gamma_{\max}}{\gamma_{\min}}$.  
\end{lemma}

Lemma~\ref{lemma:chebyshev} shows that Chebyshev scheduling minimizes the worst-case contraction factor across all off-diagonal elements with $\gamma\in[\gamma_{\min},\gamma_{\max}]$. In particular, the optimal constant step size discussed above is recovered as the special case $m=1$. 
As the period $m$ increases, the average contraction factor per iteration approaches $\frac{\sqrt{\kappa}-1}{\sqrt{\kappa}+1}$, compared with $\frac{\kappa-1}{\kappa+1}$ for the optimal constant step size. 
The Chebyshev acceleration obtained here is also well known in iterative methods for solving linear systems~\cite{young1953richardson,adsuara2017equivalence,golub1961chebyshev}.

\begin{figure}
        \centering
        \includegraphics[width=\columnwidth]{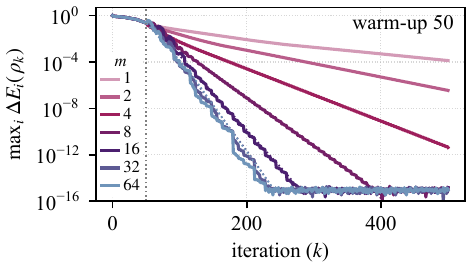}
        \caption{\textbf{Chebyshev scheduling improves convergence rate.} The maximum differences between the energies obtained from the DBF and the corresponding eigenstate energies are shown as a function of the number of iterations for different periods $m$.
        The task is to find the first four lowest energy states of the TFIM Hamiltonian. The initial states are uniform combination of all the energy eigenstates and the weights are fixed to their optimal value through inverse gap ratios. All runs start with 50 round of warm up with $m=1$ to avoid early time instabilities.  The product formula order is fixed to 4 for all cases.} 
        \label{fig:ChebyshevScheMain}
    \end{figure}

We numerically investigate the performance of the Chebyshev scheduling.
Figure~\ref{fig:ChebyshevScheMain} shows the convergence curves for Chebyshev schedules with different periods $m$ for the product-formula DBF.
As a test model, we use a $10$-qubit open chain transverse field Ising model (TFIM)~\cite{pfeuty1970}, described by $
H_{\text{TFIM}} = -\sum_i J_iZ_i Z_{i+1} - \sum_i h_iX_i  
$, where $J_i$ and $h_i$ denote the interaction strengths and transverse fields, respectively.
Our numerical task is to prepare the $r=4$ lowest-energy eigenstates of the TFIM with uniform couplings, using initial states that are uniform superpositions of all energy eigenstates. 
Each run begins with 50 warm-up iterations using a constant step size $s^\star$. 
To ensure a consistent comparison across different scheduling periods, we fix the product-formula order to four in all simulations. 

The 64-periodic schedule reaches machine precision after 212 iterations, whereas the error for the constant-step-size schedule remains above $10^{-2}$ even after 500 iterations. 
Furthermore, for $m=16,32,64$, the error does not decrease monotonically at every iteration; rather, it decreases over each complete scheduling cycle.
These results suggest that Chebyshev scheduling remains effective beyond the linear regime and that the use of a low-order product formula does not significantly compromise the energy convergence.
We remark that, despite its theoretical guarantees, the practical applicability of Chebyshev scheduling to the discretized and product-formula flows is limited by potential stability issues. We examine this limitation in Appendix~\ref{app:num_stepsize}.

\textit{Adaptive scheduling.--}
We next propose an adaptive scheduling strategy that requires no spectral information such as $\gamma_{\min}$ and $\gamma_{\max}$, unlike the optimal constant-step and the Chebyshev scheduling.

In the adaptive scheme, the weights $p_i$ and step size $s$ are estimated from measured energies. 
Initially, we choose linearly spaced weights and estimate $\gamma_{\min}$ and $\gamma_{\max}$ assuming a uniformly spaced spectrum in $[-1,1]$. 
These initial estimates are then refined during the evolution: every $L$ steps, we measure $\braket{\phi_i|H|\phi_i}$ for $i=1,\ldots,r$ to estimate the target eigenvalues and update the spectral gaps. 
For stability, we impose a lower bound of $10^{-3}$ on the estimated gaps and fix the spectral width at $2$, assuming $\|H\|_{op}\le 1$ where $\|\cdot\|_{op}$ denotes the operator norm. 
Since the $(r+1)$-th eigenvalue is unavailable, we also estimate the final gap using the smallest measured gap.

\begin{figure}[t]
  \centering
  \includegraphics[width=\columnwidth]{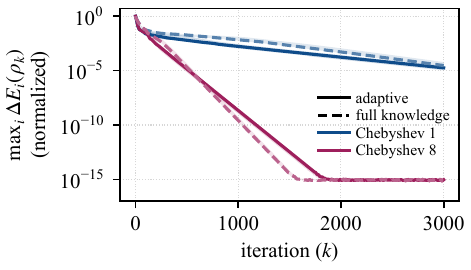}
  \caption{\textbf{Adaptive strategy retrieves target states without prior knowledge of the spectrum.} Adaptive (solid) and full-knowledge (dashed) strategies for the 10-qubit transverse-field Ising chain
  with an unknown scale factor $c$ tracking the $r = 10$ lowest eigenstates. The adaptive strategy estimates the weights
  and step sizes from the measured energies; the full-knowledge run uses the exact spectrum. We use Chebyshev scheduling with $m =1$ and $8$ (blue and pink, respectively) and order 4 product formulas. Lines are medians and bands interquartile
  ranges over 21 sets of Haar-random initial states.}
  \label{fig:adaptive}
\end{figure}

We numerically evaluate the adaptive strategy using the open-chain 10-qubit TFIM, targeting its $r=10$ lowest-energy eigenstates. 
We rescale the spectrum to $[-1,1]$ and multiply it by a randomly sampled factor $c\sim \mathcal{U}[0.8,1]$, using only the assumption $|\lambda_i|\le 1$. 
Each run starts from $r$ Haar-random orthonormal states, and we consider 21 independent initial sets. 
The first 100 iterations use a constant step size $(m=1)$. For comparison, we also simulate DBFs using the optimal weights and step sizes obtained from the exact spectral gaps. 
Figure~\ref{fig:adaptive} shows the median and interquartile range across the 21 runs.
The adaptive strategy achieves nearly the same convergence rate as the strategy with full spectral information.
With an 8-periodic Chebyshev schedule, convergence is nearly eight times faster, reaching machine precision in approximately 1700 iterations.

Overall, our numerical results demonstrate that the DBF can converge to the target eigenstates without prior spectral information by adaptively updating the weights and step sizes. 
Nevertheless, further investigation is needed to fully assess the performance and robustness of the adaptive strategy. 
In particular, the effects of stability constraints, the trade-off between product-formula order and Chebyshev scheduling period, and errors in the estimated values of $\braket{\phi_i|H|\phi_i}$ warrant systematic analysis.
These questions motivate further numerical and theoretical studies, while our results provide encouraging evidence for adaptive scheduling.

\section{Discussion}

In this work, we develop a framework based on a double-bracket flow (DBF) for the simultaneous computation of multiple excited states, without requiring prior preparation of lower-energy eigenstates. We establish a necessary condition for convergence to the target excited states and a sufficient condition for exponential convergence. The theoretical results are further supported by numerical simulations. 
In addition, we establish the optimality of Chebyshev scheduling in the near-target regime and validate its performance numerically. These results establish theoretical guarantees for the simultaneous computation of multiple excited states and provide a systematic approach to accelerating convergence.

Several directions for future research arise from the present work. First, a systematic comparison of the proposed framework with existing methods for excited-state computation, particularly Krylov subspace methods~\cite{lanczos1950,davidson1975theiterative,knyazev2001toward} and DMRG~\cite{white1992densitymatrix,schollwock2011thedensitymatrix,dorando2007targeted}, would be of considerable interest. Although these methods share the objective of computing multiple eigenstates, they rely on fundamentally different computational approaches. A comparative analysis of computational complexity, convergence, and sensitivity to initial conditions would help elucidate the relative advantages and limitations of the DBF framework.

Second, the potential implementation of the proposed framework on quantum computers merits further investigation. In particular, a comprehensive resource analysis is required to determine whether the avoidance of prior lower-energy eigenstate preparation can offset the cost of implementing the DBF. Such an analysis should account for circuit depth~\cite{gluza2026double,son2025quantum,kolotouros2026quantum}, gate complexity, and the product-formula order required to achieve a prescribed accuracy. Comparisons with existing quantum algorithms for excited-state preparation would further clarify the practical prospects of the proposed approach.

Third, further theoretical investigation is warranted to obtain sharper convergence guarantees. Our numerical results indicate that exponential convergence may persist even when the initial energy excess exceeds the sufficient threshold established in this work. A more precise characterization of the basin of attraction of the target fixed point, together with less restrictive sufficient conditions for exponential convergence, therefore constitutes a natural direction for future research. Further extensions could also address Hamiltonians with degenerate or closely spaced eigenvalues, as well as the preparation of arbitrary subsets of eigenstates rather than exclusively the lowest-energy eigenstates.

\section*{Methods}
\appendix

\newcommand{\EMprefix}{EM}
\newtheorem{theoremEM}{Theorem}
\setcounter{equation}{0}
\setcounter{theoremEM}{0}
\renewcommand{\thetheoremEM}{\EMprefix\arabic{theoremEM}}
\renewcommand{\theequation}{\EMprefix\arabic{equation}}

\paragraph*{\textbf{Algorithms for discretized and product-formula DBFs}}
Here, we provide detailed descriptions of the discretized and product-formula DBF algorithms in Algorithms~\ref{alg:dbf} and~\ref{alg:dbf-pf}, respectively.
\begin{algorithm}[h]
   \caption{Discretized DBF}
   \label{alg:dbf}
  \begin{algorithmic}[1]
     \REQUIRE Hamiltonian $H$; reference state $\rho_{\mathrm{ref}} = \sum_{j=1}^{r} p_j |{\phi_j}\rangle\langle{\phi_j}|$
       with orthonormal states $\ket{\phi_j}$ and weights $p_1 > p_2 > \dots > p_r > 0$; number of iterations $K$;
       steps $s_0, \dots, s_{K-1}$ 
     \ENSURE Approximations $\ket{\phi_j^{(K)}}$ of the $r$ lowest eigenstates of $H$ for $j = 1, \dots, r$
     \STATE $U_0 \gets I$
     \STATE $\rho_0 \gets \rho_{\mathrm{ref}}$
    \FOR{$k = 0, \dots, K-1$}
      \STATE $V_k \gets e^{s_k [\rho_k, H]}$
       \STATE $\rho_{k+1} \gets V_{k}\, \rho_{k}\, V_{k}^\dagger$
      \STATE $U_{k+1} \gets V_k U_{k}$
    \ENDFOR
     \STATE \textbf{Return} $\ket{\phi_j^{(K)}} = U_K \ket{\phi_j}$ for $j = 1, \dots, r$
   \end{algorithmic}
 \end{algorithm}

\begin{algorithm}[h]
  \caption{Product-formula DBF}
  \label{alg:dbf-pf}
  \begin{algorithmic}[1]
    \REQUIRE Hamiltonian $H$; reference state $\rho_{\mathrm{ref}} = \sum_{j=1}^{r} p_j |{\phi_j}\rangle\langle{\phi_j}|$
      with orthonormal states $\ket{\phi_j}$ and weights $p_1 > p_2 > \dots > p_r > 0$; number of iterations $K$;
      steps $s_0, \dots, s_{K-1}$; order $\ell$ of the product formula $\PC_\ell(\rho, H, s)$ for
      $e^{s[\rho, H]}$, a product of evolutions $e^{-i \tau H}$ and $e^{i \tau \rho}$ with
      $|{\tau}| \propto \sqrt{s}$
    \ENSURE Approximations $\ket{\phi^{(K)}_j}$ of the $r$ lowest eigenstates of $H$ for $j = 1, \dots, r$
    \STATE $U_0 \gets I$
    \STATE $\rho_0\gets \rho{}$
    \FOR{$k = 0, \dots, K-1$}
      \STATE $V_k \gets \PC_\ell(\rho_k, H, s_k)$
      \STATE $\rho_{k+1} \gets  V_k\rho_{k}V_k^\dagger$
      \STATE $U_{k+1} \gets V_k U_{k}$
    \ENDFOR
    \STATE \textbf{Return} $\ket{\phi^{(K)}_j} = U_K \ket{\phi_j}$ for $j = 1, \dots, r$
  \end{algorithmic}
\end{algorithm}

\begin{figure*}[t]
  \centering
  \includegraphics[width=0.9\textwidth]{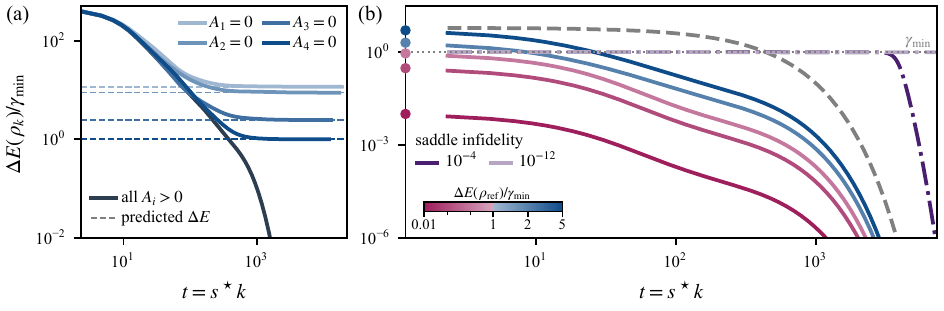}
  \caption{\textbf{The role of the initial state.} 
  The first 4 eigen energies of random 10-qubit TFIMs are tracked with  inverse-gap
  weights, optimal step $s^\star$ and fourth-order product formula~\cite{chen2022efficient}. $\Delta E(\rho_{k})$ is the excess energy after $k$
  iterations and $t = s^\star k$ the flow time. \textbf{(a):} $\Delta E(\rho_{k})/\gamma_{\min}$ for one Hamiltonian from
  Haar-random initial states (all $A_i > 0$) and from the same states with $\ket{\lambda_i}$ removed ($A_i = 0$). Dashed lines show the predicted
  confined fixed points. \textbf{(b):} $\Delta E(\rho_{k})/\gamma_{\min}$ for the same Hamiltonian from warm starts (pink,
  $\Delta E(\rho_{\mathrm{ref}}) < \gamma_{\min}$), cold starts (blue, $\Delta E(\rho_{\mathrm{ref}}) > \gamma_{\min}$) and near-saddle starts (purple). Dots on the left edge mark $\Delta E(\rho_{\mathrm{ref}})$;
  the dashed gray line is a decay $e^{-2\gamma_{\min} t}$. The
  legend gives $\Delta E(\rho_{\mathrm{ref}})/\gamma_{\min}$ of each curve, or approximate infidelity to the saddle, for the near-saddle curves.   
  }
  \label{fig:mainNumerics}
\end{figure*}

\medskip
\paragraph*{ \textbf{ Preparing a single excited state with degenerate weights.}---}

When the goal is to prepare only the $r$-th eigenstate $\ket{\lambda_r}$, distinct weights $p_i$ for the lower $r-1$ eigenstates are unnecessary, as they need not be individually resolved. 
Therefore, we can choose
\begin{equation}
\label{eq:EM-weights}
p_1=\dots=p_{r-1}=\frac{2}{2r-1},\qquad p_r=\frac{1}{2r-1}.
\end{equation}
The reference states with degenerate weights are collectively mapped onto the subspace of the $r-1$ lowest-energy eigenstates, while the remaining reference state is assigned to the target eigenstate $\ket{\lambda_r}$.
Consequently, the convergence analysis effectively involves only two subspaces: the lower-energy subspace and the target eigenspace.
In addition, since the weights are degenerate, some $\gamma_{j,j+1}$ vanish and $\gamma_{\min}$ is given by the smallest nonzero $\gamma_{j,j+1}$.
For the weights in Eq.~\eqref{eq:EM-weights}, this yields $\gamma_{\min}=\delta\lambda_{\min}/(2r-1)$ with $\delta\lambda_{\min}=\min\{\lambda_r-\lambda_{r-1},\,\lambda_{r+1}-\lambda_r\}$, which depends only on the two spectral gaps adjacent to the target level.

The following result establishes an exponential convergence rate for this approach in terms of the infidelity $1-F_r(t)$.
The proof is provided in Appendix~\ref{app:two-block-single}.

\begin{theoremEM}[Preparing a single excited state with degenerate weights, informal]
\label{theorem:single}
Let the state $\rho$ evolve according to one of the three flows: the continuous-time DBF, the discretized DBF, or the product-formula flow, defined in Eqs.~\eqref{eq:C},~\eqref{eq:D} and~\eqref{eq:PF}, respectively.
Using the weights of $\rho_{\mathrm{ref}}$ in Eq.~\eqref{eq:EM-weights}, we obtain
\begin{equation}
\label{eq:EM-kappa}
\gamma_{\min}=\frac{\delta\lambda_{\min}}{2r-1},\qquad \kappa=\frac{2\,\Omega}{\delta\lambda_{\min}}.
\end{equation}
If $\Delta E(\rho_{\mathrm{ref}})<\gamma_{\min}$, the $r$th state obeys $1-F_r(t)\le\Delta E(\rho(t))/\gamma_{\min}\le c_0\,e^{-2c\gamma_{\min}t}$, and the span of the other $r-1$ states converges to the span of $\{\ket{\lambda_i}\}_{i=1}^{r-1}$.
\end{theoremEM}

The ratio defined in Eq.~\eqref{eq:EM-kappa} is independent of $r$. In contrast, the corresponding ratio grows at least quadratically with $r$ when distinct weights are used to resolve all $r$ states. The trade-off is that the lower eigenstates are prepared only as a subspace. Theorem~\ref{theorem:single} is a special case of Theorem~\ref{thmA:equal} in Appendix~\ref{app:two-block}, which allows arbitrary equal weights.

\medskip
\paragraph*{\textbf{Numerical validation of theoretical findings.}---}

We perform numerical simulations to validate the theoretical results established in the main text. Specifically, we investigate two questions: (i) whether the necessary condition in Eq.~\eqref{eq:A_i} is satisfied and (ii) whether the sufficient condition in Theorem~\ref{theorem:convergence-rate} correctly identifies a regime of exponential convergence. 

To this end, we focus on the product-formula DBF, which necessarily entails both discretization and product-formula approximation errors.
As a model for our numerical simulations, we consider a $10$-qubit transverse field Ising model (TFIM)~\cite{pfeuty1970} with open chain boundary conditions, described by $
H_{\text{TFIM}} = -\sum_i J_iZ_i Z_{i+1} - \sum_i h_iX_i  
$, where $J_i$ and $h_i$ denote the interaction strengths and transverse fields, respectively.
Moreover, we choose the weights $p_i$ such that
$
p_i-p_{i+1} \propto (\lambda_{i+1}-\lambda_{i})^{-1}
$ and use  $s^\star= 2/(\gamma_{\max}+ \gamma_{\min})$.
Note that, as shown in Appendices~\ref{app:optimal_step_schedules} and~\ref{app:weights}, the choice of weights $p_i$ and the fixed step size $s^\star$ are optimal near a fixed point.

First, we numerically examine whether the flow converges to a fixed point for initial states satisfying the necessary overlap condition. Specifically, this condition requires that the projector onto the initial subspace, $P_r = \sum_i^r \ket{\phi_i^{(0)}}\bra{\phi_i^{(0)}}$, has nonzero overlap with each of the lowest $r$ energy eigenstates.
This is quantified by $A_i$ in Eq.~\eqref{eq:A_i}: convergence to the target requires $A_i>0$, whereas $A_i=0$ precludes it.
Figure~\ref{fig:mainNumerics}(a) shows that, for Haar-random initial states with $A_i>0$ (black curves), the energy excess $\Delta E$ decays to zero, indicating convergence to the target fixed point. In contrast, for initial states constructed by removing the component along the corresponding target eigenstate from Haar-random states, such that $A_i=0$, the energy excess stagnates; this shows that the flow fails to converge to the target fixed point. These results support the necessary condition and highlight the importance of choosing initial states that satisfy the necessary overlap condition.

Next, we numerically investigate how the convergence rate depends on the initial states. Theorem~\ref{theorem:convergence-rate} guarantees exponential convergence at a rate of $2\gamma_{\min}$ when the initial energy excess is below $\gamma_{\min}$.
To this end, we engineer initial states by rotating the target states in randomly chosen directions, allowing us to control the initial energy excess relative to $\gamma_{\min}$ through the rotation angle.
Figure~\ref{fig:mainNumerics}(b) shows that two initial states below the threshold (pink curves) converge at rates faster than $2\gamma_{\min}$, which is consistent with the theoretical guarantee. 

We observe the convergence of three initial states whose energy excesses are up to five times larger than the threshold (blue curves).
These states also converge faster than $2\gamma_{\min}$, despite substantially violating the sufficient condition. This shows that the condition in Theorem~\ref{theorem:convergence-rate} is not necessary for exponential convergence. In contrast, when the initial states are chosen close to saddle points (purple curves), convergence is significantly slower due to the small energy gradients in their vicinity. 
Nevertheless, at later stages of the evolution, the convergence rates reach the theoretical bound, suggesting that the slow convergence is primarily associated with proximity to saddle points.
These results further confirm that the sufficient condition $\Delta E(\rho_{\mathrm{ref}})<\gamma_{\min}$ is conservative, implying that, in practice, the method exhibits exponential convergence over a substantially broader range of initial conditions. A more detailed numerical study of the role of initial states in convergence rate is provided in Appendix~\ref{app:num_initial states}.

\medskip
\paragraph*{\textbf{ Proof sketch for Lemma~\ref{lemma:chebyshev}.}---} 
Under the linearization of the DBF around the target fixed point $\rho^{*}$, the off-diagonal elements, $|(\rho-\rho^{*})_{ij}|$, are suppressed or amplified by the multiplicative factor $|1-s\gamma_{ij}|$ at each iteration. 
Therefore, over a cycle of $m$ steps the contraction factor for each off-diagonal element is given by $P(\gamma_{ij}) = \prod_{k=1}^m |1-s_k\gamma_{ij}|$. 
Moreover, the maximum contraction factor over all off-diagonal elements determines the local convergence rate of both the energy and the fidelities with respect to the target eigenstates.
Consequently, we seek a schedule that minimizes the maximum per-cycle contraction factor. Equivalently, this amounts to finding a degree-$m$ polynomial $P^\star$ that minimizes its infinity norm over the interval $[\gamma_{\min},\gamma_{\max}]$, subject to $P^\star(0)=1$.
We show that the unique minimizer is a suitably shifted and normalized Chebyshev polynomial of the first kind of degree $m$ on the interval $[\gamma_{\min},\gamma_{\max}]$. 
The optimal step sizes are therefore given by the reciprocals of its roots. 
See Appendix~\ref{app:chebyshev} for the complete proof.

\medskip
\medskip
\textit{Note added.}
While finishing this manuscript, Ref.~\cite{rahman2026doublebracket} appeared which independently employs double-bracket flows for the preparation of multiple excited states and investigates their convergence properties and scheduling. Our work shares this general approach but provides a comprehensive convergence analysis, including both necessary conditions for convergence and sufficient conditions for exponential convergence. We also investigate a broader range of scheduling strategies and support our theoretical findings with numerical simulations.

\medskip
\medskip
\textit{Acknowledgments.}
RAS and MS acknowledge support from the Swiss National Science Foundation [grant number 200021-219329].
ZH acknowledges support from the Sandoz Family Foundation-Monique de Meuron program for Academic Promotion.

\medskip
\medskip
\textit{AI usage statement.}
The authors acknowledge the use of generative AI tools
(Claude, Claude Code), through the Claude Team Plan for Scientist. They were used to improve existing results, write scientific code, iterate on early variant of the manuscript, and to proof read. All AI-assisted content was reviewed, verified, and revised by the authors,
who take full responsibility for the content of this work.

\bibliography{bibliography}

\clearpage
\newpage
\onecolumngrid
\part{Appendix}

\parttoc

\appendix

\setcounter{secnumdepth}{2}

\setcounter{equation}{0}
\renewcommand{\theequation}{\thesection\arabic{equation}}

\renewcommand{\thetheorem}{\Alph{section}.\arabic{theorem}}
\renewcommand{\thelemma}{\Alph{section}.\arabic{lemma}}
\renewcommand{\theprop}{\Alph{section}.\arabic{prop}}

\newtheorem{theoremA}{Theorem}[section]
\newtheorem{propositionA}[theoremA]{Proposition}
\newtheorem{lemmaA}[theoremA]{Lemma}
\newtheorem{definitionA}[theoremA]{Definition}
\newtheorem{corollaryA}[theoremA]{Corollary}
\newtheorem{remarkA}[theoremA]{Remark}

\renewcommand{\thetheoremA}{\Alph{section}.\arabic{theoremA}}

\newpage

\section{Related works}
\label{sec:known-methods}
In this section, we review the existing methods on for ground state preparation and some of their limitations.

\subsubsection{Variational Quantum Deflation (VQD) Algorithm} 
VQD~\cite{Higgott2019variationalquantum} finds excited states sequentially by modifying the VQE cost function to penalize overlap with previously found lower-energy states. The modified cost function is given by the following. 
\begin{equation}
    \mathcal{L}(\vec{\theta}) = \langle \psi(\vec{\theta}) | H | \psi(\vec{\theta}) \rangle + \sum_{i=1}^{r-1} \beta_i |\langle \psi_i | \psi(\vec{\theta}) \rangle|^2,    
\end{equation}
where $\vec{\theta}$ is the vector of parameters, $ | \psi(\vec{\theta}) \rangle $ is a parameterized state produced by the parametrized quantum circuit, and  $| \psi_i \rangle $'s are previously found lower energy states.  
The scalar $\beta_i$ needs to be chosen large enough to push the energy of the penalized states above the target excited state, i.e., 
$\beta_i > E_{r} - E_{i}$ 
with $E_i = \langle \psi_i | H | \psi_i \rangle$. 
    
In other words, standard VQE always finds the lowest energy state. By adding the penalty term $\beta_i |\langle \psi_i | \psi(\vec{\theta}) \rangle|^2$, we are mathematically constructing an effective Hamiltonian $H_{\rm{eff}} = H + \sum_i \beta_i |\psi_i\rangle\langle\psi_i|$. If our trial state $|\psi(\vec{\theta})\rangle$ starts to have large overlaps with the lower states, the loss function is modified by increasing the penalty terms and the classical optimizer is forced to search elsewhere.

\subsubsection{Subspace-Search Variational Quantum Eigensolver (SSVQE)} 
SSVQE~\cite{Nakanishi2019Subspacesearch} prepares a set of $k$ mutually orthogonal input states and evolves them through the same parameterized unitary $U(\vec{\theta})$ to find the $k$-th excited state. By forcing orthogonality at the input, the output states remain orthogonal, converging to the lowest eigenvector when minimizing a weighted sum of their energies. Its weighted cost function is the variational counterpart of the energy $E(\rho)$ of our proposed DBF flow.

There is another similar method called Subspace Search Quantum Imaginary Time Evolution (SSQITE)~\cite{Cianci2024subspacesearchquantumimaginarytimeevolution} which calculates excited states using quantum devices by integrating key elements of the subspace search variational quantum eigensolver (SSVQE) and the variational quantum imaginary time evolution (VarQITE) method.

\subsubsection{Folded Spectrum VQE} 
These methods~\cite{Tazi2024foldedspectrumvqe}, target excited states directly by redefining the variational objective. Rather than minimizing the total energy, they minimize the energy variance around a target energy shift $\omega$, which is called the fold point and is chosen to be close to the energy of the desired eigenstate. Then the folded Hamiltonian is given by 
\begin{equation}
    {H}_{\rm{FS}} = \left(H - \omega \mathbb{I}\right)^2.
\end{equation}

The classical optimizer updates the parameters $\vec{\theta}$ of the parametrized circuit to minimize the expectation value of this squared operator:
\begin{equation}
    \mathcal{L}(\vec{\theta}) = \langle \psi(\vec{\theta})| \left(H - \omega \mathbb{I}\right)^2 | \psi(\vec{\theta}) \rangle = \langle H^2 \rangle - 2\omega \langle H \rangle + \omega^2.
\end{equation}

The main idea is that, the eigenvalues of the folded Hamiltonian $H_{\rm{FS}}$ are exactly $(E_n - \omega)^2$ with $E_n$'s being the energy eigenvalues of $H$. Hence, the lowest eigenvalue of $H_{\rm{FS}}$ occurs at the true eigenstate whose energy $E_n$ is nearest to $\omega$. This mathematically \textit{folds} the spectrum at $\omega$, transforming an interior excited state into the absolute ground state of the folded Hamiltonian.

\subsubsection{Quantum Subspace Expansion (QSE)} 
Quantum Subspace Expansion (QSE)~\cite{McClean2017hybridquantumclassicalhierarchy} is a hybrid quantum-classical method designed to extract excited states and mitigate errors around a reference quantum state. 
The reference state $|\psi\rangle$ is the base state generated by the quantum circuit before the subspace expansion operators are applied. Practically, it can the approximate ground state obtained from a converged Variational Quantum Eigensolver (VQE) method. 
Instead of re-optimizing a variational circuit for every excited state, QSE takes a converged ground-state approximation and generates a local subspace around it by applying a predefined set of operators. The Hamiltonian is then projected into this expanded, non-orthogonal subspace and diagonalized classically in that subspace.

Mathematically, given a reference state $|\psi\rangle$ and a set of excitation operators $\{O_i\}$ (typically low-weight Pauli strings or fermion operators), the expanded basis is defined as:
\begin{equation}
    |\psi_i\rangle = O_i |\psi\rangle.
\end{equation}
The performance of the QSE depends on the reference state. The algorithm assumes that $|\psi\rangle$ already captures the macroscopic entanglement and correlation structure of the target Hamiltonian and, therefore QSE cannot repair a reference state that fundamentally misses the required symmetries, and it only explores the neighborhood around it.

To find the excited states, the Schrödinger equation is projected into the subspace spanned by $\{|\psi_i\rangle\}$.
The matrix elements of the Hamiltonian $H'$ and the overlap matrix $S$ (required since the basis states are non-orthogonal) must be measured on the quantum computer:
\begin{align}
    H_{ij} &= \langle\psi_i|H|\psi_j\rangle = \langle\psi|O_i^\dagger H O_j|\psi\rangle, \\
    S_{ij} &= \langle\psi_i|\psi_j\rangle = \langle\psi|O_i^\dagger O_j|\psi\rangle.
\end{align}
Once $H$ and $S$ are reconstructed classically, the generalized eigenvalue equation is solved:
\begin{equation}
    H \vec{c}_k = E_k S \vec{c}_k \,,
\end{equation}
where the eigenvalues $E_k$ are the estimated state energies, and the eigenvectors $\vec{c}_k$ provide the linear combinations of the subspace basis vectors.

\subsubsection{Quantum Krylov Subspace Methods} 
Quantum Krylov Subspace Methods~\cite{Cortes2022quantumkrylovsubspacealgorithms} estimate ground and excited-state energies by projecting a system's Hamiltonian onto a non-orthogonal subspace generated by a quantum computer. Instead of relying on non-linear parameter optimization, quantum Krylov subspace methods construct a basis set $\{|\psi_k\rangle\} = \{f({H})^k |\psi_0\rangle\}$, where $f({H})$ is typically the real-time evolution operator $e^{-i{H}\tau}$ or imaginary-time propagator $e^{-\tau{H}}$. The quantum hardware is utilized strictly to measure the off-diagonal matrix elements of the Hamiltonian, $H'_{ij} = \langle \psi_i|H| \psi_j\rangle$, and the overlap matrix, $S_{ij} = \langle \psi_i | \psi_j\rangle$. Finally, a classical generalized eigenvalue problem, $H'\mathbf{c} = E{S}\mathbf{c}$, is solved to extract the excited state energies $E$.

\subsubsection{Quantum Imaginary Time Evolution (QITE)}
Quantum Imaginary Time Evolution (QITE) is designed to project a system toward its ground state by approximating the non-unitary imaginary time evolution operator, $e^{-\tau {H}}$, using a sequence of local unitary circuits on a quantum computer. To adapt this for excited state preparation, QITE is typically hybridized with a Quantum Lanczos (QLanczos) framework to construct a Krylov subspace, or it employs deflation techniques to sequentially penalize and project out lower-energy states from the Hamiltonian~\cite{Motta2020determiningeigenstatesandthermalstates}.

\subsubsection{Quantum Phase Estimation (QPE) for Excited States}
In this method, for finding the excited state, the circuit architecture remains identical to ground-state estimation, shifting the entire complexity burden to the initial state preparation.

Let $H$ be the system Hamiltonian with unknown eigenstates $|E_k\rangle$ and energies $E_k$. The time-evolution operator $U = e^{-iHt}$ is applied via controlled-$U^{2^j}$ gates onto a target register initialized in a non-eigenstate trial wavefunction $|\Phi\rangle$.
The trial state is decomposed in the exact eigenbasis of $H$:
\begin{equation}
    |\Phi\rangle = \sum_{k} c_k |E_k\rangle \quad,
\end{equation}
where $c_k = \langle E_k | \Phi \rangle$. Then, conditional time evolution encodes the phases $\phi_k = \frac{E_k t}{2\pi}$ onto the auxiliary clock register, generating the entangled state:
\begin{equation}
    |\Psi_{\rm{inter}}\rangle = \sum_{k} c_k |\phi_k\rangle \otimes |E_k\rangle\,.
\end{equation}
Applying the inverse Quantum Fourier Transform ($\rm{QFT}^\dagger$) maps the phase states $|\phi_k\rangle$ to their binary bitstring representations $|\tilde{\phi}_k\rangle$. Upon measurement of the clock register, the total wavefunction collapses.

By the Born rule, the probability $P(E_n)$ of successfully measuring the specific excited-state energy $E_n$ is strictly bounded by the spectral overlap of the initial trial state:
\begin{equation}
    P(E_n) = |c_n|^2 = |\langle E_n | \Phi \rangle|^2
\end{equation}
Simultaneously, a successful measurement projects the target register into the pure excited eigenstate $|E_n\rangle$.

For strongly correlated or high-lying excited states, classical approximations yield wavefunctions whose overlaps collapse exponentially with system size ($|c_n|^2 \to 0$). This scales the sampling complexity as $\mathcal{O}(1/|c_n|^2)$. Additionally, high-energy regimes exhibit dense eigenstate clustering. Resolving a target excited state separated from an adjacent state by a narrow gap $\Delta E$ requires a coherent simulation runtime scaling as $\mathcal{O}(1/\Delta E)$, mandating deep circuit depths and low logical error rates.

\subsubsection{Double-bracket imaginary-time evolution with a modified Hamiltonian}
A direct way to use double-bracket flows for excited states is to apply double-bracket quantum imaginary-time evolution, which prepares ground states, to one of the modified Hamiltonians above. For example, for the folded-spectrum method one prepares the ground state of $H_{\rm FS}=(H-\omega\mathbb{I})^2$. This approach requires no new analysis. However, it inherits the cost of the modified Hamiltonian: squaring an operator with $L$ terms of weight $k$ gives of order $L^2$ terms of weight $2k$. Moreover, $\omega$ requires an estimate of the target energy, unlike our proposed method, which uses $H$ itself and no spectral information.

\section{Problem setting and notation}
\label{app:setting}

This appendix introduces the definitions of the main text and the elementary facts that the proofs in the following appendices use repeatedly. This section contains no new results.

\medskip

\textbf{Hamiltonian and states.} The Hamiltonian is $H=\sum_{a=1}^{d}\lambda_a\ket{\lambda_a}\!\bra{\lambda_a}$ with $\lambda_1<\lambda_2<\dots<\lambda_d$, and $\Omega=\lambda_d-\lambda_1$ is its spectral width. The state $\rho$ has the form of Eq.~\eqref{eq:rho}, that is, $\rho=\sum_{i=1}^{r}p_i\ket{\phi_i}\!\bra{\phi_i}$ with orthonormal states $\ket{\phi_i}$ and weights $p_1>p_2>\dots>p_r>0$ that sum to one. We assume $r<d$ and set $p_i=0$ for $i>r$. Whenever we need a basis of the whole space, we complete the states by orthonormal vectors $\ket{\phi_{r+1}},\dots,\ket{\phi_d}$ that span the kernel of $\rho$. No result depends on how these vectors are chosen. We write
\begin{equation}
\label{eq:app-gaps}
\delta\lambda_{j+1}=\lambda_{j+1}-\lambda_j>0\quad(1\le j\le d-1),\qquad \delta p_j=p_j-p_{j+1}>0\quad(1\le j\le r),
\end{equation}
for the gaps of the energies and of the weights, where $\delta p_r=p_r$.

\medskip

\textbf{Energy, the target state, and the energy excess.} The energy of a state is $E(\rho)=\mathrm{Tr}(\rho H)$. The target state is $\rho^{*}=\sum_{i=1}^{r}p_i\ket{\lambda_i}\!\bra{\lambda_i}$ of Eq.~\eqref{eq:target}, with the energy $E^{*}=\mathrm{Tr}(\rho^{*}H)=\sum_{i=1}^{r}p_i\lambda_i$. As in the main text, the energy excess of a state $\rho$ is
\begin{equation}
\label{eq:app-dE}
\Delta E(\rho)=E(\rho)-E^{*} .
\end{equation}
We always write the state in the argument of $\Delta E$. That is, $\Delta E(\rho_{\mathrm{ref}})$ is the energy excess of the reference state, $\Delta E(\rho(t))$ is the energy excess at the flow time $t$, and $\Delta E(\rho_k)$ is the energy excess after $k$ steps.

\medskip

\textbf{Fidelities.} The fidelity of any state $\ket{\psi}$ with the $i$th target state is  $f_i(\ket{\psi})=|\langle\lambda_i|\psi\rangle|^2$. Hence, for the state $\rho$ with the form of Eq.~\eqref{eq:rho}, the fidelity of the $i$th state $\ket{\phi_i}$ is given by 
\begin{equation}
    F_i (\rho) = f_i (\ket{\phi_i}) = |\langle\lambda_i|\phi_i\rangle|^2 \,.
\end{equation}

\medskip

\textbf{Energy scales.} For two levels $i<j$ we use the quantities of Eq.~\eqref{eq:gamma-ij},
\begin{equation}
\label{eq:app-gamma}
\gamma_{ij}=(p_i-p_j)(\lambda_j-\lambda_i),\qquad \gamma_{\max}=\gamma_{1d}=p_1\Omega,\qquad \gamma_{\min}=\min_{1\le j\le r}\gamma_{j,j+1}=\min_{1\le j\le r}\delta p_j\,\delta\lambda_{j+1},
\end{equation}
and $\kappa=\gamma_{\max}/\gamma_{\min}$. For every pair $i<j$ with $i\le r$ we have
\begin{equation}
\label{eq:gamma-range}
\gamma_{\min}\le\gamma_{ij}\le\gamma_{\max}.
\end{equation}
Indeed, $p_i-p_j\ge p_i-p_{i+1}$ and $\lambda_j-\lambda_i\ge\lambda_{i+1}-\lambda_i$ give $\gamma_{ij}\ge\gamma_{i,i+1}\ge\gamma_{\min}$, and $p_i-p_j\le p_1$ and $\lambda_j-\lambda_i\le\Omega$ give $\gamma_{ij}\le\gamma_{\max}$.

\medskip

\textbf{The three flows.} We write $W=[\rho,H]$, which is anti-Hermitian. The continuous-time DBF of Eq.~\eqref{eq:C} is $\partial_t\rho(t)=[W(t),\rho(t)]$. The discretized DBF of Eq.~\eqref{eq:D} is $\rho_{k+1}=e^{s_kW_k}\rho_ke^{-s_kW_k}$ with $W_k=[\rho_k,H]$ and step sizes $s_k>0$. The product-formula DBF of Eq.~\eqref{eq:PF} is $\rho_{k+1}=V_k\rho_kV_k^\dagger$ with $V_k=e^{i\tau_kH}e^{i\theta_k\rho_k}e^{-i\tau_kH}e^{-i\theta_k\rho_k}$, angles $\tau_k,\theta_k>0$, and step size $s_k=\tau_k\theta_k$. The last factor of $V_k$ commutes with $\rho_k$. It therefore does not affect $\rho_{k+1}$, and it only multiplies the state $\ket{\phi_i^{(k)}}$ by the phase $e^{-i\theta_kp_i}$. In the proofs we thus use $V_k=e^{i\tau_kH}e^{i\theta_k\rho_k}e^{-i\tau_kH}$. All three flows rotate $\rho$ by a unitary. Hence the weights $p_i$ never change, the states $\ket{\phi_i}$ remain orthonormal, and $\rho$ keeps the form of Eq.~\eqref{eq:rho} at all times.

\medskip

\textbf{Flow time, the state and the fidelities at flow time $t$.} For the continuous-time DBF, the flow time $t$ is the evolution time, and $\rho(t)=\sum_{i=1}^{r}p_i\ket{\phi_i(t)}\!\bra{\phi_i(t)}$ is the solution of Eq.~\eqref{eq:C}. For the discretized and the product-formula DBF, the flow time after $K$ steps is
\begin{equation}
\label{eq:app-flowtime}
t=\sum_{k=0}^{K-1}s_k ,
\end{equation}
and the state at this flow time is the state after $K$ steps, $\rho(t)=\rho_K=\sum_{i=1}^{r}p_i\ket{\phi_i^{(K)}}\bra{\phi_i^{(K)}}$. With this convention, a statement about $\rho(t)$ and $\Delta E(\rho(t))$ applies to all three flows, and for the discretized and the product-formula DBF, it reads
\begin{equation}
\label{eq:app-dE-discrete}
\Delta E(\rho(t))=\Delta E(\rho_K)=E(\rho_K)-E^{*}\qquad\textup{for}\qquad t=\sum_{k=0}^{K-1}s_k .
\end{equation}
The fidelity of the $i$th state at the flow time $t$ is $F_i(t)=f_i(\ket{\phi_i(t)})$ for the continuous-time DBF and $F_i(t)=f_i(\ket{\phi_i^{(K)}})$ for the discretized and the product-formula DBF.

\medskip

\textbf{Step-size conditions.} The condition of Eq.~\eqref{eq:step-conditions} is $s_k<\bar{s}$, with
\begin{subequations}
\label{eq:app-sbar}
\begin{align}
    \textup{discretized DBF:} \;\; & \bar{s} =\frac{2}{\gamma_{\max}} \, ,
    \label{eq:app-sbar-D}
    \\
    \textup{product-formula DBF with \,} \theta_k/\tau_k=\Omega/p_1: \; \; &\bar{s}=\frac{12}{7\gamma_{\max}} \, ,
    \label{eq:app-sbar-PF}
    \\
    \textup{product-formula DBF with \,} \theta_k =\tau_k: \; \; &\bar{s}=\frac{12}{7\max(\Omega, 1)^2} \, .
    \label{eq:app-sbar-PFeq}
\end{align}
\end{subequations}
The first two values are those of Eq.~\eqref{eq:step-conditions}. The third value covers the product-formula DBF with equal angles, $\theta_k=\tau_k=\sqrt{s_k}$. Theorems~\ref{thmA:energy-D} and~\ref{thmA:energy-PF} explain where these values come from.

\medskip

\textbf{Norm and two bounds.} The Hilbert-Schmidt norm of an operator $X$ is $\|X\|_{HS}=[\mathrm{Tr}(X^\dagger X)]^{1/2}$. It is invariant under unitary rotations, $\|UXU^\dagger\|_{HS}=\|X\|_{HS}$, it satisfies $|\mathrm{Tr}(XY)|\le\|X\|_{HS}\|Y\|_{HS}$, and the norm of an integral is at most the integral of the norm. For every operator $X$ and every state $\rho$ of the form of Eq.~\eqref{eq:rho},
\begin{equation}
\label{eq:comm}
\|[H,X]\|_{HS}\le\Omega\,\|X\|_{HS},\qquad \|[\rho,X]\|_{HS}\le p_1\,\|X\|_{HS}.
\end{equation}
To see the first bound, we write $X_{ab}=\bra{\lambda_a}X\ket{\lambda_b}$. Then $[H,X]_{ab}=(\lambda_a-\lambda_b)X_{ab}$, so that $\|[H,X]\|_{HS}^2=\sum_{a,b}(\lambda_a-\lambda_b)^2|X_{ab}|^2\le\Omega^2\sum_{a,b}|X_{ab}|^2$. The second bound follows in the same way in the basis $\{\ket{\phi_i}\}_{i=1}^{d}$, in which $\rho$ is diagonal with eigenvalues between $0$ and $p_1$.

\medskip

\textbf{Matrix elements of $W$.} In the eigenbasis of $H$ and in the eigenbasis of $\rho$, respectively,
\begin{equation}
\label{eq:Wab}
\bra{\lambda_a}W\ket{\lambda_b}=(\lambda_b-\lambda_a)\,\bra{\lambda_a}\rho\ket{\lambda_b},\qquad \bra{\phi_i}W\ket{\phi_k}=(p_i-p_k)\,\bra{\phi_i}H\ket{\phi_k}.
\end{equation}

\medskip

\textbf{Cyclic identity.} For any three operators $A$, $B$, and $C$,
\begin{equation}
\label{eq:cyclic}
\mathrm{Tr}\left([A,B]\,C\right)=\mathrm{Tr}\left([B,C]\,A\right)=\mathrm{Tr}\left([C,A]\,B\right).
\end{equation}

\section{Non-increasing energy of the states for the three flows}
\label{app:energy}
\label{app_sec:fluctuation-refrigeration-relation-proof}

This appendix proves the relations of Eqs.~\eqref{eq:E-C} and~\eqref{eq:E-D}, and derives the step-size conditions of Eq.~\eqref{eq:step-conditions}. The relations are fluctuation-refrigeration relations. That is, each flow lowers the energy at a rate set by $\|[\rho,H]\|_{HS}^2$, which measures how far $\rho$ is from commuting with $H$. A state that is far from being diagonal in the eigenbasis of $H$ decreases the energy quickly, and the cooling stops only when $\rho$ commutes with $H$. We treat the three flows in turn. For the discretized and the product-formula DBF, the relation contains a factor $1-s_k/\bar s$, which is the price of a finite step.

\subsection{Continuous-time DBF}

In this section, we will show the fluctuation-refrigeration relation for the continuous-time DBF.  
The energy at time $t$ is given by
\begin{equation}
    E(\rho(t)) = \mathrm{Tr}\left(\rho(t) H\right) \,,
\end{equation}
where $\rho(t)$ is evolved based on Eq.~\eqref{eq:C}. 

\begin{theoremA}[Energy decrease, continuous-time DBF]
\label{thmA:energy-C}
\label{app:theorem-energy-fluctuation-continuous}
Let $\rho(t)$ evolve under the continuous-time DBF of Eq.~\eqref{eq:C}. Then the energy is non-increasing, and
\begin{equation}
\label{eq:thmC}
\partial_tE\left(\rho(t)\right)=-\|[\rho(t),H]\|_{HS}^2 .
\end{equation}
\end{theoremA}
\begin{proof}
We can calculate the derivative of the average energy at time $t$ using the following,
\begin{equation}
    \partial_t E(\rho(t)) = \mathrm{Tr}\left( \partial_t \rho(t) H\right).
\end{equation}
Given that in DBF we have $\partial_t \rho(t) = [W(t), \rho(t)]$, the derivative can be simplified as follows:
\begin{equation}
    \partial_t E(\rho(t)) = \mathrm{Tr}\left( [W(t), \rho(t)] H\right) = \mathrm{Tr}\left( [\rho(t), H] W(t)\right) = \mathrm{Tr}\left( W^2(t)\right) = - \mathrm{Tr}\left( W(t) W^\dagger(t)\right) = - \| [\rho(t), H] \|_{HS}^2. 
\end{equation}
Where, in the second equality, we use the fact that for any $A$, $B$, and $C$, we have $\mathrm{Tr}\left( [A,B] C\right) = \mathrm{Tr}\left( [B, C] A\right) = \mathrm{Tr}\left( [C, A] B\right)$. And we use $W^\dagger(t) = - W(t)$. 
\end{proof}

\subsection{Discretized DBF}

In this section, we will show the fluctuation-refrigeration relation for the discretized DBF, and the energy at step $k$ is given by
\begin{equation}
    E(\rho_k) = \mathrm{Tr}\left( \rho_k H\right) \,,
\end{equation}
where $\rho_k$ is evolved based on Eq.~\eqref{eq:D}.

\begin{theoremA}[Energy decrease, discretized DBF]
\label{thmA:energy-D}
\label{app:theorem-energy-fluctuation-discrete}
The average energy of the state after one step, with step size $s_k>0$, of the discretized DBF of Eq.~\eqref{eq:D} satisfies
\begin{equation}
\label{eq:thmD}
    E(\rho_{k+1}) \le E(\rho_k) - s_k \left( 1 - \frac{s_k \gamma_{\max}}{2} \right) \| W_k \|_{HS}^2 \, ,
\end{equation}
where $W_k = [\rho_k, H]$ and $\gamma_{\max}=p_1\Omega$. 
In particular, the energy does not increase if $s_k \le \bar{s} = \frac{2}{\gamma_{\max}}$, and it decreases if $s_k<\bar{s}$ and $W_k \ne 0$.
\end{theoremA}
\begin{proof}
    Let us set $s \geq 0$ and define $\rho(t) = e^{t W} \rho e^{-t W}$. Then, using Taylor's expansion and the remainder integral to the first order, we can write $E(\rho(s))$ as follows: 
    \begin{equation}
        E(\rho(s)) = E(\rho(0)) + s \partial_t E(\rho(t))|_{t=0} + \int_{0}^{s} (s - t)\, \partial^2_t E(\rho(t))\,\mathrm{d}t.
    \label{ap_eq:reminder-integral-discrete}
    \end{equation}
    In order to use~\eqref{ap_eq:reminder-integral-discrete}, first we need to compute the first and second derivatives of $E(\rho(t)) = \mathrm{Tr}\left( \rho(t) H \right)$ with respect to $t$, which are given by the following,
    \begin{align}
        \partial_t E(\rho(t)) &= \mathrm{Tr}\left( \partial_t \rho(t) H\right) = \mathrm{Tr}\left( [W, \rho(t)] H\right) = \mathrm{Tr}\left( [\rho(t), H] W\right), \\
        \partial_t^2 E(\rho(t)) &= \mathrm{Tr}\left( \partial_t \rho(t) [H , W]\right) = \mathrm{Tr}\left( [W, \rho(t)] [H, W]\right).
    \end{align}
    We can simplify $\partial_t E(\rho(t))|_{t=0} = \mathrm{Tr}\left( [\rho, H] W\right) = \mathrm{Tr}\left( W^2\right) = - \| W\|_{HS}^2$. Next, for $\partial_t^2 E(\rho(t))$ we use $\partial_t^2 E(\rho(t)) \leq | \partial_t^2 E(\rho(t)) |$ and try to upper bound its absolute value. 
    \begin{align}
        |\partial_t^2 E(\rho(t))| & = |\mathrm{Tr}\left( [W, \rho(t)] [H, W]\right)| \leq \| [W, \rho(t)]\|_{HS} \| [H, W]\|_{HS} \leq p_1 \| W\|_{HS} \Omega \| W\|_{HS} = \gamma_{\max} \| W\|_{HS}^2,
    \end{align}
    where in the first inequality we use the fact that for any matrices $A$ and $B$ we have $|\mathrm{Tr}(AB)| \leq \| A\|_{HS} \| B \|_{HS}$. In the second inequality, we bound $ \| [W, \rho(t)]\|_{HS} $ and $ \| [H, W]\|_{HS} $ with $p_1 \| W \|_{HS}$ and $\Omega \| W \|_{HS}$ respectively.

    Putting all together in~\eqref{ap_eq:reminder-integral-discrete}, we have 
    \begin{align}
        E(\rho(s)) &\leq E(\rho(0)) - s \| W \|_{HS}^2 + \gamma_{\max} \| W\|_{HS}^2 \int_{0}^{s} (s - t)  \mathrm{d}t \\
        & = E(\rho(0)) - s \| W \|_{HS}^2 + \frac{s^2}{2} \gamma_{\max} \| W\|_{HS}^2  \, .
    \end{align}
    Finally, we can set $E(\rho(s)) = E(\rho_{k+1})$, $E(\rho(0)) = E(\rho_k)$, and $s = s_k$ to write
    \begin{equation}
        E(\rho_{k+1}) \le E(\rho_k) - s_k \| W_k \|_{HS}^2 + \frac{s_k^2}{2} \gamma_{\max} \| W_k\|_{HS}^2 = E(\rho_k) - s_k  \left(1 - \frac{s_k \gamma_{\max}}{2} \right) \| W_k \|_{HS}^2\,.
    \end{equation}
    Now, we have the step-size condition $s_k < \bar{s} = 2/\gamma_{\max}$ to ensure non-increasing energies at each step. 
\end{proof}

\subsection{Product-formula DBF}

In this part, we will study the fluctuation-refrigeration relation for the product formula DBF, and find the condition on the step size to have non-increasing energy at each step. The energy at step $k$ is given by 
\begin{equation}
    E(\rho_k) = \mathrm{Tr}(\rho_k H)\, ,
\end{equation}
where $\rho_k$ is evolved based on Eq.~\eqref{eq:PF}.

\begin{theoremA}[Energy decrease, product-formula DBF]
\label{thmA:energy-PF}
\label{app:theorem-energy-fluctuation-PF}
The average energy of the state after one step, with angles $\tau_k,\theta_k>0$ and $s_k=\tau_k\theta_k$, of the product-formula DBF of Eq.~\eqref{eq:PF} satisfies
\begin{equation}
\label{eq:thmPF}
E (\rho_{k+1}) \le E(\rho_k) - s_k \left( 1 - \frac{\theta_k^2 p_1^2}{6} - \frac{\tau_k^2 \Omega^2}{6} - \frac{\tau_k\theta_k\,\Omega\,p_1}{4}\right) \|W_k\|_{HS}^2 .
\end{equation}
Here, $W_k = [\rho_k, H]$, $p_1$ is the largest weight, $\Omega$ is the spectral width of $H$, and $\gamma_{\max}=p_1\Omega$. 
\begin{enumerate}
    \item For the angle ratio $\theta_k/\tau_k=\Omega/p_1$, this relation can be simplified as
    \begin{equation}
    \label{eq:thmPF-ratio}
        E(\rho_{k+1}) \le E(\rho_k) - s_k\left(1-\frac{7\,s_k\gamma_{\max}}{12}\right) \|W_k\|_{HS}^2 \,.
    \end{equation}
    Then, the energy does not increase if $s_k < \bar{s}=\frac{12}{7\gamma_{\max}}$, and it decreases if $s_k < \bar{s}$ and $W_k \ne 0$.
    \item For the equal angles $\theta_k = \tau_k =\sqrt{s_k}$, this relation can be simplified as
    \begin{equation}
    \label{eq:thmPF-equal}
        E(\rho_{k+1}) \le E(\rho_k) - s_k\left(1-\frac{7\,s_k \max(\Omega,1)^2}{12}\right) \|W_k\|_{HS}^2 \,,
    \end{equation}
    then, the energy does not increase if $s_k < \bar{s}=\frac{12}{7 \max(\Omega, 1)^2}$, and it decreases if $s_k < \bar{s}$ and $W_k \ne 0$.
\end{enumerate}

\end{theoremA}
\begin{proof}
    Let us set $\tau, \theta \geq 0$ and define $A(x) = e^{-ixH} \sigma e^{ix H}$, $B(x) = e^{-ix \sigma} H e^{ix \sigma}$, and $W = [\sigma , H]$. Then, similar to the proof of Theorem~\ref{thmA:energy-D}, we can use the remainder integral for $A(x)$ and $B(x)$:
    \begin{align}
        A(\tau) &= A(0) + \tau \partial_x A(x) |_{x=0} + \int_{0}^{\tau} (\tau-x) \partial_x^2 A(x) \mathrm{d}x 
        \label{app_eq:pf-proof-integral-1}, \\
        B(\theta) &= B(0) + \theta \partial_x B(x) |_{x=0} + \int_{0}^{\theta} (\theta-x) \partial_x^2 B(x) \mathrm{d}x.
    \label{app_eq:pf-proof-integral-2}
    \end{align}
    Let us define $ A_2(\tau) = \int_{0}^{\tau} (\tau-x) \partial_x^2 A(x) \mathrm{d}x$ and $B_2(\theta) = \int_{0}^{\theta} (\theta-x) \partial_x^2 B(x) \mathrm{d}x$, and since $A(0) = \sigma$ and $B(0) = H$, we can simplify~\eqref{app_eq:pf-proof-integral-1} and~\eqref{app_eq:pf-proof-integral-2}  as follows:
    \begin{align}
        A(\tau) = e^{-i\tau H} \sigma e^{i\tau H} &= \sigma + \tau \partial_x A(x) |_{x=0} + A_2(\tau), 
        \label{app_eq:pf-proof-A}
        \\
        B(\theta) = e^{-i\theta \sigma} H e^{i\theta \sigma} &= H + \theta \partial_x B(x) |_{x=0} + B_2(\theta).
        \label{app_eq:pf-proof-B}
    \end{align}
    Here, we calculate the first and second derivatives of $A$ and $B$. They are given by the following:
    \begin{align}
        \partial_xA(x) &= i e^{-ixH} [\sigma , H] e^{ix H}, \\
        \partial_xB(x) &= i e^{-ix \sigma} [H, \sigma] e^{ix \sigma}, \\
        \partial_x^2A(x) &= e^{-ixH} [H , W] e^{ix H}, \\
        \partial_x^2B(x) &= e^{-ix\sigma} [W , \sigma] e^{ix \sigma}.
    \end{align}
    Note that we can bound the Hilbert–Schmidt norm of $\partial_x^2A(x)$ and $\partial_x^2B(x)$ as follows. 
    \begin{align}
        \| \partial_x^2A(x) \|_{HS} &\leq \Omega \| W \|_{HS}, 
        \label{app_eq:pf-proof-derivative-A2-bound}
        \\
        \| \partial_x^2B(x) \|_{HS} &\leq p_1 \| W \|_{HS},
        \label{app_eq:pf-proof-derivative-B2-bound}
    \end{align}
    where, for both, we use the unitary invariance of the norm. Also for $\| \partial_x^2A(x) \|_{HS}$ we use $\| [H, W]\|_{HS} \leq \Omega \| W\|_{HS}$ and for $\| \partial_x^2B(x) \|_{HS}$ we use $\| [\sigma, W]\|_{HS} \leq p_1 \| W\|_{HS}$.
    
    Now, we can find the energy $E(\sigma(\tau, \theta )) = \mathrm{Tr}(\sigma(\tau, \theta) H)$, where $\sigma(\tau, \theta) = e^{i\tau H} e^{i\theta \sigma} e^{-i\tau H} \sigma e^{i\tau H} e^{-i\theta \sigma} e^{-i\tau H}$. Putting these two together and using the cyclic property of the trace, we have 
    \begin{equation}
        E(\sigma(\tau, \theta)) = \mathrm{Tr} \left( (e^{-i\tau H} \sigma e^{i\tau H}) (e^{-i\theta \sigma}  H e^{i\theta \sigma})\right) = \mathrm{Tr} \left( A(\tau) B(\theta)\right) \,.
        \label{app_eq:pf-proof-energy-x}
    \end{equation}
    Then, using~\eqref{app_eq:pf-proof-A} and~\eqref{app_eq:pf-proof-B}, in the trace in~\eqref{app_eq:pf-proof-energy-x} we have nine terms. Since $\partial_xA(x)|_{x=0} = i [\sigma , H] = iW$ and $\partial_xB(x)|_{x=0} = i [H, \sigma] = -iW$, the energy can be simplified as 
    \begin{align}
        E(\sigma(\tau, \theta)) = \mathrm{Tr}\left( \sigma H -i\theta \sigma W + \sigma B_2(\theta) + i\tau WH + \tau \theta W^2 + i\tau  W B_2(\theta) + A_2(\tau)H - i\theta A_2(\tau)W + A_2(\tau) B_2(\theta)\right) \,.
    \end{align}
    The first term is $E(\sigma) = \mathrm{Tr}(\sigma H)$. The second and fourth terms are zero since $\mathrm{Tr}(\sigma W) = \mathrm{Tr}(H [\sigma, \sigma]) = 0$ and $\mathrm{Tr}(WH) = \mathrm{Tr}([H, H] \sigma) = 0$. The third and seventh terms are also zero since $\mathrm{Tr}(\sigma B_2(\theta)) = \int_{0}^{\theta} (\theta-x) \mathrm{Tr}\left( \sigma \partial_x^2 B(x) \right) \mathrm{d}x = \int_{0}^{\theta} (\theta-x) \mathrm{Tr}\left( \sigma e^{-ix\sigma} [W , \sigma] e^{ix \sigma} \right) \mathrm{d}x = \int_{0}^{\theta} (\theta-x) \mathrm{Tr}\left( \sigma [W , \sigma] \right) \mathrm{d}x = 0$ and we can similarly show that $\mathrm{Tr}(A_2(\tau) H)=0$. 
    The fifth term can be simplified as $\tau \theta \mathrm{Tr}\left( W^2\right) = -\tau \theta \| W\|_{HS}^{2}$. We can upper bound the absolute value of the last term as $ |\mathrm{Tr}\left( A_2(\tau) B_2(\theta) \right)| \leq \| A_2(\tau)\|_{HS} \| B_2(\theta)\|_{HS}$. Then, using~\eqref{app_eq:pf-proof-derivative-A2-bound} and~\eqref{app_eq:pf-proof-derivative-B2-bound}, $\| A_2(\tau)\|_{HS}$ and $\| B_2(\theta)\|_{HS}$ are the following: 
    \begin{align}
        \| A_2(\tau)\|_{HS} &\leq \int_{0}^{\tau} (\tau-x) \| \partial_x^2 A(x) \|_{HS} \mathrm{d}x \leq \frac{\tau^2}{2} \Omega \| W \|_{HS}, \\
        \| B_2(\theta)\|_{HS}&\leq \int_{0}^{\theta} (\theta-x) \| \partial_x^2 B(x) \|_{HS} \mathrm{d}x \leq \frac{\theta^2}{2} p_1 \| W \|_{HS}.
    \end{align}
    In the first inequality for both of them, we use the fact that for the integral of any matrix $X$ we have $\| \int X \|_{HS} \leq \int \| X\|_{HS}$. Now, we have $|\mathrm{Tr}\left( A_2(\tau) B_2(\theta) \right)| \leq \frac{\tau^2 \theta^2}{4} p_1 \Omega\| W\|_{HS}^2$.
    
    Next, we simplify the sixth and eighth terms. For the sixth term, $i\tau \mathrm{Tr}\left( W B_2(\theta)\right)$ we have  
    \begin{align}
        i\tau\mathrm{Tr}\left( W B_2(\theta)\right) &= i\tau \int_{0}^{\theta} (\theta-x)  \mathrm{Tr} \left( W \partial_x^2 B(x)\right) \mathrm{d}x \\
        & = i\tau \int_{0}^{\theta} (\theta-x)  \mathrm{Tr} \left( W e^{-ix\sigma} [W , \sigma] e^{ix \sigma}\right) \mathrm{d}x \\
        & = i\tau \int_{0}^{\theta} (\theta-x)  \mathrm{Tr} \left(W (e^{-ix\sigma} [W , \sigma] e^{ix \sigma} - [W , \sigma] )\right) \mathrm{d}x \\
        & = i\tau \int_{0}^{\theta} (\theta-x)  \mathrm{Tr} \left(W \int_{0}^{x} -i e^{-ix'\sigma} [\sigma, [\sigma, W]] e^{ix' \sigma} \mathrm{d}x' \right) \mathrm{d}x \,,
    \end{align}
    where in the third equality we use the fact that $\mathrm{Tr}\left( W [W, \sigma]\right) = 0$ to add that term, and in the final equality we use the zeroth-order remainder integral, i.e., $e^{-ix\sigma} [W, \sigma] e^{ix \sigma} =  [W,\sigma] + \int_{0}^{x} -i e^{-ix'\sigma} [\sigma, [\sigma, W]] e^{ix' \sigma} \mathrm{d}x'$. 
    Furthermore, we can bound the absolute value of $\mathrm{Tr}\left( W B_2(\theta)\right)$ as follows:
    \begin{align}
        | i\tau \mathrm{Tr}\left( W B_2(\theta)\right) | &\leq \tau \int_{x = 0}^{\theta} (\theta-x) \, \left|\mathrm{Tr} \left(W \int_{x'=0}^{x} -i e^{-ix'\sigma} [\sigma, [\sigma, W]] e^{ix' \sigma} \mathrm{d}x' \right) \right| \mathrm{d}x \\
        &\leq \tau \int_{x=0}^{\theta} (\theta-x) \,  \| W\|_{HS} \, \| \int_{x'=0}^{x} -i e^{-ix'\sigma} [\sigma, [\sigma, W]] e^{ix' \sigma} \mathrm{d}x' \|_{HS} \, \mathrm{d}x \\
        &\leq \tau \| W\|_{HS} \int_{x=0}^{\theta} (\theta-x) \, \int_{x'=0}^{x} \| e^{-ix'\sigma} [\sigma, [\sigma, W]] e^{ix' \sigma} \|_{HS} \, \mathrm{d}x' \mathrm{d}x \\
        & \leq \tau \| W\|_{HS} \int_{x=0}^{\theta} (\theta-x) \,  \int_{x'=0}^{x} \| [\sigma, [\sigma, W]]\|_{HS} \, \mathrm{d}x' \mathrm{d}x \\
        & \leq \tau \| W\|_{HS} \int_{x=0}^{\theta} (\theta-x) \,  \int_{x'=0}^{x} p_1^2 \| W\|_{HS} \, \mathrm{d}x' \mathrm{d}x \\
        & = \frac{\tau \theta^3 p_1^2}{6} \, \| W\|_{HS}^2 .
    \end{align}
    In the first inequality, we use the fact that $\theta - x \geq 0$. In the second inequality, we use $\left|\mathrm{Tr} \left(W \int_{x'=0}^{x} -i e^{-ix'\sigma} [\sigma, [\sigma, W]] e^{ix' \sigma} \mathrm{d}x' \right) \right| \leq  \| W\|_{HS} \, \| \int_{x'=0}^{x} -i e^{-ix'\sigma} [\sigma, [\sigma, W]] e^{ix' \sigma} \mathrm{d}x' \|_{HS}$. In the third inequality, we use the fact that for the integral of any matrix $X$ we have $\| \int X \|_{HS} \leq \int \| X\|_{HS}$. And finally, we use $\| [\sigma, [\sigma, W]]\|_{HS} \leq p_1^2 \| W\|_{HS}$. 
    Using similar steps and arguments, we can bound the absolute value of the eighth term as $| i\theta \mathrm{Tr}\left( A_2(\tau) W\right) | \leq \frac{\theta \tau^3 \Omega^2}{6} \, \| W\|_{HS}^2$ since we have $\| [H, [H, W]]\|_{HS} \leq \Omega^2 \| W\|_{HS}$.
    Moreover, putting all nine terms together, we have 
    \begin{equation}
        E(\sigma(\tau, \theta)) \leq E(\sigma) - \tau \theta \| W\|_{HS}^2 \left( 1 - \frac{\theta^2 p_1^2}{6} - \frac{\tau^2 \Omega^2}{6} - \frac{\tau \theta p_1 \Omega}{4} \right) \,.
    \end{equation}
    Finally, we can set $E(\sigma(\tau, \theta)) = E(\rho_{k+1})$, $E(\sigma) = E(\rho_k)$, $\tau = \tau_k$, and $\theta = \theta_k$, and write 
    \begin{equation}
        E(\rho_{k+1}) \leq E(\rho_{k}) - s_k \|W_k\|_{HS}^2 \left( 1 - \frac{\theta_k^2 p_1^2}{6} - \frac{\tau_k^2 \Omega^2}{6} - \frac{\tau_k \theta_k p_1 \Omega}{4} \right) \,,
    \end{equation}
    where we use $\tau_k \theta_k = s_k$. This inequality is Eq.~\eqref{eq:thmPF}.

    It remains to evaluate the bracket of Eq.~\eqref{eq:thmPF} for the two choices of the angles and to find the step sizes for which the energy does not increase. We abbreviate the bracket as
    \begin{equation}
    \label{eq:PF-bracket}
        \chi_k = 1 - \frac{\theta_k^2 p_1^2}{6} - \frac{\tau_k^2 \Omega^2}{6} - \frac{\tau_k \theta_k\, \Omega\, p_1}{4} \, ,
    \end{equation}
    so that Eq.~\eqref{eq:thmPF} reads $E(\rho_{k+1})\le E(\rho_k)-s_k\,\chi_k\,\|W_k\|_{HS}^2$.
    \begin{enumerate}
        \item \textit{Angle ratio $\theta_k/\tau_k=\Omega/p_1$.} The ratio gives $\theta_k p_1=\tau_k\Omega$. With $s_k=\tau_k\theta_k$ and $\gamma_{\max}=p_1\Omega$, the three products in Eq.~\eqref{eq:PF-bracket} are equal,
        \begin{align}
            \theta_k^2 p_1^2 & =\left(\theta_k p_1\right)\left(\tau_k\Omega\right)=s_k\gamma_{\max},\\ \tau_k^2\Omega^2&=\left(\tau_k\Omega\right)\left(\theta_k p_1\right)=s_k\gamma_{\max},\\ \tau_k\theta_k\,\Omega\,p_1&=s_k\gamma_{\max} \, .
        \end{align}
        Hence
        \begin{equation}
            \chi_k = 1-s_k\gamma_{\max}\left(\frac16+\frac16+\frac14\right)=1-\frac{7\,s_k\gamma_{\max}}{12}=1-\frac{s_k}{\bar s}\, ,
        \end{equation}
        with $\bar s=\frac{12}{7\gamma_{\max}}$, which gives Eq.~\eqref{eq:thmPF-ratio}. The factor $\chi_k$ is positive for $s_k<\bar s$. The energy therefore does not increase if $s_k<\bar s$, and it decreases if in addition $W_k\ne0$.
        \item \textit{Equal angles $\theta_k=\tau_k$.} The relation $s_k=\tau_k\theta_k$ gives $\theta_k=\tau_k=\sqrt{s_k}$, so that
        \begin{align}
            \theta_k^2 p_1^2 &=s_k\,p_1^2,\\ \tau_k^2\Omega^2&=s_k\,\Omega^2,\\ \tau_k\theta_k\,\Omega\,p_1&=s_k\,p_1\Omega \, .
        \end{align}
        Inserting these products into Eq.~\eqref{eq:PF-bracket} gives
        \begin{equation}
        \label{eq:PF-equal-exact}
            \chi_k = 1-s_k\left(\frac{p_1^2}{6}+\frac{\Omega^2}{6}+\frac{p_1\Omega}{4}\right)=1-\frac{s_k}{12}\left(2p_1^2+2\Omega^2+3p_1\Omega\right) .
        \end{equation}
        To obtain a condition that does not involve the weights, we bound the last bracket. The weights sum to one, so that $p_1\le1\le\max(\Omega,1)$, and $\Omega\le\max(\Omega,1)$. Each of the three terms is therefore at most a multiple of $\max(\Omega,1)^2$,
        \begin{equation}
        \label{eq:PF-equal-bound}
            2p_1^2+2\Omega^2+3p_1\Omega\ \le\ \left(2+2+3\right)\max(\Omega,1)^2=7\max(\Omega,1)^2 \, .
        \end{equation}
        With Eq.~\eqref{eq:PF-equal-exact}, we obtain
        \begin{equation}
            \chi_k\ \ge\ 1-\frac{7\,s_k\max(\Omega,1)^2}{12}=1-\frac{s_k}{\bar s} \, ,
        \end{equation}
        with $\bar s=\frac{12}{7\max(\Omega,1)^2}$, which gives Eq.~\eqref{eq:thmPF-equal}. As in the first case, the energy does not increase if $s_k<\bar s$, and it decreases if in addition $W_k\ne0$.
    \end{enumerate}

\end{proof}

Theorems~\ref{thmA:energy-D} and~\ref{thmA:energy-PF} address a single iteration of the algorithm. Equations~\eqref{eq:thmD} and~\eqref{eq:thmPF} therefore hold independently for every iteration with its corresponding step size and angles, even when these values change between iterations.

\section{The energy landscape}
\label{app:energy-landscape}

This appendix proves the statements about the energy landscape. First, the target state $\rho^{*}$ of Eq.~\eqref{eq:target} is the unique global minimum of the energy $E(\rho)=\mathrm{Tr}(\rho H)$ over all states of the form of Eq.~\eqref{eq:rho}. Second, the quantity $\gamma_{ij}$ of Eq.~\eqref{eq:gamma-ij} is the energy excess of the fixed point that is obtained from $\rho^{*}$ by exchanging two weights, and no fixed point other than $\rho^{*}$ has an excess energy below $\gamma_{\min}$. The obstacle is that the energy depends on all $r$ states at once. We remove this obstacle by writing the energy excess as a sum over the nested subspaces that are spanned by the states with the largest weights. Each term of this sum involves a single subspace, and the variational principle for a subspace bounds it. All statements of this appendix concern a single state and involve none of the flows. The subspace decomposition is also the starting point of the proofs of Lemma~\ref{lemma:gradient} in Appendix~\ref{app:lemma1} and of Theorem~\ref{theorem:convergence-rate} in Appendix~\ref{app:thm2}.

\subsection{Energy excess as a sum over subspaces}
\label{app:landscape-subspaces}

We first express the energy excess $\Delta E(\rho)=E(\rho)-E^{*}$ of a state $\rho=\sum_{i=1}^{r}p_i\ket{\phi_i}\bra{\phi_i}$ of the form of Eq.~\eqref{eq:rho} through subspaces. For $j=1,\dots,r$, let $V_j$ be the subspace spanned by the $j$ states with the largest weights in $\rho$,
\begin{equation}
\label{eq:Vj}
V_j=\mathrm{span}\{\ket{\phi_1},\ket{\phi_2},\dots,\ket{\phi_j}\} .
\end{equation}
We denote the projector onto $V_j$ by $P_j$ and the projector onto the span of the $j$ lowest eigenstates of $H$ by $\Pi_j$,
\begin{equation}
\label{eq:Pj}
P_j=\sum_{i=1}^{j}\ket{\phi_i}\bra{\phi_i},\qquad \Pi_j=\sum_{i=1}^{j}\ket{\lambda_i}\bra{\lambda_i} .
\end{equation}
The energy excess of the subspace $V_j$ is the difference between the energy of $V_j$ and the energy of the lowest $j$-dimensional subspace $\mathrm{span}\{\ket{\lambda_1},\dots,\ket{\lambda_j}\}$,
\begin{equation}
\label{app_eq:energy-excess}
\varepsilon_j=\mathrm{Tr}\left(P_jH\right)-\mathrm{Tr}\left(\Pi_jH\right)=\sum_{i=1}^{j}\left(\bra{\phi_i}H\ket{\phi_i}-\lambda_i\right) .
\end{equation}

In order to write $\Delta E(\rho)$ as a sum of the variables $\varepsilon_j$, we write each weight as a sum of the weight gaps $\delta p_j=p_j-p_{j+1}$ of Eq.~\eqref{eq:app-gaps}. Since $p_{r+1}=0$, the sum telescopes,
\begin{equation}
\label{eq:p-telescope}
p_i=\sum_{j=i}^{r}\left(p_j-p_{j+1}\right)=\sum_{j=i}^{r}\delta p_j .
\end{equation}
We insert Eq.~\eqref{eq:p-telescope} into the energy and exchange the order of the two sums, which run over the pairs $1\le i\le j\le r$,
\begin{align}
E(\rho)&=\sum_{i=1}^{r}p_i\bra{\phi_i}H\ket{\phi_i}\nonumber\\
&=\sum_{i=1}^{r}\sum_{j=i}^{r}\delta p_j\bra{\phi_i}H\ket{\phi_i}\nonumber\\
&=\sum_{j=1}^{r}\delta p_j\sum_{i=1}^{j}\bra{\phi_i}H\ket{\phi_i} .
\label{app_eq:energy-at-t-Vj}
\end{align}
Repeating the same two steps for the target state $\rho^{*}$, in which instead of $\ket{\phi_i}$ it has $\ket{\lambda_i}$, gives 
\begin{equation}
\label{app_eq:energy-infty-at-t-Vj}
E^{*}=\sum_{i=1}^{r}p_i\lambda_i=\sum_{j=1}^{r}\delta p_j\sum_{i=1}^{j}\lambda_i .
\end{equation}
Subtracting Eq.~\eqref{app_eq:energy-infty-at-t-Vj} from Eq.~\eqref{app_eq:energy-at-t-Vj} and using Eq.~\eqref{app_eq:energy-excess}, we obtain
\begin{equation}
\label{eq:dE-subspaces}
\Delta E(\rho)=\sum_{j=1}^{r}\delta p_j\sum_{i=1}^{j}\left(\bra{\phi_i}H\ket{\phi_i}-\lambda_i\right)=\sum_{j=1}^{r}\delta p_j\,\varepsilon_j .
\end{equation}
The energy excess of the state is thus a sum of the energy excesses of the subspaces $V_1,\dots,V_r$, weighted by the positive weight gaps.

\subsection{Energy excess of a single subspace}
\label{app:landscape-one-subspace}

Eq.~\eqref{eq:dE-subspaces} reduces the energy landscape to a property of a single subspace. This property does not involve $\rho$, and we state it for an arbitrary subspace.

\begin{lemmaA}[Energy excess of a subspace]
\label{lemA:subspace-excess}
Let $H$ have the eigenvalues $\lambda_1<\dots<\lambda_d$ with eigenstates $\ket{\lambda_a}$, let $1\le j<d$, and let $V$ be a $j$-dimensional subspace with the projector $P_V$. Let $\Pi_j=\sum_{a=1}^{j}\ket{\lambda_a}\bra{\lambda_a}$, and define the energy excess of $V$ as
\begin{equation}
\label{eq:eps-V}
\varepsilon(V)=\mathrm{Tr}\left(P_VH\right)-\sum_{a=1}^{j}\lambda_a .
\end{equation}
Then
\begin{equation}
\label{eq:sub-a}
\varepsilon(V)\ \ge\ \delta\lambda_{j+1}\left[\,j-\mathrm{Tr}\left(P_V\Pi_j\right)\right]\ \ge\ 0 ,
\end{equation}
where $\delta\lambda_{j+1}=\lambda_{j+1}-\lambda_j$. Moreover, $\varepsilon(V)=0$ if and only if $V=\mathrm{span}\{\ket{\lambda_1},\dots,\ket{\lambda_j}\}$.
\end{lemmaA}
\begin{proof}
We express both sides of Eq.~\eqref{eq:sub-a} as a function of the populations $w_a$ of the energy levels $\ket{\lambda_a}$ in subspace $V$, i.e.,
\begin{equation}
w_a=\bra{\lambda_a}P_V\ket{\lambda_a}\,,
\end{equation}
where $0\le w_a\le1$ and $\sum_{a=1}^{d}w_a=\mathrm{Tr}\left(P_V\right)=j$. 
The populations satisfy $0\le w_a\le1$ because $P_V$ is a projector. The total population of the levels above $j$ equals the population that is missing from the $j$ lowest levels,
\begin{equation}
\label{eq:missing-weight}
\sum_{a=j+1}^{d}w_a=j-\sum_{a=1}^{j}w_a=\sum_{a=1}^{j}\left(1-w_a\right)=j-\mathrm{Tr}\left(P_V\Pi_j\right) ,
\end{equation}
where the last equality uses $\mathrm{Tr}(P_V\Pi_j)=\sum_{a\le j}w_a$.

We now bound the energy excess of subspace $V$. 
Note that we can write 
\begin{equation}
    \mathrm{Tr}(P_V H)=\sum_{a}\lambda_aw_a \,
\end{equation}
and then Eq.~\eqref{eq:eps-V} becomes
\begin{align}
\varepsilon(V)&=\sum_{a=1}^{d}\lambda_a\,w_a-\sum_{a=1}^{j}\lambda_a\nonumber\\
&=\sum_{a=j+1}^{d}\lambda_a\,w_a-\sum_{a=1}^{j}\lambda_a\left(1-w_a\right)\nonumber\\
&\ge\lambda_{j+1}\sum_{a=j+1}^{d}w_a-\lambda_j\sum_{a=1}^{j}\left(1-w_a\right)\nonumber\\
&=\left(\lambda_{j+1}-\lambda_j\right)\left[\,j-\mathrm{Tr}\left(P_V\Pi_j\right)\right] .
\end{align}
In the third line, we use $\lambda_a\ge\lambda_{j+1}$, for $a \ge j+1$, and $w_a\ge0$ in the first sum, and $\lambda_a\le\lambda_j$, for $a \le j$, and $1-w_a\ge0$ in the second sum. The last line uses Eq.~\eqref{eq:missing-weight} for both sums. This proves the first inequality of Eq.~\eqref{eq:sub-a}. The second inequality holds because $\delta\lambda_{j+1}>0$ and because the bracket equals $\sum_{a>j}w_a\ge0$ by Eq.~\eqref{eq:missing-weight}.

Finally, we identify the case in Eq.~\eqref{eq:sub-a} with the equality. If $\varepsilon(V)=0$, Eq.~\eqref{eq:sub-a} gives $\mathrm{Tr}(P_V\Pi_j)=j$. Since $P_V$ and $\Pi_j$ are projectors of rank $j$,
\begin{equation}
\|P_V-\Pi_j\|_{HS}^2=\mathrm{Tr}\left(P_V\right)+\mathrm{Tr}\left(\Pi_j\right)-2\,\mathrm{Tr}\left(P_V\Pi_j\right)=2j-2j=0 ,
\end{equation}
so that $P_V=\Pi_j$. Conversely, $P_V=\Pi_j$ gives $\mathrm{Tr}(P_VH)=\sum_{a\le j}\lambda_a$ and hence $\varepsilon(V)=0$.
\end{proof}

Lemma~\ref{lemA:subspace-excess} is the variational principle for the sum of the $j$ lowest eigenvalues, together with a bound on the deviation of $V$ from the lowest subspace. The quantity $j-\mathrm{Tr}(P_V\Pi_j)$ vanishes exactly when $V$ is the span of the $j$ lowest eigenstates, and it equals one when $V$ is spanned by $j-1$ of the $j$ lowest eigenstates and one state above the level $j$.

\subsection{The global minimum and the fixed points}
\label{app:landscape-fixed-points}

We now combine Eq.~\eqref{eq:dE-subspaces} with Lemma~\ref{lemA:subspace-excess}. The following proposition shows the statements on the unique minimum of the energy. It also gives intuition for $\gamma_{ij}$ and $\gamma_{\min}$. Recall from Eq.~\eqref{eq:placements} that the fixed points are the states $\rho_\sigma=\sum_{a=1}^{d}p_{\sigma(a)}\ket{\lambda_a}\bra{\lambda_a}$, where $\sigma$ is a permutation of the $d$ levels and $p_i=0$ for $i>r$.

\begin{propositionA}[Energy landscape]
\label{propA:landscape}
Let $H$ have the eigenvalues $\lambda_1<\dots<\lambda_d$, and let the weights satisfy $p_1>\dots>p_r>0$ with $p_i=0$ for $i>r$. Let $\Delta E(\rho)=E(\rho)-E^{*}$ be the energy excess relative to the target state $\rho^{*}=\sum_{i=1}^{r}p_i\ket{\lambda_i}\bra{\lambda_i}$ of Eq.~\eqref{eq:target}, and let $\gamma_{ij}=(p_i-p_j)(\lambda_j-\lambda_i)$ and $\gamma_{\min}=\min_{1\le j\le r}\gamma_{j,j+1}$ as in Eqs.~\eqref{eq:gamma-ij} and~\eqref{eq:gamma-min}.
\begin{enumerate}
\item[\textup{(a)}] \textit{Unique global minimum.} Every state $\rho$ of the form of Eq.~\eqref{eq:rho} satisfies $\Delta E(\rho)\ge0$, with equality if and only if $\rho=\rho^{*}$. In that case, $\ket{\phi_i}=\ket{\lambda_i}$ up to a phase for every $i\le r$.
\item[\textup{(b)}] \textit{The energy excess given by $\gamma_{ij}$.} For two levels $i<j$ with $i\le r$, let $\rho_{\sigma_{i,j}}$ be the fixed point that is obtained from $\rho^{*}$ by exchanging the weights on the levels $i$ and $j$. Then $\Delta E(\rho_{\sigma_{i,j}})=\gamma_{ij}$.
\item[\textup{(c)}] \textit{The energy excess given by $\gamma_{\min}$.} Every fixed point $\rho_\sigma\ne\rho^{*}$ satisfies $\Delta E(\rho_\sigma)\ge\gamma_{\min}$. Equality holds for the exchange of two adjacent levels $j$ and $j+1$ with $\gamma_{j,j+1}=\gamma_{\min}$. That is, the lowest fixed point above the target lies exactly $\gamma_{\min}$ above it.
\end{enumerate}
\end{propositionA}
\begin{proof}
\textit{(a) Unique global minimum.} By Eq.~\eqref{eq:dE-subspaces}, $\Delta E(\rho)=\sum_{j=1}^{r}\delta p_j\,\varepsilon_j$ with $\varepsilon_j=\varepsilon(V_j)$. Every weight gap is positive, $\delta p_j>0$, and Lemma~\ref{lemA:subspace-excess} gives $\varepsilon_j\ge0$. Hence
\begin{equation}
\label{eq:dE-nonneg}
\Delta E(\rho)=\sum_{j=1}^{r}\delta p_j\,\varepsilon_j\ \ge\ 0 .
\end{equation}
Suppose now that $\Delta E(\rho)=0$. Since all terms of the sum are nonnegative and $\delta p_j>0$, every $\varepsilon_j$ vanishes. By Lemma~\ref{lemA:subspace-excess}, $V_j=\mathrm{span}\{\ket{\lambda_1},\dots,\ket{\lambda_j}\}$, that is, $P_j=\Pi_j$ for every $j\le r$. With $P_0=\Pi_0=0$, this gives
\begin{equation}
\ket{\phi_i}\bra{\phi_i}=P_i-P_{i-1}=\Pi_i-\Pi_{i-1}=\ket{\lambda_i}\bra{\lambda_i}\qquad(1\le i\le r) .
\end{equation}
Thus $\ket{\phi_i}=\ket{\lambda_i}$ up to a phase for every $i\le r$, and $\rho=\rho^{*}$. Conversely, $\Delta E(\rho^{*})=0$ by definition.

\smallskip

\textit{(b) The energy excess given by $\gamma_{ij}$.} The exchange moves the weight $p_i$ from the level $i$ to the level $j$ and the weight $p_j$ from the level $j$ to the level $i$, where $p_j=0$ if $j>r$. All other levels keep their weights. The energy therefore changes by
\begin{align}
\Delta E(\rho_{\sigma_{i,j}})&=\left(p_i\lambda_j+p_j\lambda_i\right)-\left(p_i\lambda_i+p_j\lambda_j\right)\nonumber\\
&=p_i\left(\lambda_j-\lambda_i\right)-p_j\left(\lambda_j-\lambda_i\right)\nonumber\\
&=\left(p_i-p_j\right)\left(\lambda_j-\lambda_i\right)=\gamma_{ij} .
\label{eq:swap-energy}
\end{align}

\smallskip

\textit{(c) The energy excess given by $\gamma_{\min}$.} Let $\rho_\sigma\ne\rho^{*}$ be a fixed point, and let $a_i$ be the level that carries the weight $p_i$ in $\rho_\sigma$, so that $\ket{\phi_i}=\ket{\lambda_{a_i}}$ for $i\le r$. Let $i$ be the smallest index with $a_i\ne i$. Such an index $i\le r$ exists because $\rho_\sigma\ne\rho^{*}$. The weights $p_1,\dots,p_{i-1}$ occupy the levels $1,\dots,i-1$, so that $a_i>i$. The subspace $V_i=\mathrm{span}\{\ket{\lambda_{a_1}},\dots,\ket{\lambda_{a_i}}\}$ then contains the eigenstate $\ket{\lambda_{a_i}}$, which is orthogonal to the $i$ lowest eigenstates. Hence
\begin{equation}
i-\mathrm{Tr}\left(P_i\Pi_i\right)=i-(i-1)=1 ,
\end{equation}
and Lemma~\ref{lemA:subspace-excess} gives $\varepsilon_i\ge\delta\lambda_{i+1}$. Since all terms of Eq.~\eqref{eq:dE-subspaces} are nonnegative, we may keep only the term $j=i$,
\begin{equation}
\Delta E(\rho_\sigma)=\sum_{j=1}^{r}\delta p_j\,\varepsilon_j\ \ge\ \delta p_i\,\varepsilon_i\ \ge\ \delta p_i\,\delta\lambda_{i+1}=\gamma_{i,i+1}\ \ge\ \gamma_{\min} .
\end{equation}
The value $\gamma_{\min}$ is attained. Indeed, let $j\le r$ be an index with $\gamma_{j,j+1}=\gamma_{\min}$. By part (b), the fixed point $\rho_{\sigma_{j,j+1}}$ that exchanges the weights on the adjacent levels $j$ and $j+1$ has the energy excess $\gamma_{j,j+1}=\gamma_{\min}$.
\end{proof}

\medskip

\section{Necessary condition for convergence to the target state}
\label{app:thm1}

In this appendix we state and prove Theorem~\ref{thmA:convergence}, which makes precise the statements on the convergence of the three flows and on the necessary condition for reaching the target state $\rho^{*}$.
The theorem says where the three flows converge.

\subsection{Statement of the theorem}
\label{app:thm1-theorem-proof}
\label{theorem:convergence}

\begin{theoremA}[Fixed points, convergence, and necessary condition]
\label{thmA:convergence}
Let $H$ have the eigenvalues $\lambda_1<\dots<\lambda_d$, and let $\rho$ evolve from the reference state $\rho_{\mathrm{ref}}$ of the form of Eq.~\eqref{eq:rho}, with weights $p_1>\dots>p_r>0$, under the continuous-time DBF of Eq.~\eqref{eq:C}, the discretized DBF of Eq.~\eqref{eq:D}, or the product-formula DBF of Eq.~\eqref{eq:PF}. For the last two flows, let $\bar s$ be the step-size upper bound of Eq.~\eqref{eq:app-sbar}, which Theorems~\ref{thmA:energy-D} and~\ref{thmA:energy-PF} derive. Let $\rho_\sigma$ be the states of Eq.~\eqref{eq:placements} and $\rho^{*}$ the target state of Eq.~\eqref{eq:target}.

\begin{enumerate}
    \item \textbf{Fixed points:} For the continuous-time DBF, and for the discretized and the product-formula DBF with any step size $0<s_k<\bar{s}$, a state of the form of Eq.~\eqref{eq:rho} is a fixed point if and only if it commutes with $H$. There exist $d!/(d-r)!$ fixed points $\rho_\sigma$ of Eq.~\eqref{eq:placements}.
    \item \textbf{Convergence:} For the continuous-time DBF, and for the discretized and product-formula DBF with step sizes $s_{\min}\le s_k\le s_{\max}$ for two constants $0<s_{\min}\le s_{\max}<\bar{s}$, every reference state converges to one of the fixed points $\rho_\sigma$. In particular, this statement holds for a constant step size $0<s<\bar s$.
    \item \textbf{Necessary condition:} Let $P_r=\sum_{j=1}^{r}\ket{\phi_j}\bra{\phi_j}$ be the projector onto the span of the $r$ reference states. Then $\rho$ converges to $\rho^{*}$ only if $A_i = \bra{\lambda_i}P_r\ket{\lambda_i}>0$ for all $i\le r$. That is, if $A_i = \bra{\lambda_i}P_r\ket{\lambda_i}=0$ for some $i\le r$, then $\rho\ket{\lambda_i}=0$ at all times, and $\rho$ does not converge to $\rho^{*}$. This statement holds for all step sizes $s_k>0$.
\end{enumerate}
\end{theoremA}
\begin{proof}
    We prove the three statements in the following subsections. Throughout the proof, $W=[\rho,H]$ and $W_k=[\rho_k,H]$. In Appendix~\ref{app:thm1-fixed} we prove the \textit{fixed points} statement. Appendix~\ref{app:thm1-convergence} shows the \textit{convergence} statement. And Appendix~\ref{app:thm1-necessary} provides the proof of the \textit{necessary condition} statement. 
\end{proof}

\subsection{Proof of statement 1: fixed points}
\label{app:thm1-fixed}

We show that the flows stop exactly at the states that commute with $H$ in three steps. 
First, a state that commutes with $H$ is a fixed point. 
Second, a fixed point commutes with $H$. 
Third, the states that commute with $H$ are the states $\rho_\sigma$.

First, suppose that $\rho$ commutes with $H$, that is, $W=0$. Then none of the flows moves the state, whatever the step size is.
\begin{itemize}
    \item \textit{Continuous-time DBF.} Equation~\eqref{eq:C} gives $\partial_t\rho=[W,\rho]=0$.
    \item \textit{Discretized DBF.} The unitary $e^{s_kW_k}$ is the identity, so that Eq.~\eqref{eq:D} gives $\rho_{k+1}=\rho_k$.
    \item \textit{Product-formula DBF.} Since $\rho_k$ commutes with $H$, the unitary $e^{i\theta_k\rho_k}$ commutes with $e^{-i\tau_kH}$. Hence $V_k=e^{i\tau_kH}e^{i\theta_k\rho_k}e^{-i\tau_kH}=e^{i\theta_k\rho_k}$, which commutes with $\rho_k$, and Eq.~\eqref{eq:PF} gives $\rho_{k+1}=V_k\rho_kV_k^\dagger=\rho_k$.
\end{itemize}

Second, suppose that $\rho$ is a fixed point. The state does not change, so its energy does not change. Using the energy relations from Appendix~\ref{app:energy}, we can show the following.
\begin{itemize}
    \item \textit{Continuous-time DBF.} Theorem~\ref{thmA:energy-C} gives $0=\partial_tE(\rho)=-\|W\|_{HS}^2$. Hence $W=0$.
    \item \textit{Discretized and product-formula DBF.} With $\rho_{k+1}=\rho_k$, using Eqs.~\eqref{eq:thmD},~\eqref{eq:thmPF-ratio}, and~\eqref{eq:thmPF-equal}, we have
    \begin{equation}
        0=E(\rho_{k+1})-E(\rho_k)\le-s_k\left(1-\frac{s_k}{\bar s}\right)\|W_k\|_{HS}^2 .
    \end{equation}
    For $0<s_k<\bar s$, the factor $s_k(1-s_k/\bar s)$ is positive. Hence $\|W_k\|_{HS}=0$.
    
    The condition $s_k<\bar{s}$ is needed in this step. Eqs.~\eqref{eq:thmD},~\eqref{eq:thmPF-ratio}, and~\eqref{eq:thmPF-equal} are inequalities, so a constant energy forces $W_k=0$ only if the factor in front of $\|W_k\|_{HS}^2$ is positive.
\end{itemize}

Third, we identify fixed points which are the states that commute with $H$. 
Since $H$ is non-degenerate, $W=0$ holds if and only if $\rho$ is diagonal in the eigenbasis $\{ \ket{\lambda_i}\}_{i=1}^{d}$ of $H$. 
The diagonal entries of a state of the form of Eq.~\eqref{eq:rho}, these are the weights $p_1,\dots,p_r$ and $p_i = 0$ for $i>r$. 
The states that commute with $H$ are therefore the states $\rho_\sigma=\sum_ap_{\sigma(a)}\ket{\lambda_a}\bra{\lambda_a}$ of Eq.~\eqref{eq:placements}, where $\sigma$ is a permutation of the $d$ eigenvalues. 
Two permutations give the same state if and only if they differ only in the arrangement of the $d-r$ zeros, so there are $d!/(d-r)!$ different fixed points. 
In the following, we label a fixed point by its diagonal entries $q_a=p_{\sigma(a)}$ and write
\begin{equation}
\label{eq:placement}
\rho_q=\sum_{a=1}^{d}q_a\ket{\lambda_a}\bra{\lambda_a}.
\end{equation}
The target $\rho^{*}$ is the fixed point with $q_a=p_a$ for all $a$.

This proves the fixed points statement.

\subsection{Proof of statement 2: convergence to a fixed point}
\label{app:thm1-convergence}

We show that $\|[\rho,H]\|_{HS}$ tends to zero along every trajectory and that the trajectory cannot move between different fixed points.

Using Theorems~\ref{thmA:energy-C},~\ref{thmA:energy-D}, and~\ref{thmA:energy-PF}, the energy never increases and is bounded from below, so the energy converges.
However, this fact alone does not show that the state converges too. 
We therefore proceed with the proof in four steps as follows. 
The commutator $W$ tends to zero along the trajectory (Step~1). A state with a small commutator is close to a fixed point (Step~2). Consecutive states are close to each other (Step~3). Since different fixed points are a fixed distance apart, the trajectory cannot move from one fixed point to another, and it converges (Step~4).

\smallskip
\uline{Step 1:}
Here we will show that $W$ decreases and goes to zero along a trajectory for all three flows.
\begin{itemize}
\item \textit{Continuous-time DBF.} Theorem~\ref{thmA:energy-C} states that $\partial_tE(\rho(t))=-\|W(t)\|_{HS}^2$. 
The energy is thus non-increasing and lower bounded by $\lambda_1$, so it has a limit at $t\to\infty$. 
Integrating the derivative of the energy, we have 
\begin{equation}
\label{eq:thm1-int}
\int_t^\infty\|W(t')\|_{HS}^2\,\mathrm{d}t'=E(\rho(t))-\lim_{t'\to\infty}E(\rho(t'))\,.
\end{equation}
Where at $t \to \infty$ the RHS becomes zero. Then, to show that the integrand also goes to zero, we lower bound its derivative. 
Using Eq.~\eqref{eq:C}, we have $\partial_tW=[\partial_t\rho,H]=[[W,\rho],H]$, and Eq.~\eqref{eq:comm} gives $\|\partial_tW\|_{HS}\le\Omega\,p_1\|W\|_{HS}=\gamma_{\max}\|W\|_{HS}$. Therefore
\begin{equation}
\partial_t\|W\|_{HS}^2=2\,\mathrm{Re}\,\mathrm{Tr}\left(W^\dagger\,\partial_tW\right)\ge-2\,\|W\|_{HS}\,\|\partial_tW\|_{HS}\ge-2\gamma_{\max}\|W\|_{HS}^2 \,,
\end{equation}
which after taking its integral from $t$ to $t'$, we have $\|W(t')\|_{HS}^2\ge e^{-2\gamma_{\max}(t'-t)}\,\|W(t)\|_{HS}^2$ for $t'\ge t$. 
Inserting this bound into Eq.~\eqref{eq:thm1-int} gives
\begin{equation}
\frac{\|W(t)\|_{HS}^2}{2\gamma_{\max}}\le E(\rho(t))-\lim_{t'\to\infty}E(\rho(t')) .
\end{equation}
Hence $\|W(t)\|_{HS}\to0$ for $t\to\infty$.

\item \textit{Discretized and product-formula DBF.} Using Eqs.~\eqref{eq:thmD},~\eqref{eq:thmPF-ratio}, and~\eqref{eq:thmPF-equal} and the condition $s_{\min}\le s_k\le s_{\max}$, with $s_{\min} > 0$ and $s_{\max} < \bar{s}$, gives $s_k(1-s_k/\bar s)\ge s_{\min}(1-s_{\max}/\bar s)>0$, so that
\begin{equation}
s_{\min}\left(1-\frac{s_{\max}}{\bar s}\right)\|W_k\|_{HS}^2\le E(\rho_k)-E(\rho_{k+1}).
\end{equation}
We sum this inequality over the steps $k=0,\dots,K-1$. In the RHS sum, every state satisfies $E(\rho)\ge\lambda_1$, hence
\begin{equation}
\label{eq:thm1-sum}
s_{\min}\left(1-\frac{s_{\max}}{\bar s}\right)\sum_{k=0}^{K-1}\|W_k\|_{HS}^2\le E(\rho_0)-E(\rho_K)\le E(\rho_0)-\lambda_1 .
\end{equation}
The bound on $\| W_k\|_{HS}^2$ does not depend on $K$. 
The series $\sum_k\|W_k\|_{HS}^2$ therefore converges, and its terms tend to zero. That is, $\|W_k\|_{HS}\to0$ for $k\to\infty$.

\end{itemize}

\smallskip
\uline{Step 2:}
In this step we show that a state $\rho$ with a small commutator $W$ is close to a fixed point.
we write $\mathrm{dist}(\rho)=\min_q\|\rho-\rho_q\|_{HS}$ for the distance of a state $\rho$ from the nearest fixed point. 
We show that every state of the form of Eq.~\eqref{eq:rho} satisfies
\begin{equation}
\label{eq:thm1-dist}
\mathrm{dist}(\rho)\le\frac{\sqrt2\,\|W\|_{HS}}{\min_{a\ne b}|\lambda_a-\lambda_b|} \, .
\end{equation}
Let $\rho_D=\sum_a\rho_{aa}\ket{\lambda_a}\bra{\lambda_a}$ be the diagonal part of $\rho$ in the eigenbasis of $H$, where $\rho_{ab}=\bra{\lambda_a}\rho\ket{\lambda_b}$ and $W_{ab}=\bra{\lambda_a}W\ket{\lambda_b}$. 

The proof of Eq.~\eqref{eq:thm1-dist} has three parts.
First, $\rho$ is close to $\rho_D$. Second, $\rho_D$ is close to a fixed point $\rho_q$, which we construct explicitly. Third, we combine the two distances.

\textit{(i) The state is close to its diagonal part.} By Eq.~\eqref{eq:Wab}, the off-diagonal elements of $\rho$ are $\rho_{ab}=W_{ab}/(\lambda_b-\lambda_a)$, so that
\begin{equation}
\|\rho-\rho_D\|_{HS}^2=\sum_{a\ne b}|\rho_{ab}|^2=\sum_{a\ne b}\frac{|W_{ab}|^2}{(\lambda_a-\lambda_b)^2}\le\frac{\|W\|_{HS}^2}{\min_{a\ne b}(\lambda_a-\lambda_b)^2} \, .
\end{equation}

\textit{(ii) The diagonal part is close to a fixed point.} We choose the fixed point that places the largest weight on the level with the largest diagonal entry of $\rho_D$, the second largest weight on the level with the second largest diagonal entry, and so on. To this end, we arrange the levels as $a_1,\dots,a_d$ such that the diagonal entries $u_j=\rho_{a_ja_j}$ satisfy $u_1\ge u_2\ge\dots\ge u_d$. We let $\rho_q$ be the fixed point with $q_{a_j}=p_j$ for all $j$, where $p_j=0$ for $j>r$ as before. We show that
\begin{equation}
\label{eq:thm1-diag}
\|\rho_D-\rho_q\|_{HS}^2=\sum_{j=1}^{d}(u_j-p_j)^2\le\|\rho-\rho_D\|_{HS}^2 \,.
\end{equation}
We first rewrite the RHS. Since $\rho-\rho_D$ has only off-diagonal elements and $\rho_D$ has only diagonal elements, $\|\rho\|_{HS}^2=\|\rho_D\|_{HS}^2+\|\rho-\rho_D\|_{HS}^2$. With $\|\rho\|_{HS}^2=\mathrm{Tr}(\rho^2)=\sum_jp_j^2$ and $\|\rho_D\|_{HS}^2=\sum_ju_j^2$, this gives $\|\rho-\rho_D\|_{HS}^2=\sum_jp_j^2-\sum_ju_j^2$. Expanding the square on the LHS of Eq.~\eqref{eq:thm1-diag}, we find that Eq.~\eqref{eq:thm1-diag} is equivalent to
\begin{equation}
\label{eq:thm1-key}
\sum_{j=1}^{d}u_j\,(p_j-u_j)\ge0 \,.
\end{equation}

To prove Eq.~\eqref{eq:thm1-key}, we first show that the sum of the $k$ largest diagonal entries is at most the sum of the $k$ largest weights,
\begin{equation}
\label{eq:thm1-partial}
S_k=\sum_{j=1}^{k}u_j\ \le\ T_k=\sum_{j=1}^{k}p_j \,,
\end{equation}
for $1\le k\le d$, with equality for $k=d$, where both sides are equal to $\mathrm{Tr}(\rho)=1$. To see this, we expand the diagonal entries in the eigenbasis $\{\ket{\phi_c}\}_{c=1}^{d}$ of $\rho$ of Appendix~\ref{app:setting}, in which $\rho\ket{\phi_c}=p_c\ket{\phi_c}$. This gives $u_j=\sum_cp_c\,|\langle\lambda_{a_j}|\phi_c\rangle|^2$, and hence
\begin{equation}
S_k=\sum_{c=1}^{d}w_c\,p_c,\qquad w_c=\sum_{j=1}^{k}|\langle\lambda_{a_j}|\phi_c\rangle|^2 \,.
\end{equation}
Each $w_c$ lies between $0$ and $1$, because the sum over all $j$ from $1$ to $d$ gives $\langle\phi_c|\phi_c\rangle=1$. Moreover, $\sum_cw_c=k$, because each $\ket{\lambda_{a_j}}$ is normalized. Since the weights are non-increasing, $p_c\ge p_k$ for $c\le k$ and $p_c\le p_k$ for $c>k$. Therefore
\begin{equation}
S_k-T_k=\sum_{c\le k}(w_c-1)\,p_c+\sum_{c>k}w_c\,p_c\le p_k\Big[\sum_{c\le k}(w_c-1)+\sum_{c>k}w_c\Big]=p_k\Big(\sum_{c}w_c-k\Big)=0 \,,
\end{equation}
where we used $w_c-1\le0$ in the first sum and $w_c\ge0$ in the second sum. This proves Eq.~\eqref{eq:thm1-partial}.

We now obtain Eq.~\eqref{eq:thm1-key} by summation by parts. Since $p_j-u_j=(T_j-S_j)-(T_{j-1}-S_{j-1})$ with $T_0=S_0=0$,
\begin{equation}
\sum_{j=1}^{d}u_j\,(p_j-u_j)=\sum_{k=1}^{d-1}(u_k-u_{k+1})\,(T_k-S_k)+u_d\,(T_d-S_d).
\end{equation}
Every term on the RHS is non-negative. Indeed, $u_k-u_{k+1}\ge0$ by the ordering of the levels, $T_k-S_k\ge0$ by Eq.~\eqref{eq:thm1-partial}, and $T_d-S_d=0$. This proves Eq.~\eqref{eq:thm1-key} and hence Eq.~\eqref{eq:thm1-diag}.

\textit{(iii) Combining the two distances.} The operator $\rho-\rho_D$ has only off-diagonal elements and $\rho_D-\rho_q$ has only diagonal elements. Hence, by parts (i) and (ii),
\begin{equation}
\|\rho-\rho_q\|_{HS}^2=\|\rho-\rho_D\|_{HS}^2+\|\rho_D-\rho_q\|_{HS}^2\le2\,\|\rho-\rho_D\|_{HS}^2\le\frac{2\,\|W\|_{HS}^2}{\min_{a\ne b}(\lambda_a-\lambda_b)^2} .
\end{equation}
This proves Eq.~\eqref{eq:thm1-dist}. Together with Step~1, the distance from the nearest fixed point tends to zero along the trajectory of each of the three flows.

\smallskip
\uline{Step 3:} 
In this step, we show that the consecutive states are close to each other.
This step concerns the discretized and the product-formula DBFs with discrete steps. For the continuous-time DBF, its role is played by the continuity of $\rho(t)$ in $t$.
\begin{itemize}
\item \textit{Discretized DBF.} We write the change of the state in one step as an integral over the step,
\begin{equation}
\rho_{k+1}-\rho_k=\int_0^{s_k}\partial_x\left(e^{xW_k}\rho_k\,e^{-xW_k}\right)\mathrm{d}x=\int_0^{s_k}e^{xW_k}\,[W_k,\rho_k]\,e^{-xW_k}\,\mathrm{d}x .
\end{equation}
The norm of the integrand is $\|[W_k,\rho_k]\|_{HS}\le p_1\|W_k\|_{HS}$ by Eq.~\eqref{eq:comm}. Hence
\begin{equation}
\|\rho_{k+1}-\rho_k\|_{HS}\le s_k\,p_1\|W_k\|_{HS}\le\bar s\,p_1\|W_k\|_{HS} .
\end{equation}
\item \textit{Product-formula DBF.} Let $A(\tau_k)=e^{-i\tau_kH}\rho_k\,e^{i\tau_kH}$. Since $e^{i\theta_k\rho_k}$ commutes with $\rho_k$, the change of the state in one step can be written as
\begin{align}
\rho_{k+1} - \rho_k&=e^{i\tau_kH}e^{i\theta_k\rho_k}A(\tau_k)\,e^{-i\theta_k\rho_k}e^{-i\tau_kH} - \rho_k\nonumber\\
&=e^{i\tau_kH}\rho_k\,e^{-i\tau_kH} - \rho_k +e^{i\tau_kH}e^{i\theta_k\rho_k}\left[A(\tau_k)-\rho_k\right]e^{-i\theta_k\rho_k}e^{-i\tau_kH} .
\end{align}
The operators $e^{i\tau_kH}\rho_k\,e^{-i\tau_kH}-\rho_k$ and $A(\tau_k)-\rho_k$ both have a norm of at most $\tau_k\|W_k\|_{HS}$. For the first operator, this bound follows from $e^{i\tau_kH}\rho_k\,e^{-i\tau_kH}-\rho_k=\int_0^{\tau_k}i\,e^{ixH}[H,\rho_k]\,e^{-ixH}\,\mathrm{d}x$, and the second operator is treated in the same way. Hence
\begin{equation}
\|\rho_{k+1}-\rho_k\|_{HS}\le2\,\tau_k\|W_k\|_{HS} .
\end{equation}
The angles $\tau_k$ are bounded because $s_k<\bar s$. They are $\tau_k=\sqrt{s_kp_1/\Omega}$ for the angle ratio $\theta_k/\tau_k=\Omega/p_1$ and $\tau_k=\sqrt{s_k}$ for equal angles.
\end{itemize}
With Step~1, both bounds give $\|\rho_{k+1}-\rho_k\|_{HS}\to0$.

\smallskip
\uline{Step 4:} In the last step, we show that the trajectory stays near one fixed point. 
There are finitely many fixed points. Let $2\delta>0$ be the smallest distance $\|\rho_q-\rho_{q'}\|_{HS}$ between two different fixed points, i.e., $\min_{q\ne q'} \|\rho_q-\rho_{q'}\|_{HS} = 2 \delta $. 
A state with $\mathrm{dist}(\rho)<\delta$ then has a unique nearest fixed point, because two fixed points within the distance $\delta$ of the same state would be closer to each other than $2\delta$.
\begin{itemize}
\item \textit{Continuous-time DBF.} By Steps~1 and~2, there is a time $T$ such that $\mathrm{dist}(\rho(t))<\delta/2$ for all $t\ge T$. Let $\rho_q$ be the fixed point nearest to $\rho(T)$. Suppose that at some later time the nearest fixed point is a different state $\rho_{q'}$. At that time, $\|\rho(t)-\rho_q\|_{HS}\ge\|\rho_q-\rho_{q'}\|_{HS}-\|\rho(t)-\rho_{q'}\|_{HS}>2\delta-\delta/2>\delta$. The distance $\|\rho(t)-\rho_q\|_{HS}$ is continuous in $t$ and smaller than $\delta/2$ at $t=T$, so there is a time in between at which it equals $\delta$. At this time, every other fixed point is at a distance of at least $2\delta-\delta=\delta$, so that $\mathrm{dist}(\rho(t))=\delta$. This contradicts $\mathrm{dist}(\rho(t))<\delta/2$. The nearest fixed point is therefore $\rho_q$ for all $t\ge T$, and $\|\rho(t)-\rho_q\|_{HS}=\mathrm{dist}(\rho(t))\to0$.
\item \textit{Discretized and product-formula DBF.} By Steps~1 to~3, there is a step $K$ such that $\mathrm{dist}(\rho_k)<\delta/2$ and $\|\rho_{k+1}-\rho_k\|_{HS}<\delta$ for all $k\ge K$. For such a $k$, let $\rho_q$ and $\rho_{q'}$ be the fixed points nearest to $\rho_k$ and to $\rho_{k+1}$. The triangle inequality gives
\begin{equation}
\|\rho_q-\rho_{q'}\|_{HS}\le\|\rho_q-\rho_k\|_{HS}+\|\rho_k-\rho_{k+1}\|_{HS}+\|\rho_{k+1}-\rho_{q'}\|_{HS}<\frac{\delta}{2}+\delta+\frac{\delta}{2}=2\delta .
\end{equation}
Two different fixed points are at least $2\delta$ apart, so $\rho_{q'}=\rho_q$. By induction, the nearest fixed point is the same state $\rho_q$ for all $k\ge K$. Hence $\|\rho_k-\rho_q\|_{HS}=\mathrm{dist}(\rho_k)\to0$. That is, $\rho_k$ converges to $\rho_q$.
\end{itemize}
This proves the convergence statement.

\subsection{Proof of statement 3: necessary condition}
\label{app:thm1-necessary}

We prove the contrapositive. 
Suppose that $A_i = \bra{\lambda_i}P_r\ket{\lambda_i}=0$ for some $i\le r$. 
We show that the state then never acquires a component along $\ket{\lambda_i}$.

First, the assumption means that $\rho_{\mathrm{ref}}\ket{\lambda_i}=0$. Indeed, $\bra{\lambda_i}P_r\ket{\lambda_i}=\sum_{j=1}^{r}|\langle\phi_j|\lambda_i\rangle|^2$ vanishes if and only if $\langle\phi_j|\lambda_i\rangle=0$ for all $j\le r$, and then $\rho_{\mathrm{ref}}\ket{\lambda_i}=\sum_{j=1}^{r}p_j\ket{\phi_j}\langle\phi_j|\lambda_i\rangle=0$.

Next, each flow preserves the property $\rho\ket{\lambda_i}=0$. The reason is that every state $\rho$ satisfies
\begin{equation}
\label{eq:Wlambda}
W\ket{\lambda_i}=\rho H\ket{\lambda_i}-H\rho\ket{\lambda_i}=(\lambda_i-H)\,\rho\ket{\lambda_i},
\end{equation}
so that $\rho\ket{\lambda_i}=0$ implies $W\ket{\lambda_i}=0$.
\begin{itemize}
\item \textit{Continuous-time DBF.} The vector $\ket{x(t)}=\rho(t)\ket{\lambda_i}$ obeys
\begin{equation}
\partial_t\ket{x}=[W,\rho]\ket{\lambda_i}=W\rho\ket{\lambda_i}-\rho\,W\ket{\lambda_i}=\left(W-\rho\,(\lambda_i-H)\right) \ket{x},
\end{equation}
where we used Eq.~\eqref{eq:Wlambda} in the last step. This is a linear differential equation for $\ket{x}$ with bounded coefficients. Its only solution with $\ket{x(0)}=0$ is $\ket{x(t)}=0$ for all $t$.
\item \textit{Discretized DBF.} Suppose that $\rho_k\ket{\lambda_i}=0$. Then $W_k\ket{\lambda_i}=0$ and hence $e^{-s_kW_k}\ket{\lambda_i}=\ket{\lambda_i}$. Equation~\eqref{eq:D} gives $\rho_{k+1}\ket{\lambda_i}=e^{s_kW_k}\rho_k\,e^{-s_kW_k}\ket{\lambda_i}=e^{s_kW_k}\rho_k\ket{\lambda_i}=0$. By induction, $\rho_k\ket{\lambda_i}=0$ for all $k$.
\item \textit{Product-formula DBF.} Suppose that $\rho_k\ket{\lambda_i}=0$. Then $e^{-i\theta_k\rho_k}\ket{\lambda_i}=\ket{\lambda_i}$. Since $\ket{\lambda_i}$ is an eigenstate of $H$,
\begin{equation}
V_k^\dagger\ket{\lambda_i}=e^{i\tau_kH}e^{-i\theta_k\rho_k}e^{-i\tau_kH}\ket{\lambda_i}=e^{-i\tau_k\lambda_i}\,e^{i\tau_kH}e^{-i\theta_k\rho_k}\ket{\lambda_i}=e^{-i\tau_k\lambda_i}\,e^{i\tau_kH}\ket{\lambda_i}=\ket{\lambda_i}.
\end{equation}
Hence $\rho_{k+1}\ket{\lambda_i}=V_k\rho_kV_k^\dagger\ket{\lambda_i}=V_k\rho_k\ket{\lambda_i}=0$. By induction, $\rho_k\ket{\lambda_i}=0$ for all $k$.
\end{itemize}

Finally, we compare with the target. In all three cases $\rho\ket{\lambda_i}=0$ at all times, whereas $\rho^{*}\ket{\lambda_i}=p_i\ket{\lambda_i}$. The Hilbert-Schmidt norm of an operator is at least the norm of the vector that the operator gives when it is applied to a normalized state. Hence
\begin{equation}
\|\rho-\rho^{*}\|_{HS}\ge\left\|(\rho-\rho^{*})\ket{\lambda_i}\right\|=p_i>0
\end{equation}
at all times, and $\rho$ does not converge to $\rho^{*}$. The necessary condition statement holds for any step sizes. This proves statement~3 and completes the proof of Theorem~\ref{thmA:convergence}.\hfill$\square$

\section{Proof of Lemma~\ref{lemma:gradient}}
\label{app:lemma1}
This section proves Lemma~\ref{lemma:gradient}, which lower bounds the energy dissipation rate $\|[\rho,H]\|_{HS}^2$ by the energy excess $\Delta E(\rho)$. Lemma~\ref{lemma:gradient} is a statement about a single state and involves none of the flows. Its difficulty is that both $\|[\rho,H]\|_{HS}^2$ and $\Delta E(\rho)$ depend on all $r$ states at once. As in Appendix~\ref{app:energy-landscape}, we resolve both quantities into contributions of the nested subspaces $V_j$ that are spanned by the states $\{ \ket{\phi_i}\}_{i=1}^{j}$ with the largest weights. The energy excess is a sum of the energy excesses of these subspaces, Eq.~\eqref{eq:dE-subspaces}. The dissipation rate is lower bounded by a sum of their energy leakages, which measure how strongly $H$ couples a subspace to its orthogonal complement. The core of the proof is a relation between the energy leakage and the energy excess of a single subspace (Lemma~\ref{lemA:subspace}). At the end of the appendix, we make precise the sufficient condition for convergence to $\rho^{*}$.

\begin{lemmaA}[Lemma~\ref{lemma:gradient}, formal]
\label{lemA:gradient}
Let $H$ have the eigenvalues $\lambda_1<\dots<\lambda_d$, and let $\rho=\sum_{i=1}^{r}p_i\ket{\phi_i}\bra{\phi_i}$ be a state of the form of Eq.~\eqref{eq:rho} with orthonormal states $\ket{\phi_i}$ and weights $p_1>\dots>p_r>0$. Let $\Delta E(\rho)=E(\rho)-E^{*}$ be its energy excess relative to the target state $\rho^{*}$ of Eq.~\eqref{eq:target}, and let $\gamma_{\min}$ be defined by Eq.~\eqref{eq:gamma-min}. Then
\begin{align}
\Delta E(\rho)&\ge0,\label{app_eq:PL_positive}\\
\|[\rho,H]\|_{HS}^2&\ge2\,\Delta E(\rho)\,\left(\gamma_{\min}-\Delta E(\rho)\right)\,.\label{app_eq:PL}
\end{align} 
\end{lemmaA}
\begin{proof}

Equation~\eqref{app_eq:PL_positive} is Proposition~\ref{propA:landscape}(a). 
The proof of Eq.~\eqref{app_eq:PL} has three steps. Step~1 (Appendix~\ref{app:lemA-subspaces}) writes $\Delta E(\rho)$ as a sum over the energy excesses of the subspaces $V_j$ and lower bounds $\|W\|_{HS}^2$ with $W=[\rho,H]$ by a sum over their energy leakages. Step~2 (Appendix~\ref{app:lemA-one-subspace}) lower bounds the energy leakage of a single subspace by its energy excess (Lemma~\ref{lemA:subspace}). Step~3 (Appendix~\ref{app:lemA-sum-of-subspaces}) inserts the bound of Step~2 into the sums of Step~1.
\end{proof}

\subsection{Step 1: Define subspaces and write \texorpdfstring{$\Delta E$}{energy excess} and \texorpdfstring{$\|W\|_{HS}^2$}{the norm of the gradient} in those subspaces}
\label{app:lemA-subspaces}
We use the subspaces of Appendix~\ref{app:landscape-subspaces}. For $j=1,\dots,r$, the subspace $V_j$ of Eq.~\eqref{eq:Vj} is spanned by the $j$ states with the largest weights $\{ \ket{\phi_i}\}_{i=1}^{j}$, and its orthogonal complement is
\begin{equation}
V_j^\perp=\mathrm{span}\{\ket{\phi_{j+1}},\ket{\phi_{j+2}},\dots,\ket{\phi_d}\},
\end{equation}
where $\ket{\phi_{r+1}},\dots,\ket{\phi_d}$ complete the states to a basis (Appendix~\ref{app:setting}). We assign two numbers to each subspace. The first number is the energy excess $\varepsilon_j$ of Eq.~\eqref{app_eq:energy-excess}. The second number is the energy leakage of the subspace $V_j$,
\begin{equation}
\label{app_eq:energy-leakage}
l_j=\sum_{i=1}^{j}\sum_{k=j+1}^{d}\left|\bra{\phi_i}H\ket{\phi_k}\right|^2 \, ,
\end{equation}
which measures how strongly $H$ maps $V_j$ into $V_j^\perp$. It vanishes exactly when $H$ maps $V_j$ into itself, that is, when $V_j$ is spanned by eigenvectors of $H$.

\smallskip
Now we write the energy excess $\Delta E$ and $\| W\|_{HS}^2$ as weighted sums of the subspace energy excess $\varepsilon_j$ and leakage $l_j$. 

\textit{Energy excess.} Equation~\eqref{eq:dE-subspaces} of Appendix~\ref{app:landscape-subspaces} expresses the energy excess through the subspaces,
\begin{equation}
\label{eq:dE-subspaces-recall}
\Delta E(\rho)=\sum_{j=1}^{r}\delta p_j\,\varepsilon_j .
\end{equation}

\smallskip

\textit{Dissipation rate.} We now derive the corresponding bound for $\|W\|_{HS}^2$. We work in the basis $\{\ket{\phi_i}\}_{i=1}^{d}$ and abbreviate $H_{ik}=\bra{\phi_i}H\ket{\phi_k}$. By Eq.~\eqref{eq:Wab}, the matrix elements of $W$ in this basis are
\begin{equation}
\bra{\phi_i}W\ket{\phi_k}=\bra{\phi_i}\left(\rho H-H\rho\right)\ket{\phi_k}=\left(p_i-p_k\right)H_{ik} .
\end{equation}
The squared Hilbert-Schmidt norm is the sum of the squared moduli of all matrix elements,
\begin{equation}
\|W\|_{HS}^2=\sum_{i=1}^{d}\sum_{k=1}^{d}\left(p_i-p_k\right)^2\left|H_{ik}\right|^2 .
\end{equation}
The terms with $i=k$ vanish, and the terms $(i,k)$ and $(k,i)$ are equal because $|H_{ik}|=|H_{ki}|$. Hence
\begin{equation}
\label{eq:W-pairs}
\|W\|_{HS}^2=2\sum_{1\le i<k\le d}\left(p_i-p_k\right)^2\left|H_{ik}\right|^2 .
\end{equation}
Next, we write one of the two factors $p_i-p_k$ as a sum of weight gaps. For $i<k$,
\begin{equation}
\label{eq:pik-telescope}
p_i-p_k=\sum_{j=i}^{k-1}\left(p_j-p_{j+1}\right)=\sum_{j=i}^{k-1}\delta p_j ,
\end{equation}
where $\delta p_j=0$ for $j>r$, because $p_j=0$ for $j>r$. We insert Eq.~\eqref{eq:pik-telescope} into Eq.~\eqref{eq:W-pairs} and exchange the order of the sums. A pair $i<k$ contributes to the index $j$ if and only if $i\le j<k$, so that
\begin{align}
\|W\|_{HS}^2&=2\sum_{1\le i<k\le d}\ \sum_{j=i}^{k-1}\delta p_j\left(p_i-p_k\right)\left|H_{ik}\right|^2\nonumber\\
&=2\sum_{j=1}^{r}\delta p_j\sum_{i=1}^{j}\sum_{k=j+1}^{d}\left(p_i-p_k\right)\left|H_{ik}\right|^2 .
\label{eq:W-subspaces}
\end{align}
Finally, we bound the remaining factor $p_i-p_k$. Since the weights are ordered, $i\le j<k$ implies $p_i\ge p_j$ and $p_k\le p_{j+1}$, and therefore
\begin{equation}
p_i-p_k\ \ge\ p_j-p_{j+1}=\delta p_j\qquad(i\le j<k) .
\end{equation}
Inserting this bound into Eq.~\eqref{eq:W-subspaces} and using the definition of the energy leakage, Eq.~\eqref{app_eq:energy-leakage}, we obtain
\begin{equation}
\label{eq:W-leakage}
\|W\|_{HS}^2\ \ge\ 2\sum_{j=1}^{r}\delta p_j^2\sum_{i=1}^{j}\sum_{k=j+1}^{d}\left|H_{ik}\right|^2=2\sum_{j=1}^{r}\delta p_j^2\,l_j \,.
\end{equation}
Equations~\eqref{eq:dE-subspaces-recall} and~\eqref{eq:W-leakage} reduce the lemma to a relation between the energy leakage $l_j$ and the energy excess $\varepsilon_j$ of a single subspace.

\subsection{Step 2: Bound the energy excess and leakage of a single subspace}
\label{app:lemA-one-subspace}
The relation between the energy leakage and the energy excess does not involve $\rho$, and we state it for an arbitrary subspace. It complements Lemma~\ref{lemA:subspace-excess}, which lower bounds the energy excess of a subspace.

\begin{lemmaA}[Energy leakage of a subspace]
\label{lemA:subspace}
Let $H$ have the eigenvalues $\lambda_1<\dots<\lambda_d$ with eigenstates $\ket{\lambda_a}$, let $1\le j<d$, and let $V$ be a $j$-dimensional subspace with the projector $P_V$. Let $\{\ket{x_i}\}_{i=1}^{j}$ be an orthonormal basis of $V$, and let $\{\ket{y_k}\}_{k=1}^{d-j}$ be an orthonormal basis of its orthogonal complement $V^\perp$. Define the energy excess and the energy leakage of $V$ as
\begin{equation}
\label{eq:eps-l-V}
\varepsilon(V)=\sum_{i=1}^{j}\left(\bra{x_i}H\ket{x_i}-\lambda_i\right) \,,\qquad l(V)=\sum_{i=1}^{j}\sum_{k=1}^{d-j}\left|\bra{y_k}H\ket{x_i}\right|^2 \,.
\end{equation}
Then
\begin{equation}
\label{eq:sub-b}
l(V)\ \ge\ \varepsilon(V)\left(\delta\lambda_{j+1}-\varepsilon(V)\right)\,,
\end{equation}
where $\delta\lambda_{j+1}=\lambda_{j+1}-\lambda_j$.
\end{lemmaA}
\begin{proof}
The proof has five parts. We first show that $\varepsilon(V)$ and $l(V)$ do not depend on the choice of the bases. We then choose a basis of $V$ that is adapted to the $j$ lowest eigenstates of $H$, express $\varepsilon(V)$ in this basis, and bound $l(V)$ from below.

\smallskip

\textit{(i) Independence of the bases.} Both numbers depend only on $V$. Indeed,
\begin{equation}
\varepsilon(V)=\mathrm{Tr}\left(P_VH\right)-\sum_{i=1}^{j}\lambda_i,\qquad l(V)=\left\|\left(\mathbb{I}-P_V\right)HP_V\right\|_{HS}^2 .
\end{equation}
The first expression is Eq.~\eqref{eq:eps-V} of Lemma~\ref{lemA:subspace-excess}. In particular, $\varepsilon(V_j)=\varepsilon_j$ and $l(V_j)=l_j$ for the bases $\{\ket{\phi_i}\}_{i\le j}$ and $\{\ket{\phi_k}\}_{k>j}$. We are therefore free to choose convenient bases.

\smallskip

\textit{(ii) Low and high vectors.} Let $\Pi_j=\sum_{a=1}^{j}\ket{\lambda_a}\bra{\lambda_a}$. We call a vector \textit{low} if it lies in the span of the $j$ lowest eigenstates, and \textit{high} if it is orthogonal to this span,
\begin{equation}
\mathcal{V}_{\mathrm{low}}=\mathrm{span}\{\ket{\lambda_a}\}_{a=1}^{j},\qquad \mathcal{V}_{\mathrm{high}}=\mathrm{span}\{\ket{\lambda_a}\}_{a=j+1}^{d} .
\end{equation}
The Hamiltonian maps low vectors to low vectors and high vectors to high vectors. Hence, for a low vector $\ket{u}$ and a high vector $\ket{v}$,
\begin{equation}
\label{eq:low-high-cross}
\bra{u}H\ket{v}=0 .
\end{equation}
Moreover, for unit vectors $\ket{u}\in\mathcal{V}_{\mathrm{low}}$ and $\ket{v}\in\mathcal{V}_{\mathrm{high}}$,
\begin{subequations}
\label{eq:low-high}
    \begin{align}
    \bra{u}H\ket{u} &\le\lambda_j \,, \\ \bra{v}H\ket{v} &\ge\lambda_{j+1} \,,
    \end{align}
\end{subequations}
because $\bra{u}H\ket{u}$ is an average of the eigenvalues $\lambda_1,\dots,\lambda_j$ and $\bra{v}H\ket{v}$ is an average of the eigenvalues $\lambda_{j+1},\dots,\lambda_d$.

\smallskip

\textit{(iii) Choice of basis of $V$ adapted to the low space.} Let $\{\ket{x'_n}\}_{n=1}^{j}$ be an arbitrary orthonormal basis of $V$, and consider the $j\times j$ Hermitian matrix with the entries $\bra{x'_n}\Pi_j\ket{x'_m}$. Its eigenvalues lie between zero and one, because $0\le\Pi_j\le\mathbb{I}$. We write them as $\cos^2\alpha_n$ with $\alpha_n\in[0,\pi/2]$ and choose the orthonormal basis $\{\ket{x_n}\}_{n=1}^{j}$ of $V$ that diagonalizes this matrix. In this basis,
\begin{align}
\bra{x_n}\Pi_j\ket{x_m} &= \cos^2\alpha_n\,\delta_{nm},\\ \bra{x_n}(\mathbb{I}-\Pi_j)\ket{x_m} &= \delta_{nm}-\cos^2\alpha_n\,\delta_{nm}=\sin^2\alpha_n\,\delta_{nm}.
\end{align}
The low parts $\Pi_j\ket{x_n}$ of the basis vectors are thus mutually orthogonal with the lengths $\cos\alpha_n$, and the high parts $(\mathbb{I}-\Pi_j)\ket{x_n}$ are mutually orthogonal with the lengths $\sin\alpha_n$. Normalizing them gives
\begin{equation}
\label{eq:x-decomposition}
\ket{x_n}=\cos\alpha_n\ket{u_n}+\sin\alpha_n\ket{v_n},
\end{equation}
with orthonormal low vectors $\ket{u_n}\in\mathcal{V}_{\mathrm{low}}$ and orthonormal high vectors $\ket{v_n}\in\mathcal{V}_{\mathrm{high}}$. If $\cos\alpha_n=0$ for some $n$, the vector $\ket{u_n}$ is not determined by Eq.~\eqref{eq:x-decomposition}, and we choose these vectors such that $\{\ket{u_n}\}_{n=1}^{j}$ is an orthonormal basis of $\mathcal{V}_{\mathrm{low}}$. If $\sin\alpha_n=0$, the vector $\ket{v_n}$ does not appear. For every $n$ with $\sin\alpha_n>0$ we define
\begin{equation}
\label{eq:gn}
g_n=\bra{v_n}H\ket{v_n}-\bra{u_n}H\ket{u_n}\ \ge\ \lambda_{j+1}-\lambda_j=\delta\lambda_{j+1}>0 \,,
\end{equation}
where the inequality follows from Eqs.~\eqref{eq:low-high}. For $\sin\alpha_n=0$ we set $g_n=\delta\lambda_{j+1}$. These numbers enter all following expressions with the factor $\sin\alpha_n=0$.

\smallskip

\textit{(iv) Energy excess in the adapted basis.} By Eq.~\eqref{eq:x-decomposition} and Eq.~\eqref{eq:low-high-cross}, the cross terms between low and high vectors vanish, and
\begin{align}
\bra{x_n}H\ket{x_n}&=\cos^2\alpha_n\bra{u_n}H\ket{u_n}+\sin^2\alpha_n\bra{v_n}H\ket{v_n}\nonumber\\
&=\bra{u_n}H\ket{u_n}+\sin^2\alpha_n\left(\bra{v_n}H\ket{v_n}-\bra{u_n}H\ket{u_n}\right)\nonumber\\
&=\bra{u_n}H\ket{u_n}+\sin^2\alpha_n\,g_n \,,
\label{eq:xHx}
\end{align}
where the second line uses $\cos^2\alpha_n=1-\sin^2\alpha_n$. The vectors $\{\ket{u_n}\}_{n=1}^{j}$ form an orthonormal basis of $\mathcal{V}_{\mathrm{low}}$, so that
\begin{equation}
\label{eq:sum-u}
\sum_{n=1}^{j}\bra{u_n}H\ket{u_n}=\mathrm{Tr}\left(\Pi_jH\right)=\sum_{i=1}^{j}\lambda_i \,.
\end{equation}
Summing Eq.~\eqref{eq:xHx} over $n$ and subtracting Eq.~\eqref{eq:sum-u}, we obtain
\begin{equation}
\label{eq:eps-angles}
\varepsilon(V)=\sum_{n=1}^{j}\bra{x_n}H\ket{x_n}-\sum_{i=1}^{j}\lambda_i=\sum_{n=1}^{j}\sin^2\alpha_n\,g_n \,.
\end{equation}

\smallskip

\textit{(v) Energy leakage.} For each $n$ with $\sin\alpha_n>0$, we define the vector
\begin{equation}
\label{eq:yn}
\ket{y_n}=-\sin\alpha_n\ket{u_n}+\cos\alpha_n\ket{v_n} \,.
\end{equation}
These vectors are orthonormal, and they are orthogonal to every $\ket{x_m}$. Indeed, since the vectors $\ket{u_n}$ are orthonormal, the vectors $\ket{v_n}$ are orthonormal, and low vectors are orthogonal to high vectors,
\begin{align}
\langle y_n|y_m\rangle&=\left(\sin\alpha_n\sin\alpha_m+\cos\alpha_n\cos\alpha_m\right)\delta_{nm}=\delta_{nm} \,,\\
\langle y_n|x_m\rangle&=\left(-\sin\alpha_n\cos\alpha_m+\cos\alpha_n\sin\alpha_m\right)\delta_{nm}=0 \,.
\end{align}
The vectors $\ket{y_n}$ therefore lie in $V^\perp$ and can be completed to an orthonormal basis of $V^\perp$. By part (i), we may evaluate $l(V)$ in this basis. All terms of $l(V)$ are nonnegative. Keeping only the terms in which $\ket{y_n}$ is paired with $\ket{x_n}$, we obtain
\begin{equation}
\label{eq:l-diag}
l(V)\ \ge\ \sum_{n:\,\sin\alpha_n>0}\left|\bra{y_n}H\ket{x_n}\right|^2 .
\end{equation}
We evaluate these matrix elements with Eqs.~\eqref{eq:x-decomposition} and~\eqref{eq:yn}. The cross terms vanish by Eq.~\eqref{eq:low-high-cross}, so that
\begin{align}
\bra{y_n}H\ket{x_n}&=-\sin\alpha_n\cos\alpha_n\bra{u_n}H\ket{u_n}+\cos\alpha_n\sin\alpha_n\bra{v_n}H\ket{v_n}\nonumber\\
&=\sin\alpha_n\cos\alpha_n\,g_n \,.
\end{align}
Inserting this result into Eq.~\eqref{eq:l-diag} and using $\cos^2\alpha_n=1-\sin^2\alpha_n$ gives
\begin{align}
l(V)&\ge\sum_{n=1}^{j}\sin^2\alpha_n\cos^2\alpha_n\,g_n^2\nonumber\\
&=\sum_{n=1}^{j}\sin^2\alpha_n\,g_n^2-\sum_{n=1}^{j}\left(\sin^2\alpha_n\,g_n\right)^2 ,
\label{eq:l-two-sums}
\end{align}
where the terms with $\sin\alpha_n=0$ vanish. We bound the two sums separately. In the first sum, $g_n\ge\delta\lambda_{j+1}$ by Eq.~\eqref{eq:gn}, and Eq.~\eqref{eq:eps-angles} gives
\begin{equation}
\sum_{n=1}^{j}\sin^2\alpha_n\,g_n^2\ \ge\ \delta\lambda_{j+1}\sum_{n=1}^{j}\sin^2\alpha_n\,g_n=\delta\lambda_{j+1}\,\varepsilon(V) \,.
\end{equation}
In the second sum, the numbers $z_n=\sin^2\alpha_n\,g_n$ are nonnegative, so that $\sum_nz_n^2\le(\sum_nz_n)^2$, and Eq.~\eqref{eq:eps-angles} gives
\begin{equation}
\sum_{n=1}^{j}\left(\sin^2\alpha_n\,g_n\right)^2\ \le\ \Big(\sum_{n=1}^{j}\sin^2\alpha_n\,g_n\Big)^2=\varepsilon(V)^2 \,.
\end{equation}
Inserting both bounds into Eq.~\eqref{eq:l-two-sums}, we arrive at
\begin{equation}
l(V)\ \ge\ \delta\lambda_{j+1}\,\varepsilon(V)-\varepsilon(V)^2=\varepsilon(V)\left(\delta\lambda_{j+1}-\varepsilon(V)\right) ,
\end{equation}
which is Eq.~\eqref{eq:sub-b}.
\end{proof}

\subsection{Step 3: Lower bound on \texorpdfstring{$\|W\|_{HS}^2$}{gradient norm}}
\label{app:lemA-sum-of-subspaces}

We now apply Lemmas~\ref{lemA:subspace-excess} and~\ref{lemA:subspace} to $V=V_j$ for each $j\le r$, for which $\varepsilon(V_j)=\varepsilon_j$ and $l(V_j)=l_j$, and insert the result into the sums of Step~1.

First, we bound each term of the energy excess by the whole sum. By Eq.~\eqref{eq:sub-a}, $\varepsilon_j\ge0$ for every $j\le r$. All terms of Eq.~\eqref{eq:dE-subspaces-recall} are therefore nonnegative, and each of them is at most the sum. That is, for every $j\le r$,
\begin{equation}
\label{eq:term-bound}
0\ \le\ \delta p_j\,\varepsilon_j\ \le\ \sum_{j'=1}^{r}\delta p_{j'}\,\varepsilon_{j'}=\Delta E(\rho)\,.
\end{equation}

Second, we bound the energy leakage of each subspace. We multiply Eq.~\eqref{eq:sub-b} for $V=V_j$ by $\delta p_j>0$,
\begin{align}
\delta p_j\,l_j&\ge\delta p_j\,\varepsilon_j\left(\delta\lambda_{j+1}-\varepsilon_j\right)\nonumber\\
&=\varepsilon_j\left(\delta p_j\,\delta\lambda_{j+1}-\delta p_j\,\varepsilon_j\right)\nonumber\\
&=\varepsilon_j\left(\gamma_{j,j+1}-\delta p_j\,\varepsilon_j\right) ,
\end{align}
where the last line uses $\gamma_{j,j+1}=\delta p_j\,\delta\lambda_{j+1}$. In the bracket, $\gamma_{j,j+1}\ge\gamma_{\min}$ by the definition of $\gamma_{\min}$, and $\delta p_j\,\varepsilon_j\le\Delta E(\rho)$ by Eq.~\eqref{eq:term-bound}. Since $\varepsilon_j\ge0$, both replacements can only lower the right-hand side, whatever the sign of the bracket is. Hence
\begin{equation}
\label{eq:delta-p-l}
\delta p_j\,l_j\ \ge\ \varepsilon_j\left(\gamma_{\min}-\Delta E(\rho)\right) .
\end{equation}

Third, we sum over the subspaces. We start from Eq.~\eqref{eq:W-leakage}, insert Eq.~\eqref{eq:delta-p-l} multiplied by $\delta p_j>0$, and use Eq.~\eqref{eq:dE-subspaces-recall},
\begin{align}
\frac12\|W\|_{HS}^2&\ge\sum_{j=1}^{r}\delta p_j\left(\delta p_j\,l_j\right)\nonumber\\
&\ge\sum_{j=1}^{r}\delta p_j\,\varepsilon_j\left(\gamma_{\min}-\Delta E(\rho)\right)\nonumber\\
&=\left(\gamma_{\min}-\Delta E(\rho)\right)\sum_{j=1}^{r}\delta p_j\,\varepsilon_j\nonumber\\
&=\left(\gamma_{\min}-\Delta E(\rho)\right)\Delta E(\rho) .
\end{align}
With $W=[\rho,H]$, this inequality is Eq.~\eqref{app_eq:PL}, which completes the proof of Lemma~\ref{lemA:gradient}. \hfill$\square$

\subsection{The sufficient condition for convergence to the global energy minimum}
\label{app:lemma1-consequences}

Lemma~\ref{lemA:gradient} shows that the dissipation rate is strictly positive whenever $0<\Delta E(\rho)<\gamma_{\min}$. A state in this range is therefore not a fixed point, in agreement with Proposition~\ref{propA:landscape}(c). Then $\Delta E(\rho)<\gamma_{\min}$ is a sufficient condition for convergence to $\rho^{*}$. The following corollary shows this conclusion.

\begin{corollaryA}[Sufficient condition for convergence to $\rho^{*}$]
\label{corA:criterion}
Let $\rho$ evolve from the reference state $\rho_{\mathrm{ref}}$ under one of the three flows under the conditions of statement~2 of Theorem~\ref{thmA:convergence}. Then $\rho$ converges to the target state $\rho^{*}$ if and only if the energy excess satisfies $\Delta E(\rho(t))<\gamma_{\min}$ at some finite flow time $t$. In particular, $\Delta E(\rho_{\mathrm{ref}})<\gamma_{\min}$ is sufficient for convergence to $\rho^{*}$.
\end{corollaryA}
\begin{proof}
By statement~2 of Theorem~\ref{thmA:convergence}, $\rho$ converges to a fixed point $\rho_\sigma$, and by Theorems~\ref{thmA:energy-C} to~\ref{thmA:energy-PF} the energy does not increase along the trajectory. If $\Delta E(\rho(t))<\gamma_{\min}$ at some flow time $t$, then
\begin{equation}
\Delta E(\rho_\sigma)\ \le\ \Delta E(\rho(t))\ <\ \gamma_{\min} ,
\end{equation}
and Proposition~\ref{propA:landscape}(c) gives $\rho_\sigma=\rho^{*}$. Conversely, if $\rho$ converges to $\rho^{*}$, then $\Delta E(\rho(t))\to0$, so that $\Delta E(\rho(t))<\gamma_{\min}$ from some flow time on.
\end{proof}

Corollary~\ref{corA:criterion} is a condition on the trajectory. As a condition on the initial state,$\Delta E(\rho_{\mathrm{ref}})<\gamma_{\min}$ is sufficient but not necessary, since a trajectory that starts above the lowest saddle can fall below it later. 
The corollary states only that the trajectory converges. Theorem~\ref{theorem:convergence-rate}, which we prove in Appendix~\ref{app:thm2}, uses Lemma~\ref{lemA:gradient} to show that the convergence is exponential.

\section{Formal version and proof of Theorem~\ref{theorem:convergence-rate}}
\label{app_sec:exponential-convergence-proof}
\label{app:thm2}
This section states the formal version of Theorem~\ref{theorem:convergence-rate}, including the explicit constant $c$, and proves it. The theorem has one statement for each of the three flows. Each statement is the chain of inequalities of Eq.~\eqref{eq:rate}, with one intermediate quantity added. The first inequality of the chain, $1-F_i(t)\le\Delta E(\rho(t))/\gamma_{\min}$, is a property of a single state, which we prove in Lemma~\ref{lemA:fidelity}. The remaining inequalities describe the decay of the energy excess $\Delta E(\rho(t))$ along the trajectory. For each flow, we combine the energy relation of Appendix~\ref{app:energy} with Lemma~\ref{lemma:gradient} and obtain a closed inequality for the energy excess, which we then solve. The lower bound $2\Delta E(\rho)\,(\gamma_{\min}-\Delta E(\rho))$ of Lemma~\ref{lemma:gradient} vanishes at the target and again at $\Delta E(\rho)=\gamma_{\min}$. We therefore follow the ratio $\Delta E(\rho)/(\gamma_{\min}-\Delta E(\rho))$, which decays at a rate that does not depend on the energy excess. As a result, the initial energy excess enters only the prefactor $c_0$ and not the rate. After the proof, we derive the number of steps required to reach a target infidelity.

\subsection{Formal version of Theorem~\ref{theorem:convergence-rate}}
We first fix what the quantities in Eq.~\eqref{eq:rate} mean for each of the three flows. 
As in the main text, $f_i(\ket{\phi})=|\langle\lambda_i|\phi\rangle|^2$ is the fidelity of a state $\ket{\phi}$ with the target eigenstate $\ket{\lambda_i}$, and $\Delta E(\rho)=E(\rho)-E^{*}$ is the energy excess of a state $\rho$ relative to the target state $\rho^{*}$ of Eq.~\eqref{eq:target}. For the continuous-time DBF, $t$ is the evolution time and $\rho(t)=\sum_{i}p_i\ket{\phi_i(t)}\bra{\phi_i(t)}$ is the solution of Eq.~\eqref{eq:C} with $\rho(0)=\rho_{\mathrm{ref}}$, so that
\begin{equation}
\label{eq:dE-of-t-C}
\Delta E(\rho(t))=\mathrm{Tr}\left(\rho(t)H\right)-E^{*},\qquad F_i(t)=f_i(\ket{\phi_i(t)}) .
\end{equation}
For the discretized and the product-formula DBF, the flow time takes the discrete values $t=\sum_{k=0}^{K-1}s_k$ with $K=0,1,2,\dots$, and the state at this flow time is the state $\rho_K=\sum_{i}p_i\ket{\phi_i^{(K)}}\bra{\phi_i^{(K)}}$ after $K$ steps of Eq.~\eqref{eq:D} or Eq.~\eqref{eq:PF} with $\rho_0=\rho_{\mathrm{ref}}$. That is,
\begin{equation}
\label{eq:dE-of-t-D}
\rho(t)=\rho_K,\qquad \Delta E(\rho(t))=\Delta E(\rho_K)=\mathrm{Tr}\left(\rho_KH\right)-E^{*},\qquad F_i(t)=f_i(\ket{\phi_i^{(K)}})\qquad\textup{for}\qquad t=\sum_{k=0}^{K-1}s_k .
\end{equation}

\begin{theoremA}[Theorem~\ref{theorem:convergence-rate}, formal]
\label{thmA:rates}
Let $H$ have the eigenvalues $\lambda_1<\dots<\lambda_d$, and let $\rho$ evolve from the reference state $\rho_{\mathrm{ref}}$ of the form of Eq.~\eqref{eq:rho}, with weights $p_1>\dots>p_r>0$, under the continuous-time DBF of Eq.~\eqref{eq:C}, the discretized DBF of Eq.~\eqref{eq:D}, or the product-formula DBF of Eq.~\eqref{eq:PF}. Let $\gamma_{\min}$ and $\gamma_{\max}=p_1\Omega$ be defined by Eqs.~\eqref{eq:gamma-min} and~\eqref{eq:app-gamma}, and let $\Delta E(\rho(t))$ and $F_i(t)$ be given by Eq.~\eqref{eq:dE-of-t-C} for the continuous-time DBF and by Eq.~\eqref{eq:dE-of-t-D} for the other two flows. Suppose that the initial energy excess satisfies $\Delta E(\rho_{\mathrm{ref}})<\gamma_{\min}$, and let
\begin{equation}
\label{eq:c0}
c_0=\frac{\Delta E(\rho_{\mathrm{ref}})}{\gamma_{\min}-\Delta E(\rho_{\mathrm{ref}})} .
\end{equation}
\begin{enumerate}
\item \textbf{Continuous-time DBF:} For all $t\ge0$ and all $i\le r$,
\begin{equation}
\label{eq:rate-C}
1-F_i(t)\le\frac{\Delta E(\rho(t))}{\gamma_{\min}}\le\frac{\Delta E(\rho(t))}{\gamma_{\min}-\Delta E(\rho(t))}\le c_0\,e^{-2\gamma_{\min}t}.
\end{equation}
\item \textbf{Discretized DBF:} Here $\bar s=2/\gamma_{\max}$.
\begin{enumerate}
\item \textit{Every step.} If $0< s_k<\bar s$ for all $k$, then at every step
\begin{equation}
\label{eq:rate-step}
\frac{\Delta E(\rho_{k+1})}{\gamma_{\min}-\Delta E(\rho_{k+1})}\le\frac{\Delta E(\rho_k)}{\gamma_{\min}-\Delta E(\rho_k)}\;\frac{1}{1+2s_k\left(1-s_k/\bar s\right)\gamma_{\min}} .
\end{equation}
\item \textit{After $K$ steps.} If, in addition, $s_k\le s_{\max}$ for all $k$ with a constant $s_{\max}<\bar s$, then for all $K\ge0$ and all $i\le r$, with the flow time $t = \sum_{k=0}^{K-1} s_k$,
\begin{equation}
\label{eq:rate-c}
1-F_i(t)\le\frac{\Delta E(\rho_K)}{\gamma_{\min}}\le\frac{\Delta E(\rho_K)}{\gamma_{\min}-\Delta E(\rho_K)}\le c_0\,e^{-2c\,\gamma_{\min}t},
\end{equation}
with the constant $c$ given by
\begin{equation}
\label{eq:rate-const}
    c=\frac{1-s_{\max}/\bar s}{1+2s_{\max}\left(1-s_{\max}/\bar s\right)\gamma_{\min}} .
\end{equation}
The constant $c$ satisfies $0<c<1$, tends to one for $s_{\max}\to0$, and satisfies $c\ge1/4$ if $s_{\max}\le\bar s/2$.
\end{enumerate}
\item \textbf{Product-formula DBF:} Statement~2 holds with $\bar s=\frac{12}{7\gamma_{\max}}$ for the angle ratio $\theta_k/\tau_k=\Omega/p_1$, and with $\bar s=  \frac{12}{7 \max(\Omega, 1)^2}$ for equal angles $\tau_k=\theta_k = \sqrt{s_k}$.
\end{enumerate}
\end{theoremA}
\begin{proof}
Every chain consists of three inequalities. The first inequality of Eqs.~\eqref{eq:rate-C} and~\eqref{eq:rate-c} is Lemma~\ref{lemA:fidelity} of Appendix~\ref{app:infidelity}, applied to the state at the flow time $t$. The second inequality holds because $0<\gamma_{\min}-\Delta E(\rho(t))\le\gamma_{\min}$. The third inequality is the dynamical part, which we prove for one flow at a time in Appendices~\ref{app:thm2-C}, \ref{app:thm2-D}, and~\ref{app:thm2-PF}.
\end{proof}

\subsection{Upper bounding infidelities by the energy excess}
\label{app:infidelity}

The first inequality in each statement of Theorem~\ref{thmA:rates} does not involve the dynamics. It is a property of every state of the form of Eq.~\eqref{eq:rho}, which we now prove.

\begin{lemmaA}[Energy excess upper bounding the infidelities]
\label{lemA:fidelity}
For every state $\rho=\sum_{i=1}^{r}p_i\ket{\phi_i}\bra{\phi_i}$ of the form of Eq.~\eqref{eq:rho} and every $i\le r$,
\begin{equation}
\label{eq:fidelity}
1-f_i(\ket{\phi_i})\le\frac{\Delta E(\rho)}{\min\{\gamma_{i-1,i},\gamma_{i,i+1}\}}\le\frac{\Delta E(\rho)}{\gamma_{\min}},
\end{equation}
where $f_i(\ket{\phi_i}) = |\langle\lambda_i | \phi_i \rangle|^2$, and the minimum is $\gamma_{1,2}$ for $i=1$.
\end{lemmaA}
\begin{proof}
The state $\ket{\phi_i}$ lies in the subspace $V_i$ and is orthogonal to the subspace $V_{i-1}$ of Eq.~\eqref{eq:Vj}. We bound its infidelity by the misalignment of these two subspaces with the lowest eigenstates, and then bound the misalignment by the energy excess. As in Eq.~\eqref{eq:Pj}, $P_j=\sum_{i'=1}^{j}\ket{\phi_{i'}}\bra{\phi_{i'}}$ is the projector onto $V_j$ and $\Pi_j=\sum_{a=1}^{j}\ket{\lambda_a}\bra{\lambda_a}$ is the projector onto $\mathrm{span}\{\ket{\lambda_1},\dots,\ket{\lambda_j}\}$, with $P_0=\Pi_0=0$.

\smallskip

\textit{(i) Splitting the infidelity.} Since $\ket{\lambda_i}\bra{\lambda_i}=\Pi_i-\Pi_{i-1}$, the fidelity is $f_i(\ket{\phi_i})=\bra{\phi_i}\Pi_i\ket{\phi_i}-\bra{\phi_i}\Pi_{i-1}\ket{\phi_i}$, and
\begin{equation}
\label{eq:infidelity-split}
1-f_i(\ket{\phi_i})=\bra{\phi_i}\left(\mathbb{I}-\Pi_i\right)\ket{\phi_i}+\bra{\phi_i}\Pi_{i-1}\ket{\phi_i}.
\end{equation}
The first term is the population of $\ket{\phi_i}$ above the level $i$, and the second term is its population below the level $i$.

\smallskip

\textit{(ii) Bounding each term by a trace.} The state $\ket{\phi_i}$ lies in $V_i$, so that $P_i-\ket{\phi_i}\bra{\phi_i}$ is a projector and hence positive semidefinite. Since $\mathbb{I}-\Pi_i$ is positive semidefinite as well,
\begin{equation}
\bra{\phi_i}\left(\mathbb{I}-\Pi_i\right)\ket{\phi_i}\ \le\ \mathrm{Tr}\left(P_i\left(\mathbb{I}-\Pi_i\right)\right)=i-\mathrm{Tr}\left(P_i\Pi_i\right) .
\end{equation}
The state $\ket{\phi_i}$ is orthogonal to $V_{i-1}$, so that $\mathbb{I}-P_{i-1}-\ket{\phi_i}\bra{\phi_i}$ is a projector, and
\begin{equation}
\bra{\phi_i}\Pi_{i-1}\ket{\phi_i}\ \le\ \mathrm{Tr}\left(\left(\mathbb{I}-P_{i-1}\right)\Pi_{i-1}\right)=(i-1)-\mathrm{Tr}\left(P_{i-1}\Pi_{i-1}\right) .
\end{equation}
Inserting both bounds into Eq.~\eqref{eq:infidelity-split} gives
\begin{equation}
    1-f_i(\ket{\phi_i})\ \le\ \left[\,i-\mathrm{Tr}\left(P_i\Pi_i\right)\right]+\left[(i-1)-\mathrm{Tr}\left(P_{i-1}\Pi_{i-1}\right)\right] .
    \label{eq:infidelities-bound-trace}
\end{equation}

\smallskip

\textit{(iii) Bounding the traces by the energy excesses of the subspaces.} Equation~\eqref{eq:sub-a} of Lemma~\ref{lemA:subspace-excess}, applied to $V=V_j$ with $\varepsilon(V_j)=\varepsilon_j$, reads $\varepsilon_j\ge\delta\lambda_{j+1}\,[\,j-\mathrm{Tr}(P_j\Pi_j)]$. That is,
\begin{equation}
\label{eq:trace-eps}
    j-\mathrm{Tr}\left(P_j\Pi_j\right)\ \le\ \frac{\varepsilon_j}{\delta \lambda_{j+1}} .
\end{equation}
We use Eq.~\eqref{eq:trace-eps} for $j=i$ and for $j=i-1$ in Eq.~\eqref{eq:infidelities-bound-trace},
\begin{equation}
\label{eq:fid-eps}
1-f_i(\ket{\phi_i})\ \le\ \frac{\varepsilon_i}{\delta\lambda_{i+1}}+\frac{\varepsilon_{i-1}}{\delta\lambda_i},
\end{equation}
where the second term is absent for $i=1$.

\smallskip

\textit{(iv) Comparing with the energy excess of the state.} By Eq.~\eqref{eq:dE-subspaces}, $\Delta E(\rho)=\sum_{j}\delta p_j\,\varepsilon_j$ with nonnegative terms. Keeping only the terms $j=i-1$ and $j=i$, and using $\delta p_j=\gamma_{j,j+1}/\delta\lambda_{j+1}$, we find
\begin{align}
\Delta E(\rho)&\ge\delta p_{i-1}\,\varepsilon_{i-1}+\delta p_i\,\varepsilon_i\nonumber\\
&=\gamma_{i-1,i}\,\frac{\varepsilon_{i-1}}{\delta\lambda_i}+\gamma_{i,i+1}\,\frac{\varepsilon_i}{\delta\lambda_{i+1}}\nonumber\\
&\ge\min\{\gamma_{i-1,i},\gamma_{i,i+1}\}\left(\frac{\varepsilon_{i-1}}{\delta\lambda_i}+\frac{\varepsilon_i}{\delta\lambda_{i+1}}\right)\nonumber\\
&\ge\min\{\gamma_{i-1,i},\gamma_{i,i+1}\}\left(1-f_i(\ket{\phi_i})\right),
\end{align}
where the last step is Eq.~\eqref{eq:fid-eps}. This inequality is the first inequality of Eq.~\eqref{eq:fidelity}. The second inequality follows from $\gamma_{i-1,i}\ge\gamma_{\min}$ and $\gamma_{i,i+1}\ge\gamma_{\min}$.
\end{proof}
Lemma~\ref{lemA:fidelity} shows that the fidelity of the $i$th state is controlled by the two energy gaps next to the level $i$ and the two weight gaps next to $p_i$. It holds for every state, whatever its energy excess.

\subsection{Proof of statement 1: continuous-time DBF}
\label{app:thm2-C}
In this part, we derive a closed differential inequality for the energy excess and solve it for the ratio $\Delta E(\rho(t))/\left(\gamma_{\min}-\Delta E(\rho(t))\right)$. Within this subsection, we abbreviate
\begin{equation}
x(t)=\Delta E(\rho(t)),\qquad y(t)=\frac{x(t)}{\gamma_{\min}-x(t)},
\end{equation}
so that $x(0)=\Delta E(\rho_{\mathrm{ref}})$ and $y(0)=c_0$.

\smallskip

\textit{(i) A closed inequality for the energy excess.} Since $E^{*}$ is a constant, $\partial_tx(t)=\partial_tE(\rho(t))$. Theorem~\ref{thmA:energy-C} expresses this derivative through $W(t)=[\rho(t),H]$, and Lemma~\ref{lemma:gradient}, applied to the state $\rho(t)$, bounds $\|W(t)\|_{HS}^2$ from below,
\begin{equation}
\label{eq:logistic}
\partial_tx(t)=-\|W(t)\|_{HS}^2\ \le\ -2\,x(t)\left(\gamma_{\min}-x(t)\right).
\end{equation}

\smallskip

\textit{(ii) The energy excess stays below $\gamma_{\min}$.} By Theorem~\ref{thmA:energy-C} the energy does not increase, and by Proposition~\ref{propA:landscape}(a) the energy excess is nonnegative. Hence
\begin{equation}
0\ \le\ x(t)\ \le\ x(0)=\Delta E(\rho_{\mathrm{ref}})\ <\ \gamma_{\min}\qquad\textup{for all } t\ge0 .
\end{equation}
The denominator $\gamma_{\min}-x(t)$ is therefore positive, and the ratio $y(t)$ is well defined and nonnegative at all times.

\smallskip

\textit{(iii) A linear inequality for the ratio.} We differentiate the ratio and insert Eq.~\eqref{eq:logistic},
\begin{align}
\partial_ty(t)&=\frac{\partial_tx(t)\left(\gamma_{\min}-x(t)\right)+x(t)\,\partial_tx(t)}{\left(\gamma_{\min}-x(t)\right)^2}\nonumber\\
&=\frac{\gamma_{\min}\,\partial_tx(t)}{\left(\gamma_{\min}-x(t)\right)^2}\nonumber\\
&\le-\frac{2\gamma_{\min}\,x(t)\left(\gamma_{\min}-x(t)\right)}{\left(\gamma_{\min}-x(t)\right)^2}\nonumber\\
&=-2\gamma_{\min}\,y(t) .
\label{eq:ratio-ode}
\end{align}
The nonlinear factor $\gamma_{\min}-x(t)$ of Eq.~\eqref{eq:logistic} has canceled, and the ratio obeys a linear differential inequality with the constant rate $2\gamma_{\min}$.

\smallskip

\textit{(iv) Solving the inequality.} By Eq.~\eqref{eq:ratio-ode},
\begin{equation}
\partial_t\left(e^{2\gamma_{\min}t}\,y(t)\right)=e^{2\gamma_{\min}t}\left(\partial_ty(t)+2\gamma_{\min}\,y(t)\right)\ \le\ 0 .
\end{equation}
The function $e^{2\gamma_{\min}t}\,y(t)$ therefore does not increase, so that $e^{2\gamma_{\min}t}\,y(t)\le y(0)=c_0$. That is,
\begin{equation}
\label{eq:ratio-solution}
\frac{\Delta E(\rho(t))}{\gamma_{\min}-\Delta E(\rho(t))}\ \le\ c_0\,e^{-2\gamma_{\min}t}\qquad\textup{for all } t\ge0 ,
\end{equation}
which is the last inequality of Eq.~\eqref{eq:rate-C}.

\smallskip

\textit{(v) The complete chain.} Lemma~\ref{lemA:fidelity}, applied to the state $\rho(t)$ with $f_i(\ket{\phi_i(t)})=F_i(t)$, gives the first inequality of Eq.~\eqref{eq:rate-C}. The second inequality follows from part (ii),
\begin{equation}
\frac{\Delta E(\rho(t))}{\gamma_{\min}}\ \le\ \frac{\Delta E(\rho(t))}{\gamma_{\min}-\Delta E(\rho(t))}\qquad\textup{because}\qquad 0<\gamma_{\min}-\Delta E(\rho(t))\le\gamma_{\min} .
\end{equation}
Together with Eq.~\eqref{eq:ratio-solution}, this proves statement~1.

\subsection{Proof of statement 2: discretized DBF}
\label{app:thm2-D}

The proof follows the continuous-time case, with the energy relation of one step in place of the differential equation. Within this subsection we abbreviate
\begin{equation}
\label{eq:xk-bk}
x_k=\Delta E(\rho_k),\qquad y_k=\frac{x_k}{\gamma_{\min}-x_k},\qquad b_k=s_k\left(1-\frac{s_k}{\bar s}\right),
\end{equation}
so that $x_0=\Delta E(\rho_{\mathrm{ref}})$ and $y_0=c_0$. The proof uses the discretized DBF only through two facts, which we state explicitly because statement~3 relies on them. The first fact is the energy relation of Theorem~\ref{thmA:energy-D}, which reads with $\bar s=2/\gamma_{\max}$
\begin{equation}
\label{eq:fact-energy}
E(\rho_{k+1})\ \le\ E(\rho_k)-s_k\left(1-\frac{s_k}{\bar s}\right)\|W_k\|_{HS}^2 .
\end{equation}
The second fact is the bound
\begin{equation}
\label{eq:fact-sbar}
\bar s\ \le\ \frac{2}{\gamma_{\max}} .
\end{equation}

\medskip

\textit{(a) Every step.} We show by induction that $x_k<\gamma_{\min}$ for all $k$ and that every step satisfies Eq.~\eqref{eq:rate-step}. The assumption $x_0=\Delta E(\rho_{\mathrm{ref}})<\gamma_{\min}$ starts the induction. Suppose now that $x_k<\gamma_{\min}$ for some $k$, and let $0<s_k<\bar s$.

\smallskip

\textit{Range of $b_k$.} The function $s\,(1-s/\bar s)$ is positive for $0<s<\bar s$ and takes its largest value $\bar s/4$ at $s=\bar s/2$, since
\begin{equation}
s\left(1-\frac{s}{\bar s}\right)=\frac{\bar s}{4}-\frac{1}{\bar s}\left(s-\frac{\bar s}{2}\right)^2 .
\end{equation}
With Eq.~\eqref{eq:fact-sbar}, we obtain
\begin{equation}
\label{eq:b-range}
0\ <\ b_k\ \le\ \frac{\bar s}{4}\ \le\ \frac{1}{2\gamma_{\max}} .
\end{equation}

\smallskip

\textit{One-step inequality for the energy excess.} We subtract $E^{*}$ on both sides of Eq.~\eqref{eq:fact-energy} and bound $\|W_k\|_{HS}^2$ from below with Lemma~\ref{lemma:gradient}, applied to the state $\rho_k$,
\begin{align}
x_{k+1}&\le x_k-b_k\,\|W_k\|_{HS}^2\nonumber\\
&\le x_k-2b_k\,x_k\left(\gamma_{\min}-x_k\right)\nonumber\\
&=x_k\left[1-2b_k\left(\gamma_{\min}-x_k\right)\right].
\label{eq:one-step}
\end{align}
The bracket lies between zero and one. Indeed, $0<\gamma_{\min}-x_k\le\gamma_{\min}$ and Eq.~\eqref{eq:b-range} give
\begin{equation}
0\ <\ 2b_k\left(\gamma_{\min}-x_k\right)\ \le\ \frac{\gamma_{\min}}{\gamma_{\max}}\ \le\ 1 .
\end{equation}
Together with $x_{k+1}\ge0$, which is Proposition~\ref{propA:landscape}(a), Eq.~\eqref{eq:one-step} gives
\begin{equation}
\label{eq:xk-range}
0\ \le\ x_{k+1}\ \le\ x_k\left[1-2b_k\left(\gamma_{\min}-x_k\right)\right]\ \le\ x_k\ <\ \gamma_{\min} .
\end{equation}
In particular, $x_{k+1}<\gamma_{\min}$, so that the induction hypothesis holds again at the next step.

\smallskip

\textit{One-step inequality for the ratio.} The function $x\mapsto x/(\gamma_{\min}-x)$ increases on the interval $[0,\gamma_{\min})$. By Eq.~\eqref{eq:xk-range}, we may therefore replace $x_{k+1}$ by the larger number $x_k[1-2b_k(\gamma_{\min}-x_k)]$,
\begin{equation}
\label{eq:ratio-step-1}
y_{k+1}=\frac{x_{k+1}}{\gamma_{\min}-x_{k+1}}\ \le\ \frac{x_k\left[1-2b_k\left(\gamma_{\min}-x_k\right)\right]}{\gamma_{\min}-x_k\left[1-2b_k\left(\gamma_{\min}-x_k\right)\right]} .
\end{equation}
The denominator factorizes,
\begin{equation}
\gamma_{\min}-x_k+2b_k\,x_k\left(\gamma_{\min}-x_k\right)=\left(\gamma_{\min}-x_k\right)\left(1+2b_k\,x_k\right) ,
\end{equation}
so that Eq.~\eqref{eq:ratio-step-1} becomes
\begin{equation}
\label{eq:ratio-step-2}
y_{k+1}\ \le\ \frac{x_k}{\gamma_{\min}-x_k}\;\frac{1-2b_k\left(\gamma_{\min}-x_k\right)}{1+2b_k\,x_k}=y_k\;\frac{1-2b_k\left(\gamma_{\min}-x_k\right)}{1+2b_k\,x_k} .
\end{equation}
It remains to remove $x_k$ from the last factor. We claim that
\begin{equation}
\label{eq:factor-claim}
\frac{1-2b_k\left(\gamma_{\min}-x_k\right)}{1+2b_k\,x_k}\ \le\ \frac{1}{1+2b_k\,\gamma_{\min}} .
\end{equation}
Both denominators are positive, so Eq.~\eqref{eq:factor-claim} is equivalent to $[1-2b_k(\gamma_{\min}-x_k)]\,(1+2b_k\gamma_{\min})\le1+2b_kx_k$. Expanding the left-hand side gives
\begin{align}
\left[1-2b_k\left(\gamma_{\min}-x_k\right)\right]\left(1+2b_k\,\gamma_{\min}\right)&=1+2b_k\,\gamma_{\min}-2b_k\left(\gamma_{\min}-x_k\right)-4b_k^2\,\gamma_{\min}\left(\gamma_{\min}-x_k\right)\nonumber\\
&=1+2b_k\,x_k-4b_k^2\,\gamma_{\min}\left(\gamma_{\min}-x_k\right)\nonumber\\
&\le1+2b_k\,x_k ,
\end{align}
where the last step uses $\gamma_{\min}-x_k>0$. This proves Eq.~\eqref{eq:factor-claim}. Inserting it into Eq.~\eqref{eq:ratio-step-2} and using the definitions of Eq.~\eqref{eq:xk-bk}, we obtain
\begin{equation}
\frac{\Delta E(\rho_{k+1})}{\gamma_{\min}-\Delta E(\rho_{k+1})}\ \le\ \frac{\Delta E(\rho_k)}{\gamma_{\min}-\Delta E(\rho_k)}\;\frac{1}{1+2s_k\left(1-s_k/\bar s\right)\gamma_{\min}} ,
\end{equation}
which is Eq.~\eqref{eq:rate-step}. By induction, it holds at every step.

\medskip

\textit{(b) After $K$ steps.} We apply Eq.~\eqref{eq:rate-step} to the steps $k=0,\dots,K-1$ and use $y_0=c_0$,
\begin{equation}
\label{eq:product}
\frac{\Delta E(\rho_K)}{\gamma_{\min}-\Delta E(\rho_K)}\ \le\ c_0\prod_{k=0}^{K-1}\frac{1}{1+\beta_k},\qquad \beta_k=2s_k\left(1-\frac{s_k}{\bar s}\right)\gamma_{\min}>0 .
\end{equation}
To bring the product into exponential form, we use the elementary inequality $\ln(1+\beta)\ge\beta/(1+\beta)$ for $\beta\ge0$, which holds because both sides vanish at $\beta=0$ and the derivative $1/(1+\beta)$ of the left-hand side is at least the derivative $1/(1+\beta)^2$ of the right-hand side. Exponentiating gives
\begin{equation}
\label{eq:factor-exp}
\frac{1}{1+\beta_k}\ \le\ \exp\left(-\frac{\beta_k}{1+\beta_k}\right) .
\end{equation}
We write the exponent as the step size times a rate,
\begin{align}
\frac{\beta_k}{1+\beta_k}&=2\gamma_{\min}\,s_k\;\frac{1-s_k/\bar s}{1+2s_k\left(1-s_k/\bar s\right)\gamma_{\min}}\nonumber\\
&=2\gamma_{\min}\,s_k\;\frac{1}{\left(1-s_k/\bar s\right)^{-1}+2s_k\,\gamma_{\min}} ,
\label{eq:exponent}
\end{align}
where the second line divides the numerator and the denominator by $1-s_k/\bar s>0$. Both terms of the last denominator increase with $s_k$ on the interval $(0,\bar s)$. For $s_k\le s_{\max}$, the last fraction is therefore at least its value at $s_k=s_{\max}$, which is the constant $c$ of Eq.~\eqref{eq:rate-const},
\begin{equation}
\frac{1}{\left(1-s_k/\bar s\right)^{-1}+2s_k\,\gamma_{\min}}\ \ge\ \frac{1}{\left(1-s_{\max}/\bar s\right)^{-1}+2s_{\max}\,\gamma_{\min}}=\frac{1-s_{\max}/\bar s}{1+2s_{\max}\left(1-s_{\max}/\bar s\right)\gamma_{\min}}=c .
\end{equation}
With Eqs.~\eqref{eq:factor-exp} and~\eqref{eq:exponent}, each factor of the product in Eq.~\eqref{eq:product} is thus at most $e^{-2c\gamma_{\min}s_k}$, and
\begin{equation}
\frac{\Delta E(\rho_K)}{\gamma_{\min}-\Delta E(\rho_K)}\ \le\ c_0\prod_{k=0}^{K-1}e^{-2c\,\gamma_{\min}s_k}=c_0\,\exp\Big(-2c\,\gamma_{\min}\sum_{k=0}^{K-1}s_k\Big)=c_0\,e^{-2c\,\gamma_{\min}t} ,
\end{equation}
with the flow time $t=\sum_{k=0}^{K-1}s_k$. This is the last inequality of Eq.~\eqref{eq:rate-c}. The first inequality is Lemma~\ref{lemA:fidelity}, applied to the state $\rho_K$ with $f_i(\ket{\phi_i^{(K)}})=F_i(t)$. The second inequality follows from $0<\gamma_{\min}-\Delta E(\rho_K)\le\gamma_{\min}$, which is Eq.~\eqref{eq:xk-range}.

\medskip

\textit{(c) Properties of $c$.} By Eq.~\eqref{eq:exponent} with $s_k=s_{\max}$,
\begin{equation}
c=\frac{1}{\left(1-s_{\max}/\bar s\right)^{-1}+2s_{\max}\,\gamma_{\min}} .
\end{equation}
The denominator is finite and larger than one for $0<s_{\max}<\bar s$, so that $0<c<1$, and it tends to one for $s_{\max}\to0$, so that $c\to1$. If $s_{\max}\le\bar s/2$, the two terms of the denominator satisfy
\begin{equation}
\left(1-\frac{s_{\max}}{\bar s}\right)^{-1}\le2,\qquad 2s_{\max}\,\gamma_{\min}\ \le\ \bar s\,\gamma_{\min}\ \le\ \frac{2\gamma_{\min}}{\gamma_{\max}}\ \le\ 2 ,
\end{equation}
where the second chain uses Eq.~\eqref{eq:fact-sbar}. Hence the denominator is at most four, and $c\ge1/4$. This proves statement~2.

\subsection{Proof of statement 3: product-formula DBF}
\label{app:thm2-PF}
We show that the proof of statement~2 applies without change. That proof uses the discretized DBF only through the two facts of Eqs.~\eqref{eq:fact-energy} and~\eqref{eq:fact-sbar}, that is, through the energy relation of one step with the factor $1-s_k/\bar s$ and through the bound $\bar s\le2/\gamma_{\max}$. The bound enters through $b_k\le1/(2\gamma_{\max})$ in part~(a) and through $c\ge1/4$ in part~(c). We verify both facts for the two choices of the angles.

\smallskip

\textit{Angle ratio $\theta_k/\tau_k=\Omega/p_1$.} The energy relation is Eq.~\eqref{eq:thmPF-ratio} of Theorem~\ref{thmA:energy-PF},
\begin{equation}
E(\rho_{k+1})\ \le\ E(\rho_k)-s_k\left(1-\frac{7\,s_k\gamma_{\max}}{12}\right)\|W_k\|_{HS}^2=E(\rho_k)-s_k\left(1-\frac{s_k}{\bar s}\right)\|W_k\|_{HS}^2 ,
\end{equation}
with $\bar s=\frac{12}{7\gamma_{\max}}$, which is Eq.~\eqref{eq:fact-energy}. The bound of Eq.~\eqref{eq:fact-sbar} holds because
\begin{equation}
\bar s=\frac{12}{7\gamma_{\max}} <\ \frac{2}{\gamma_{\max}} \,.
\end{equation}

\smallskip

\textit{Equal angles $\theta_k=\tau_k=\sqrt{s_k}$.} The energy relation is Eq.~\eqref{eq:thmPF-equal} of Theorem~\ref{thmA:energy-PF},
\begin{equation}
E(\rho_{k+1})\ \le\ E(\rho_k)-s_k\left(1-\frac{7\,s_k\max(\Omega,1)^2}{12}\right)\|W_k\|_{HS}^2=E(\rho_k)-s_k\left(1-\frac{s_k}{\bar s}\right)\|W_k\|_{HS}^2\qquad\textup{with}\, ,
\end{equation}
with $\bar s=\frac{12}{7\max(\Omega,1)^2}$, which is again Eq.~\eqref{eq:fact-energy}. For the bound of Eq.~\eqref{eq:fact-sbar}, we use that the largest weight satisfies $p_1\le1\le\max(\Omega,1)$ and that $\Omega\le\max(\Omega,1)$. Hence
\begin{equation}
\gamma_{\max}=p_1\,\Omega\ \le\ \max(\Omega,1)^2\qquad\textup{and}\qquad \bar s=\frac{12}{7\max(\Omega,1)^2}\ \le\ \frac{12}{7\gamma_{\max}}\ <\ \frac{2}{\gamma_{\max}} .
\end{equation}

\smallskip

Both facts thus hold for the product-formula DBF with either choice of the angles. Parts (a) to (c) of Appendix~\ref{app:thm2-D} therefore apply word for word, with $\rho_k$ generated by Eq.~\eqref{eq:PF} and with the respective value of $\bar s$. This proves statement~3 and completes the proof of Theorem~\ref{thmA:rates}. \hfill$\square$

\subsection{Number of steps needed for reaching a target infidelity}
\label{app for steps}

Let $\varepsilon$ be the target infidelity, so that we require $1-F_i(t)\le\varepsilon$ for all $r$ states. By Eqs.~\eqref{eq:rate-C} and~\eqref{eq:rate-c}, it suffices to reach $\Delta E(\rho(t))/(\gamma_{\min}-\Delta E(\rho(t)))\le\varepsilon$.

\textit{Continuous-time DBF.} By Eq.~\eqref{eq:rate-C}, the flow time $t=\ln(c_0/\varepsilon)/(2\gamma_{\min})$ suffices.

\textit{Discretized and product-formula DBF.} For a constant step size $s$, Eq.~\eqref{eq:product} gives the ratio $c_0\,[1+2s(1-s/\bar s)\gamma_{\min}]^{-K}$ after $K$ steps. The contraction per step is strongest for the step size that maximizes $s(1-s/\bar s)$, which is $s=\bar s/2$. This step size gives the factor $(1+\bar s\gamma_{\min}/2)^{-1}$ per step, and the target is reached after $K=\ln(c_0/\varepsilon)/\ln(1+\bar s\gamma_{\min}/2)$ steps. Since $\ln(1+u)\ge u/(1+u)$, we have $1/\ln(1+u)\le1+1/u$. Hence any number of steps
\begin{equation}
\label{eq:steps-general}
K\ge\left(1+\frac{2}{\bar s\,\gamma_{\min}}\right)\ln\frac{c_0}{\varepsilon}
\end{equation}
suffices. For the discretized DBF, $\bar s=2/\gamma_{\max}$ and Eq.~\eqref{eq:steps-general} reads
\begin{equation}
K\ge(1+\kappa)\,\ln\frac{c_0}{\varepsilon},
\end{equation}
with $\kappa=\gamma_{\max}/\gamma_{\min}$. For the product-formula DBF with the angle ratio $\theta_k/\tau_k=\Omega/p_1$, $\bar s=12/(7\gamma_{\max})$ gives
\begin{equation}
K\ge\left(1+\frac{7\kappa}{6}\right)\ln\frac{c_0}{\varepsilon}.
\end{equation}
For equal angles, $\bar s=12/[7\max(\Omega,1)^2]$ gives
\begin{equation}
K\ge\left(1+\frac{7\max(\Omega,1)^2}{6\,\gamma_{\min}}\right)\ln\frac{c_0}{\varepsilon},
\end{equation}
which is at least the count for the angle ratio, because $\max(\Omega,1)^2\ge\gamma_{\max}$.
All counts grow logarithmically with $1/\varepsilon$ and linearly with $\kappa$. The step size $\bar s/2$ gives the best guarantee, but it is not the fastest choice close to the target, as Appendix~\ref{app:optimal_step_schedules} shows.

\section{Preparing excited states with degenerate weights}
\label{app:two-block}
This section proves Theorem~\ref{theorem:single} of Methods. To this end, it relaxes the assumption that the weights are distinct. In Appendices~\ref{app:energy-landscape} to~\ref{app:thm2} we assumed $p_1>\dots>p_r$, which makes the flow resolve all $r$ eigenstates and makes the rate depend on all gaps between the $r+1$ lowest levels. Here we allow $p_1\ge p_2\ge\dots\ge p_r>0$. States with equal weights are then prepared only as a subspace, and the gaps inside this subspace no longer affect the rate. We first show which of the previous results survive for equal weights (Theorem~\ref{thmA:equal}). We then specialize to the weights of Eq.~\eqref{eq:EM-weights}, which prepare a single excited state at a cost that does not grow with $r$ (Corollary~\ref{corA:single}).

\subsection{Preparing several excited states with degenerate weights}

A state $\rho$ with some equal weights determines only the subspaces of equal weight, and not the individual states in them. The fixed points are again the states that commute with $H$, which are diagonal in the eigenbasis of $H$ and of which there are finitely many. The target $\rho^{*}=\sum_{i=1}^{r} p_i\ket{\lambda_i}\bra{\lambda_i}$ is unchanged. With equal weights, some weight gaps $\delta p_j=p_j-p_{j+1}$ vanish, and so do the corresponding $\gamma_{j,j+1}=\delta p_j\,\delta\lambda_{j+1}$. Throughout this appendix we redefine
\begin{equation}
\label{eq:gmin-equal}
\gamma_{\min}=\min\left\{\gamma_{j,j+1}:\ 1\le j\le r,\ \delta p_j>0\right\},
\end{equation}
which is the smallest nonzero $\gamma_{j,j+1}$ and reduces to Eq.~\eqref{eq:gamma-min} for distinct weights.

\begin{theoremA}[Sufficient condition for exponential convergence with degenerate weights]
\label{thmA:equal}
Let $H$ have the eigenvalues $\lambda_1<\dots<\lambda_d$. Let $\rho$ evolve from a reference state $\rho_{\mathrm{ref}}=\sum_{i=1}^{r}p_i\ket{\phi_i}\bra{\phi_i}$ with orthonormal states $\ket{\phi_i}$ and weights $p_1\ge\dots\ge p_r>0$ that sum to one, under the continuous-time DBF of Eq.~\eqref{eq:C}, the discretized DBF of Eq.~\eqref{eq:D}, or the product-formula DBF of Eq.~\eqref{eq:PF}. Let $\rho^{*}$ be the target state of Eq.~\eqref{eq:target}, let $\Delta E(\rho)=E(\rho)-E^{*}$, let $\gamma_{\max}=p_1\Omega$, and let $\gamma_{\min}$ be defined by Eq.~\eqref{eq:gmin-equal}. Then the following statements hold.
\begin{enumerate}
    \item \textit{Energy and convergence.} Theorems~\ref{thmA:energy-C}, \ref{thmA:energy-D}, and~\ref{thmA:energy-PF} hold without change. Statements~1 to~3 of Theorem~\ref{thmA:convergence} hold as well, with the only difference that fewer of the fixed points $\rho_\sigma$ are distinct.
    \item \textit{Energy landscape and dissipation rate.} Every state $\rho$ with these weights satisfies $\Delta E(\rho)\ge0$, with equality if and only if $\rho=\rho^{*}$. Proposition~\ref{propA:landscape}(c), Lemma~\ref{lemA:gradient}, and Corollary~\ref{corA:criterion} hold with $\gamma_{\min}$ of Eq.~\eqref{eq:gmin-equal}.
    \item \textit{Decay rate of the energy excess.} If $\Delta E(\rho_{\mathrm{ref}})<\gamma_{\min}$, the bounds of Theorem~\ref{thmA:rates} on the energy excess hold under the step-size conditions stated there. That is, the last two inequalities of Eqs.~\eqref{eq:rate-C} and~\eqref{eq:rate-c} and Eq.~\eqref{eq:rate-step} hold with $\gamma_{\min}$ of Eq.~\eqref{eq:gmin-equal}.
    \item \textit{Fidelities.} For every state $\rho$ with the weights above and every $j\le r$ with $\delta p_j>0$, the subspace $V_j$ spanned by the $j$ states with the largest weights satisfies
    \begin{equation}
    \label{eq:subspace-fidelity}
    j-\mathrm{Tr}\left(P_j\Pi_j\right)\le\frac{\Delta E(\rho)}{\gamma_{j,j+1}}\le\frac{\Delta E(\rho)}{\gamma_{\min}},
    \end{equation}
    where $P_j = \sum_{i=1}^{j} \ket{\phi_i}\bra{\phi_i}$ and $\Pi_j = \sum_{i=1}^{j} \ket{\lambda_i}\bra{\lambda_i}$ are the projectors onto $V_j$ and onto $\mathrm{span}\{\ket{\lambda_1},\dots,\ket{\lambda_j}\}$. If $\delta p_{i-1}>0$ and $\delta p_i>0$, where the first condition is absent for $i=1$, the state $\ket{\phi_i}$ satisfies
    \begin{equation}
    \label{eq:fidelity-equal-p}
    1-f_i(\ket{\phi_i})\le\frac{\Delta E(\rho)}{\min\{\gamma_{i-1,i},\gamma_{i,i+1}\}}\le\frac{\Delta E(\rho)}{\gamma_{\min}}\,.
    \end{equation}
\end{enumerate}
\end{theoremA}
\begin{proof}
We go through the proofs of the cited results and identify where the weights enter.

\smallskip

\textit{Statement 1.} The proofs of the energy relations in Appendix~\ref{app:energy} use the weights only through the bound $\|[\rho,X]\|_{HS}\le p_1\|X\|_{HS}$ of Eq.~\eqref{eq:comm}, which holds for equal weights because the eigenvalues of $\rho$ still lie between $0$ and $p_1$. The proof of Theorem~\ref{thmA:convergence} uses these relations, the fact that a state that commutes with the nondegenerate $H$ is diagonal in the eigenbasis of $H$, the fact that the fixed points are finitely many, and, in Step~2 of Appendix~\ref{app:thm1-convergence}, only that the weights are non-increasing. All of these hold for equal weights.

\smallskip

\textit{Statement 2.} The subspace decomposition of Appendix~\ref{app:landscape-subspaces} uses only the telescoping sum $p_i=\sum_{j\ge i}\delta p_j$ and therefore holds for equal weights,
\begin{equation}
\label{eq:dE-equal}
\Delta E(\rho)=\sum_{j=1}^{r}\delta p_j\,\varepsilon_j=\sum_{j:\,\delta p_j>0}\delta p_j\,\varepsilon_j ,
\end{equation}
where the terms with $\delta p_j=0$ drop out. The subspaces $V_j$ with $\delta p_j>0$ are determined by $\rho$, although the individual states with equal weights are not. Lemmas~\ref{lemA:subspace-excess} and~\ref{lemA:subspace} concern a single subspace and do not involve the weights. We now go through the three results.

For Proposition~\ref{propA:landscape}(a), Eq.~\eqref{eq:dE-equal} and $\varepsilon_j\ge0$ give $\Delta E(\rho)\ge0$. If $\Delta E(\rho)=0$, then $\varepsilon_j=0$ and hence $P_j=\Pi_j$ for every $j$ with $\delta p_j>0$. Since $\rho=\sum_{i}p_i\ket{\phi_i}\bra{\phi_i}=\sum_{j}\delta p_j\,P_j$ by the same telescoping sum, this gives
\begin{equation}
\rho=\sum_{j:\,\delta p_j>0}\delta p_j\,P_j=\sum_{j:\,\delta p_j>0}\delta p_j\,\Pi_j=\rho^{*} .
\end{equation}

For Proposition~\ref{propA:landscape}(c), let $\rho_\sigma\ne\rho^{*}$ be a fixed point. By the last equation, there is an index $j$ with $\delta p_j>0$ and $P_j\ne\Pi_j$. The subspace $V_j$ is spanned by $j$ eigenstates of $H$, at least one of which lies above the level $j$, so that $j-\mathrm{Tr}(P_j\Pi_j)\ge1$. Equation~\eqref{eq:sub-a} then gives $\varepsilon_j\ge\delta\lambda_{j+1}$, and
\begin{equation}
\Delta E(\rho_\sigma)\ \ge\ \delta p_j\,\varepsilon_j\ \ge\ \delta p_j\,\delta\lambda_{j+1}=\gamma_{j,j+1}\ \ge\ \gamma_{\min} ,
\end{equation}
where the last step is Eq.~\eqref{eq:gmin-equal}. The value $\gamma_{\min}$ is attained by exchanging the weights on two adjacent levels $j$ and $j+1$ with $\gamma_{j,j+1}=\gamma_{\min}$, as in Eq.~\eqref{eq:swap-energy}.

For Lemma~\ref{lemA:gradient}, Step~1 of its proof uses only that the weights are non-increasing, so that Eq.~\eqref{eq:W-leakage} holds for equal weights, and the terms with $\delta p_j=0$ vanish in it. In Step~3, the bound $\delta p_j\,\delta\lambda_{j+1}=\gamma_{j,j+1}\ge\gamma_{\min}$ is needed only for the terms with $\delta p_j>0$, for which it holds by Eq.~\eqref{eq:gmin-equal}. Hence
\begin{equation}
\frac12\|[\rho,H]\|_{HS}^2\ \ge\ \sum_{j:\,\delta p_j>0}\delta p_j\left(\delta p_j\,l_j\right)\ \ge\ \left(\gamma_{\min}-\Delta E(\rho)\right)\sum_{j:\,\delta p_j>0}\delta p_j\,\varepsilon_j=\left(\gamma_{\min}-\Delta E(\rho)\right)\Delta E(\rho) .
\end{equation}
The proof of Corollary~\ref{corA:criterion} uses only statement~1 and Proposition~\ref{propA:landscape}(c).

\smallskip

\textit{Statement 3.} The proofs of the third inequality of Eq.~\eqref{eq:rate-C}, of Eq.~\eqref{eq:rate-step}, and of the third inequality of Eq.~\eqref{eq:rate-c} in Appendices~\ref{app:thm2-C} to~\ref{app:thm2-PF} use only the energy relations, Lemma~\ref{lemA:gradient}, and $\Delta E(\rho)\ge0$, which hold by statements~1 and~2. The second inequality of each chain holds because $0<\gamma_{\min}-\Delta E(\rho(t))\le\gamma_{\min}$.

\smallskip

\textit{Statement 4.} For $j\le r$ with $\delta p_j>0$, we keep only the term $j$ of Eq.~\eqref{eq:dE-equal} and use Eq.~\eqref{eq:sub-a},
\begin{equation}
\Delta E(\rho)\ \ge\ \delta p_j\,\varepsilon_j\ \ge\ \delta p_j\,\delta\lambda_{j+1}\left[\,j-\mathrm{Tr}\left(P_j\Pi_j\right)\right]=\gamma_{j,j+1}\left[\,j-\mathrm{Tr}\left(P_j\Pi_j\right)\right] .
\end{equation}
Dividing by $\gamma_{j,j+1}\ge\gamma_{\min}$ gives Eq.~\eqref{eq:subspace-fidelity}. The proof of Lemma~\ref{lemA:fidelity} in Appendix~\ref{app:infidelity} uses only the two subspaces $V_{i-1}$ and $V_i$ and the terms $j=i-1$ and $j=i$ of Eq.~\eqref{eq:dE-equal}. If $\delta p_{i-1}>0$ and $\delta p_i>0$, both subspaces are determined by $\rho$ and both terms are present, so that the proof applies without change and gives Eq.~\eqref{eq:fidelity-equal-p}.
\end{proof}

The LHS of Eq.~\eqref{eq:subspace-fidelity} vanishes if and only if $V_j$ is the span of the $j$ lowest eigenstates. The flow thus sorts the states down to the resolution of the weights. It drives each subspace of equal weight to the span of the corresponding eigenstates and does nothing inside these subspaces.
In particular, $\rho=\rho^{*}$ does not imply $\ket{\phi_i}=\ket{\lambda_i}$ for a state whose weight equals that of a neighbor. For example, $p_1=p_2$ with $\ket{\phi_{1,2}}=(\ket{\lambda_1}\pm\ket{\lambda_2})/\sqrt2$ gives $\rho=\rho^{*}$.

\subsection{Preparing a single excited state with degenerate weights}
\label{app:two-block-single}

Suppose that only the $r$th eigenstate $\ket{\lambda_r}$ is needed, with $r\ge2$. We then give equal weights to the $r-1$ lower states and a smaller weight to the last one, as in Eq.~\eqref{eq:EM-weights}.

\begin{corollaryA}[Theorem~\ref{theorem:single}, formal]
\label{corA:single}
Let $r\ge2$, and let the weights of the reference state $\rho_{\mathrm{ref}}$ be those of Eq.~\eqref{eq:EM-weights},
\begin{equation}
\label{eq:single-weights}
p_1=\dots=p_{r-1}=\frac{2}{2r-1},\qquad p_r=\frac{1}{2r-1} .
\end{equation}
Let $\delta\lambda_{\min}=\min\{\lambda_r-\lambda_{r-1},\,\lambda_{r+1}-\lambda_r\}$ be the smaller of the two energy gaps next to the target level. Then the quantities of Eq.~\eqref{eq:gmin-equal} and $\kappa=\gamma_{\max}/\gamma_{\min}$ are
\begin{equation}
\label{eq:single-gammas}
\gamma_{\min}=\frac{\delta\lambda_{\min}}{2r-1},\qquad \gamma_{\max}=\frac{2\,\Omega}{2r-1},\qquad \kappa=\frac{2\,\Omega}{\delta\lambda_{\min}} .
\end{equation}
Let $\rho$ evolve from $\rho_{\mathrm{ref}}$ under one of the three flows, with the step sizes of Theorem~\ref{thmA:rates} for the discretized and the product-formula DBF, and let $c_0$, $c$, the flow time $t$, $\rho(t)$, and $F_r(t)$ be defined as in Theorem~\ref{thmA:rates}, with $c=1$ for the continuous-time DBF. If $\Delta E(\rho_{\mathrm{ref}})<\gamma_{\min}$, then
\begin{equation}
\label{eq:single-rate}
1-F_r(t)\ \le\ \frac{\Delta E(\rho(t))}{\gamma_{\min}}\ \le\ c_0\,e^{-2c\,\gamma_{\min}t},\qquad (r-1)-\mathrm{Tr}\left(P_{r-1}(t)\,\Pi_{r-1}\right)\ \le\ \frac{\Delta E(\rho(t))}{\gamma_{\min}}\ \le\ c_0\,e^{-2c\,\gamma_{\min}t} ,
\end{equation}
where $P_{r-1}(t)$ is the projector onto the span of the $r-1$ states with the larger weight at the flow time $t$. That is, the $r$th state converges to $\ket{\lambda_r}$, and the span of the other $r-1$ states converges to the span of $\{\ket{\lambda_i}\}_{i=1}^{r-1}$.
\end{corollaryA}
\begin{proof}
The weights of Eq.~\eqref{eq:single-weights} sum to one, since $\left(2(r-1)\right)/(2r-1)+1/(2r-1)=1$. Their only nonzero weight gaps are
\begin{equation}
\delta p_{r-1}=p_{r-1}-p_r=\frac{1}{2r-1},\qquad \delta p_r=p_r=\frac{1}{2r-1} .
\end{equation}
Equation~\eqref{eq:gmin-equal} therefore contains only the two terms
\begin{equation}
\gamma_{r-1,r}=\delta p_{r-1}\left(\lambda_r-\lambda_{r-1}\right)=\frac{\lambda_r-\lambda_{r-1}}{2r-1},\qquad \gamma_{r,r+1}=\delta p_r\left(\lambda_{r+1}-\lambda_r\right)=\frac{\lambda_{r+1}-\lambda_r}{2r-1} ,
\end{equation}
whose minimum is $\gamma_{\min}=\delta\lambda_{\min}/(2r-1)$. Moreover, $\gamma_{\max}=p_1\Omega=2\Omega/(2r-1)$, and the ratio of the two is $\kappa=2\Omega/\delta\lambda_{\min}$. This proves Eq.~\eqref{eq:single-gammas}.

For Eq.~\eqref{eq:single-rate}, we apply statement~4 of Theorem~\ref{thmA:equal} to the state $\rho(t)$. Since $\delta p_{r-1}>0$ and $\delta p_r>0$, Eq.~\eqref{eq:fidelity-equal-p} with $i=r$ gives $1-F_r(t)\le\Delta E(\rho(t))/\gamma_{\min}$, and Eq.~\eqref{eq:subspace-fidelity} with $j=r-1$ gives $(r-1)-\mathrm{Tr}(P_{r-1}(t)\Pi_{r-1})\le\Delta E(\rho(t))/\gamma_{\min}$. Statement~3 of Theorem~\ref{thmA:equal} bounds the energy excess,
\begin{equation}
\frac{\Delta E(\rho(t))}{\gamma_{\min}}\ \le\ \frac{\Delta E(\rho(t))}{\gamma_{\min}-\Delta E(\rho(t))}\ \le\ c_0\,e^{-2c\,\gamma_{\min}t} ,
\end{equation}
which completes the proof.
\end{proof}

Corollary~\ref{corA:single} is Theorem~\ref{theorem:single} of the End Matter, with the convergence of the lower subspace stated quantitatively. The condition number $\kappa$ of Eq.~\eqref{eq:single-gammas} does not depend on $r$, and it contains only the two gaps next to the target level. By Eq.~\eqref{eq:steps-general}, the number of steps of the discretized and of the product-formula DBF is therefore independent of $r$, and the flow time of the continuous-time DBF grows only quadratically with $r$, because $\gamma_{\min}$ shrinks as $1/(2r-1)$. In contrast, $\kappa$ grows at least quadratically with $r$ when distinct weights resolve all $r$ states (Appendix~\ref{app:weights}). The price is that the lower eigenstates are prepared only as a subspace.

\section{Linearized flow and the choice of parameters}
\label{app:linearizedFlow}
This appendix analyzes the flow close to the target and derives the statements of the main text on the decay of the off-diagonal elements and the optimal choice of step size in this regime (Lemma~\ref{lemma:chebyshev}). We start by linearizing the flow close to the target. Next, we prove that $s^\star$ of Eq.~\eqref{eq:s-opt} is the optimal fixed step size. We show the optimality of Chebyshev scheduling and prove the full statement of Lemma~\ref{lemma:chebyshev} with the contraction rates. Finally, we derive the optimal weights in the linear regime and show that optimal weight differences need to be proportional to the inverse energy gaps.

\subsection{Linearization at the target} \label{appSubsec:Linearize}

Theorem~\ref{theorem:convergence-rate} is a guarantee for the whole region $\Delta E(\rho_{\mathrm{ref}})<\gamma_{\min}$, and its constant $c$ reflects the worst case over this region. However, close to the target, the flow can be linearized, which determines the actual decay and the choice of the optimal step size.

Let $\rho=\rho^{*}+\delta\rho$ be a state of the form of Eq.~\eqref{eq:rho} close to the target, and let $\delta\rho_{ab}=\bra{\lambda_a}\delta\rho\ket{\lambda_b}$. As in the main text, we call the matrix elements $\delta\rho_{ab}=(\rho-\rho^{*})_{ab}$ with $a<b$ and $a\le r$ the \textit{off-diagonal elements}.

\begin{propositionA}[Linearized flow at the target]
\label{propA:linear}

\textup{(a)} To first order, one step of the discretized DBF with step size $s$ multiplies each off-diagonal element by a real factor,
\begin{equation}
\label{eq:linear-D}
\delta\rho_{ab}\ \to\ (1-s\gamma_{ab})\,\delta\rho_{ab},\qquad \gamma_{\min}\le\gamma_{ab}\le\gamma_{\max}.
\end{equation}
\textup{(b)} To second order, the energy excess and the infidelities are quadratic in the off-diagonal elements,
\begin{equation}
\label{eq:quadratic}
\Delta E(\rho)=\sum_{a<b,\ a\le r}\frac{\lambda_b-\lambda_a}{p_a-p_b}\,|\delta\rho_{ab}|^2,\qquad 
1-f_i=\sum_{a\ne i}\frac{|\delta\rho_{ai}|^2}{(p_i-p_a)^2}\quad(i\le r).
\end{equation}
\end{propositionA}
\begin{proof}
In the eigenbasis of $H$, for any operator $X$ we have
$[X,H]_{ab}=(\lambda_b-\lambda_a)X_{ab}$ and $[X,\rho^*]_{ab}=(p_b-p_a)X_{ab}$.
Every state on the unitary orbit near $\rho^*$ can be written as $\rho=e^{K}\rho^* e^{-K}$, where $K$ is a small anti-Hermitian matrix with $K_{ab}=0$ whenever $p_a=p_b$. Then,
\begin{equation}
\label{eq:drho-K}
\delta\rho=[K,\rho^*]+\tfrac12\big[K,[K,\rho^*]\big]+O(K^3),
\qquad \delta\rho_{ab}=(p_b-p_a)K_{ab}+O(K^2).
\end{equation}
Thus, to first order, the only nonzero entries of $\delta\rho$ are the off-diagonal elements ($p_a\ne p_b$), and $|K_{ab}|=|\delta\rho_{ab}|/|p_a-p_b|$.

(a) Since $[\rho^*,H]=0$, the DBF generator is $W=[\rho,H]=[\delta\rho,H]$, which is of first order. One step therefore gives
\begin{equation}
e^{sW}\rho\,e^{-sW}=\rho+s[W,\rho^*]+O(\delta\rho^2),
\qquad
[W,\rho^*]_{ab}=(p_b-p_a)(\lambda_b-\lambda_a)\,\delta\rho_{ab}=-\gamma_{ab}\,\delta\rho_{ab},
\end{equation}
and hence $\delta\rho_{ab}\to(1-s\gamma_{ab})\delta\rho_{ab}$. The bounds on $\gamma_{ab}$ are Eq.~\eqref{eq:gamma-range}.

(b) \emph{Energy.} The first-order term of \eqref{eq:drho-K} contributes nothing, because $\mathrm{Tr}([K,\rho^*]H)=\mathrm{Tr}(K[\rho^*,H])=0$. For the second-order term, $\mathrm{Tr}([K,X]H)=\mathrm{Tr}(X[H,K])$ gives
\begin{equation}
\Delta E(\rho)=\tfrac12\,\mathrm{Tr}\big([K,\rho^*][H,K]\big)
=\tfrac12\sum_{a,b}(p_b-p_a)(\lambda_b-\lambda_a)K_{ab}K_{ba}
=\sum_{a<b}\gamma_{ab}|K_{ab}|^2 ,
\end{equation}
using anti-Hermitianity of $K$. Substituting $|K_{ab}|=|\delta\rho_{ab}|/(p_a-p_b)$ gives the first relation in \eqref{eq:quadratic}.

\emph{Fidelity.} With $\ket{\phi_i}=e^{K}\ket{\lambda_i}$,
$\langle\lambda_i|\phi_i\rangle=1+K_{ii}+\tfrac12(K^2)_{ii}+O(K^3)$.
Because $K_{ii}$ is imaginary and $(K^2)_{ii}=-\sum_a|K_{ai}|^2$,
\begin{equation}
f_i=|\langle\lambda_i|\phi_i\rangle|^2=1-\sum_{a\ne i}|K_{ai}|^2+O(K^3),
\end{equation}
and substituting $|K_{ai}|$ gives the second relation.
\end{proof}
Proposition~\ref{propA:linear} proves the statements of the main text on the role of the quantities $\gamma_{ij}$. Each off-diagonal element decays at its own rate $\gamma_{ab}$.  The largest rate is $\gamma_{\max}=\gamma_{1d}$, and the smallest is $\gamma_{\min}$, which belongs to a pair of adjacent levels. By part (c), the energy excess and the infidelities decay with the squares of the factors of the off-diagonal elements. For the continuous-time DBF, the slowest off-diagonal element decays as $e^{-\gamma_{\min}t}$, so that $\Delta E$ and $1-f_i$ decay as $e^{-2\gamma_{\min}t}$. The rate $2\gamma_{\min}$ of Theorem~\ref{theorem:convergence-rate} is thus the exact asymptotic rate of the continuous-time DBF for a generic reference state near the target.

\subsection{Optimal step schedules} \label{app:optimal_step_schedules}
\subsubsection{Warm-up: fixed step sizes}
Equation~\eqref{eq:linear-D} quantifies the effect of step sizes on the contraction rates. The slowest decay among all off-diagonal elements is governed by the factor $\max\{|1-s\gamma_{\min}|,|1-s\gamma_{\max}|\}$, because $|1-s\gamma|$ takes its largest value on the interval $[\gamma_{\min},\gamma_{\max}]$ at one of the two ends. This maximum is smallest when the two expressions are equal. Indeed, if $|1-s\gamma_{\min}|$ is the larger one, a slightly larger step size lowers it, and if $|1-s\gamma_{\max}|$ is the larger one, a slightly smaller step size lowers it. Equality requires $1-s\gamma_{\min}=s\gamma_{\max}-1$, which gives Eq.~\eqref{eq:s-opt},
\begin{equation}
\label{eq:stepoptConst}
s^\star=\frac{2}{\gamma_{\max}+\gamma_{\min}},\qquad \max_{\gamma_{\min}\le\gamma\le\gamma_{\max}}|1-s^\star\gamma|=\frac{\gamma_{\max}-\gamma_{\min}}{\gamma_{\max}+\gamma_{\min}}=\frac{\kappa-1}{\kappa+1},
\end{equation}
By Eq.~\eqref{eq:quadratic}, the energy excess and the infidelities then asymptotically shrink by the factor $[(\kappa-1)/(\kappa+1)]^2$ per iteration.

\subsubsection{Chebyshev scheduling is the optimal non-constant schedule}
\label{app:chebyshev}

In this section, we motivate the use of non-constant schedules and prove Lemma~\ref{lemma:chebyshev} of the main text, which states that the Chebyshev schedule of Eq.~\eqref{chebyschMain} is the optimal periodic schedule close to the target. The setting is the linearized discretized DBF of Proposition~\ref{propA:linear}. The proof reduces the choice of the step sizes to a classical problem on polynomials, which is the same problem that arises for the Richardson iteration for linear systems~\cite{young1953richardson,adsuara2017equivalence}.

Consider the discretized DBF close to the target with step sizes $s_1,\dots,s_m$ that are repeated with the period $m$. By Eq.~\eqref{eq:linear-D}, one period multiplies the off-diagonal element  $\delta\rho_{ab}$ by contraction rate, $P(\gamma_{ab})$, where
\begin{equation}
\label{eq:period-polynomial}
P(\gamma)=\prod_{k=1}^{m}(1-s_k\gamma) ,
\end{equation}
and the numbers $\gamma_{ab}$ lie in the interval $[\gamma_{\min},\gamma_{\max}]$. We measure the quality of a schedule by its largest contraction factor, i.e.
\begin{equation} \label{appeq:fulknowledgecontraction}
\tilde{v}_{\max} = \left(\max_{ij} |P(\gamma_{ij})| \right ) ^{1/m} 
\end{equation}
where the power of $1/m$ indicates that $\tilde{v}_{\max}$ is the per iteration contraction factor averaged over one period of the schedule. 
 However, finding $\tilde{v}_{\max}$ involves a discrete optimization problem over a large set of $\gamma_{ij}$'s and does not have a closed form solution. We therefore consider the following problem 
$$
v_{\max} := \left ( \max_{\gamma \in [\gamma_{min},\gamma_{max}]} |P(\gamma)|\right) ^{1/m} \ .
$$
which bounds the average contraction rate of every off-diagonal element , i.e. $v_{\max} \geq \tilde{v}_{\max}$. 
Consequently, we set the optimal step schedule to be the schedule that minimizes this rate
$$
\{s^\star_k\}_{k=1}^m := \argmin_{\{s_k\}\in \mathbb{R}^m}    v_{\max} \ .
$$
The steps uniquely define the polynomial of Eq.~\eqref{eq:period-polynomial} and any degree $m$ polynomial such that $P(0)=1$ corresponds to a set of steps through its roots. Therefore, we state the problem as an optimization over the set of all such polynomials. 
$$
v_{\max}^\star = \left (\min_{P\in \PC_m ,\ P(0) =1} \max_{\gamma_{\min}\leq \gamma\leq\gamma_{\max}} |P(\gamma)| \right )^{1/m}, 
$$
where $\PC_m$ is the set of all polynomials of degree at most $m$. The optimal steps then would be the reciprocals of the roots of the optimal polynomial. Having set the problem, we present the following theorem that establishes Chebyshev polynomials as its solutions.
\begin{theoremA}[Complete statement of Lemma~\ref{lemma:chebyshev}]
\label{thmA:chebyshev}
Let $\kappa=\gamma_{\max}/\gamma_{\min}>1$ and $\eta=(\sqrt\kappa-1)/(\sqrt\kappa+1)$.

\textup{(a)} For every choice of the step sizes $s_1,\dots,s_m$,
\begin{equation}
\label{eq:ChebyCR}
\max_{\gamma_{\min}\le\gamma\le\gamma_{\max}}|P(\gamma)|\ \ge\ \frac{2\eta^m}{1+\eta^{2m}} .
\end{equation}

\textup{(b)} The Chebyshev schedule of Eq.~\eqref{chebyschMain} attains equality in Eq.~\eqref{eq:ChebyCR}, and it is the only schedule of period $m$ that does so, up to the order of the steps.

\textup{(c)} The average factor per iteration of the Chebyshev schedule, $v_{\max}^\star =  [2\eta^m/(1+\eta^{2m})]^{1/m}$, is at least $\eta$ and tends to $\eta$ for $m\to\infty$.

\textup{(d)} For $m=1$, the Chebyshev schedule is the constant step size $s^\star$ of Eq.~\eqref{eq:s-opt}, and Eq.~\eqref{eq:ChebyCR} reads $(\kappa-1)/(\kappa+1)$.
\end{theoremA}
The proof is presented at the end of this section. 
This theorem makes precise in which sense the Chebyshev schedule is optimal: it minimizes the worst case over all $\gamma$ in $[\gamma_{\min},\gamma_{\max}]$, which is the relevant measure when only $\gamma_{\min}$ and $\gamma_{\max}$ are known. 
 For a given spectrum, the numbers $\gamma_{ab}$ form a finite subset of this interval, and a schedule that is adapted to this subset can have smaller factors. However, finding such a schedule demands a full knowledge of all $\gamma_{ab}$'s and solving the more involved optimization problem of Eq.~\eqref{appeq:fulknowledgecontraction}. Therefore, the Chebyshev schedule offers a middle ground: it has optimality in a weaker sense but maintains practicality.

The definition of an average contraction rate does not mean that this method converges uniformly. In fact, some elements of the residual matrix can be amplified over a single iteration of the algorithm since individual contraction terms $|1-s^\star_k\gamma_{ij}|$ can be more than one. However, after one epoch of $m$ iterations all residuals are guaranteed to reduce at least by the contraction rate of Eq.~\eqref{eq:ChebyCR}. The order of the steps within a cycle does not change its net
effect, but it sets how far a mode can grow in between: in increasing or decreasing order, the long steps amplify
the fast modes by a large factor before the short steps damp them again. This can also amplify higher order terms and undermine the stability of the algorithm. Therefore, we take the steps in the order proposed by  Lebedev
and Finogenov~\cite{lebedev1971ordering}, which interleaves long and short steps and keeps this intermediate growth
bounded.

 So far we have established that the off-diagonal elements contract with a rate of $v_{\max}$ per iteration. Therefore, the residual matrix will have a Frobenius norm of at least $\epsilon$ after
\be \label{eq:maxIters}
K = \mathcal{O}\left(\frac{\log(\epsilon/{2})}{\log(v_{\max})}\right) 
\ee
iterations. 
The average contraction factor $v^\star_{\max}$ of the Chebyshev schedule decreases monotonically in the period $m$. For the constant step size $s^\star$, the slowest off-diagonal element  contracts by $(\kappa-1)/(\kappa+1)\approx e^{-2/\kappa}$ per iteration. For a Chebyshev schedule with a long period, the average factor per step approaches $\eta\approx e^{-2/\sqrt\kappa}$. The number of steps needed to reduce the off-diagonal elements by a given factor thus scales as $\sqrt\kappa$ instead of $\kappa$. In this sense, the schedule effectively replaces $\kappa$ by its square root. 
These results directly link the number of required iterations to $\kappa$, which is the subject of study in the next section.

\subsubsection{Proof of Lemma~\ref{lemma:chebyshev}}

 The function $P$ of Eq.~\eqref{eq:period-polynomial} is a real polynomial of degree $m$ with $P(0)=1$ and with the roots $1/s_k$. We prove the bound of part (a) for all real polynomials of degree at most $m$ with $P(0)=1$. Part (b) then follows because the minimizing polynomial has $m$ real roots. To this end, we construct a candidate polynomial using Chebyshev polynomials and prove its optimality. 
 
 The affine map
\begin{equation}
x(\gamma)=\frac{\gamma_{\max}+\gamma_{\min}-2\gamma}{\gamma_{\max}-\gamma_{\min}}
\end{equation}
sends $\gamma_{\min}$ to $1$, $\gamma_{\max}$ to $-1$, and $\gamma=0$ to
\begin{equation}
x_0=\frac{\gamma_{\max}+\gamma_{\min}}{\gamma_{\max}-\gamma_{\min}}=\frac{\kappa+1}{\kappa-1}>1 .
\end{equation}
Let $T_m$ be the Chebyshev polynomial of the first kind, $T_m(x)=\cos(m\arccos x)$ for $|x|\le1$. We define
\begin{equation}
P^\star(\gamma)=\frac{T_m\left(x(\gamma)\right)}{T_m(x_0)} .
\end{equation}
It is a polynomial of degree $m$ with $P^\star(0)=1$. Moreover, on the interval $[\gamma_{\min}, \gamma_{\max}]$ we have  $ |P(\gamma)| \leq 1/T_m(x_0)$ because $|T_m|\le1$ on $[-1,1]$. The polynomial $T_m$ takes the values $T_m(\cos(l\pi/m))=(-1)^l$ at the $m+1$ points $\cos(l\pi/m)$, $l=0,\dots,m$. Hence, $P^\star$ takes the values $\pm1/T_m(x_0)$ with alternating signs at $m+1$ points, that we denote by $\gamma_{\min}=\gamma^{(0)}<\gamma^{(1)}<\dots<\gamma^{(m)}=\gamma_{\max}$.

 Next, we prove by contradiction that every other degree $m$ polynomial that vanishes at $\gamma = 0$ has a higher maximum absolute value than $P^\star$ in the interval $[\gamma_{\min}, \gamma_{\max}]$. Suppose that a polynomial $P \neq P^\star$ of degree at most $m$ with $P(0)=1$ had a maximum absolute value smaller than $1/T_m(x_0)$. Then, the difference polynomial $\Delta P = P^\star-P$ would have the sign of $P^\star$ at each of the $m+1$ points $\gamma^{(l)}$, so it would change sign at least $m$ times in $[\gamma_{\min},\gamma_{\max}]$ and would have at least $m$ zeros in that interval. In addition, it vanishes at $\gamma=0$, which lies outside this interval. Therefore, $\Delta P$ of degree at most $m$ has $m+1$ zeros. Thus, $\Delta P = 0$ for all points and $P=P^\star$, which contradicts the assumption. 
 
 This proves part (a) with the value $1/T_m(x_0)$. If $P$ has the same worst-case factor as $P^\star$, the same counting argument, with zeros counted with their multiplicity, again gives $P=P^\star$, which is the uniqueness in part (b).

The roots of $T_m$ are $\cos[(2k-1)\pi/(2m)]$, $k=1,\dots,m$. Mapping them back with $\gamma=[\gamma_{\max}+\gamma_{\min}-(\gamma_{\max}-\gamma_{\min})x]/2$ gives the roots $1/s_k$ of $P^\star$. Since the set of the numbers $\cos[(2k-1)\pi/(2m)]$ is symmetric under a change of sign, these roots are
\begin{equation}
\label{chebysch}
\frac{1}{s^\star_k}=\frac{\gamma_{\max}+\gamma_{\min}}{2}+\frac{\gamma_{\max}-\gamma_{\min}}{2}\cos\left(\frac{(2k-1)\pi}{2m}\right),\qquad k=1,\dots,m,
\end{equation}
which is Eq.~\eqref{chebyschMain}. All roots lie in $(\gamma_{\min},\gamma_{\max})$, so $P^\star$ is of the form of Eq.~\eqref{eq:period-polynomial} with positive step sizes.

Next, we compute $\frac{1}{T_m(x_0)}$. For $x\ge1$, $T_m(x)=\tfrac12[(x+\sqrt{x^2-1})^m+(x-\sqrt{x^2-1})^m]$. Moreover, for  $x_0=(\kappa+1)/(\kappa-1)$ we have $\sqrt{x_0^2-1}=2\sqrt\kappa/(\kappa-1)$, so that $x_0+\sqrt{x_0^2-1}=(\sqrt\kappa+1)^2/(\kappa-1)=1/\eta$ and $x_0-\sqrt{x_0^2-1}=\eta$. Therefore,
\begin{equation}
\frac{1}{T_m(x_0)}=\frac{2}{\eta^{-m}+\eta^{m}}=\frac{2\eta^m}{1+\eta^{2m}},
\end{equation}
which completes parts (a) and (b). Part (c) follows from $1\le2/(1+\eta^{2m})\le2$, which gives $\eta\le[2\eta^m/(1+\eta^{2m})]^{1/m}\le2^{1/m}\eta$. For part (d), $m=1$ gives $1/s^\star_1=(\gamma_{\max}+\gamma_{\min})/2$ and $2\eta/(1+\eta^2)=(\kappa-1)/(\kappa+1)$. \hfill$\square$
\subsection{Optimal choice of weights}
\label{app:weights}

This section determines how the weights enter the convergence rate in the linear regime. We prove that the inverse-gap weights used in the numerical simulations are optimal, and derive the scaling of the number of required iterations with the number of states $r$. 

The weights enter all results through $\gamma_{\min}, \gamma_{\max}$, which set the size of the convergence region $ \Delta E(\rho_{\mathrm{ref}})<\gamma_{\min}$, the guaranteed rate $2\gamma_{\min}$, and through $\kappa=\gamma_{\max}/\gamma_{\min}$ (Appendix~\ref{app:optimal_step_schedules}). In the linear regime, $\kappa$ sets the optimal contraction rate and a lower $\kappa$ results in a faster convergence. Therefore, in this regime, we aim to find a set of weights that minimize $\kappa$ 
\begin{align}
\min_{\mathbf{p}\in \Delta^{r-1}}& \kappa =  \frac{\gamma_{\max}}{\gamma_{\min}} \nonumber \, , 
\end{align}
where $\mathbf{p}$ is the vector of the weights and $\Delta^{r-1}$ denotes the $(r-1)$-dimensional simplex. Without loss of generality, we assume the $r$ non-zero weights are decreasingly ordered, i.e. we have $\mathbf{p}\in \Delta^{r-1}, \ p_i\geq p_j \ \forall i<j $. $\gamma_{\min}, \gamma_{\max}$ are the highest and lowest numbers among $\gamma_{ij} = (p_j-p_i)(\lambda_i-\lambda_j)$. Therefore, $\gamma_{\max} = p_1\Omega$. Moreover, $\gamma_{\min}$ is among the rates set by adjacent levels $\gamma_{i,i+1}$. Thus,  we can write the problem as the following convex problem
\begin{align}
&\min_{\mathbf{p}\in \Delta^{r-1}, \gamma_{\min} \in \mathbb{R_+}} \frac{p_1\Omega}{\gamma_{\min}}  \nonumber \\
\mathrm{s.t} & \quad  \gamma_{\min} \leq \delta p_{i}\delta\lambda_{i} \quad \forall i\leq r ,
\end{align} 
where $\Omega$ is the spectral width, $\delta \lambda_i := \lambda_{i+1}-\lambda_{i}$, and $\delta p_i := p_i-p_{i+1}$, with the convention $\delta p_r = p_r$. 

In the following proposition we show that the \textit{inverse-gap weights} are the solution to the problem and we present their corresponding $\kappa$. 

\begin{propositionA}[Inverse-gap weights]
\label{propA:weights}
Among all weights $p_1>\dots>p_r>0$ that sum to one, the weights with the gaps
\begin{equation}
\label{eq:pstar}
\delta p_j=\frac{\gamma_{\min}}{\delta\lambda_{j}},\qquad \gamma_{\min}=\left(\sum_{j=1}^{r}\frac{j}{\delta\lambda_{j}}\right)^{-1},
\end{equation}
maximize $\gamma_{\min}$ and minimize $\kappa$. For these weights all $\gamma_{j,j+1}$ with $j\le r$ are equal, and
\begin{equation}
\label{eq:kappastar}
\kappa^\star=\Omega\sum_{j=1}^{r}\frac{1}{\delta\lambda_{j}} .
\end{equation}
\end{propositionA}
\begin{proof}
Since $p_i=\sum_{j=i}^{r}\delta p_j$, the normalization reads $1=\sum_ip_i=\sum_{j=1}^{r}j\,\delta p_j$, and $p_1=\sum_{j=1}^{r}\delta p_j$. By the definition of $\gamma_{\min}$, every weight gap satisfies $\delta p_j\ge\gamma_{\min}/\delta\lambda_{j}$. Hence $1\ge\gamma_{\min}\sum_jj/\delta\lambda_{j}$, which bounds $\gamma_{\min}$ from above by the value of Eq.~\eqref{eq:pstar}. In the same way, $p_1\ge\gamma_{\min}\sum_j1/\delta\lambda_{j}$, so that $\kappa=p_1\Omega/\gamma_{\min}\ge\Omega\sum_j1/\delta\lambda_{j}$. Both bounds are attained if and only if $\delta p_j\,\delta\lambda_{j}=\gamma_{\min}$ for all $j\le r$.
\end{proof}
The proposition states that if the differences between weights are proportional to the inverse energy gaps of the corresponding states, the minimum $\kappa$ is achieved. Crucially, only the gaps between the $r+1$ lowest levels affect the optimal value, and not the smallest gap of the entire spectrum.

With $\delta\lambda_{\min}$ and $\delta\lambda_{\max}$ denoting the smallest and the largest of the gaps $\delta\lambda_1,\dots,\delta\lambda_{r}$, Eq.~\eqref{eq:kappastar} gives 
$$r \frac{\Omega}{\delta\lambda_{\max}}\le\kappa^\star\le r\,\frac{\Omega}{\delta\lambda_{\min}}, $$
and 
$$2\frac{\delta\lambda_{\min}}{r(r+1)}\le\gamma_{\min}\le \,2\frac{\delta\lambda_{\max}}{r(r+1)} .$$

In the linear regime, a Chebyshev schedule with a long period reduces the number of iterations of the discretized DBF from order $\kappa$ to order $\sqrt\kappa$ (Appendix~\ref{app:chebyshev}), that is, to order $(r\,\Omega/\delta\lambda_{\min})^{1/2}$. However, for the optimal constant step size the scaling remains linear.

Moreover, by combining these results with the results of Appendix~\ref{app for steps}, we observe the scaling of various parameters of the problem with $r$. The flow time of the continuous-time DBF grows quadratically with $r$, because $\gamma_{\min}$ shrink as $r^{-2}$. For the same reason, the guaranteed region $\Delta E(\rho_{\mathrm{ref}})<\gamma_{\min}$ also shrinks as $r^{-2}$ for the optimal weights. This hints that the choice of the reference state plays a more crucial role in the convergence speed for large $r$.

\section{Numerical Experiments} \label{app:numerics}
The DBF algorithm for excited state computation has three main hyperparameters: the step $s$ of the discrete double bracket evolution, the weights $p$, and the reference orthogonal states $\ket{\phi_i}$. In this section, we numerically study the role of these parameters on the performance of the algorithm. So far, our theoretical results have provided us with guidelines about the selection of these parameters to ensure the fastest performance. More specifically, we have derived optimal values and strategies for step sizes and weights in Appendix~\ref{app:linearizedFlow}, and sufficient and necessary conditions on $\ket{\phi_i}$'s in Theorems~\ref{theorem:convergence},~\ref{theorem:convergence-rate}. Here, we numerically verify each of these analytical results in a separate section: Sections~\ref{app:num_stepsize},~\ref{app:num_weights},~\ref{app:num_initial states} explore the step schedules, the weights and the choice of $\ket{\phi_i}$'s respectively.  Our numerical studies reveal important nuances about the practicality and stability of the optimal step sizes, which should be taken into account when using product formulae for the implementation of the algorithm. A main limitation of the theoretical optimal parameters is that they require some knowledge of the energy spectrum which is not provided in practical problems. To circumvent this issue,  we also suggest  and study a practical adaptive algorithm in Section~\ref{app:num_adaptive}, which does not require any prior information about the spectrum. 

Primarily, to verify the optimality of the proposed schedules, we use a toy model of a 10-qubit transverse-field Ising Hamiltonian 
 \be \label{eq:TFIM}
H_{\text{TFIM}} = -\sum_i J_i Z_i Z_{i+1} - \sum_i h_i X_i \,.
\ee

The exact settings of the chain in each numerical experiment is shown in Table~\ref{tab:hamiltonians}.

\begin{table}[t]
  \centering
  \small
  \newcommand{\cell}[2]{\parbox[t]{#1\linewidth}{\raggedright #2\strut}}
  \begin{tabular}{@{}llll@{}}
    \toprule
    Figs. & Boundary & Couplings & Normalization \\
    \midrule
    \cell{0.17}{Figures~\ref{fig:last-gap}, ~\ref{fig:grid}}
      & open
      & \cell{0.20}{uniform, $J = h = 1$}
      & \cell{0.35}{$H / \|H\|$, $\|H\| = 12.3815$} \\
    \addlinespace
    \cell{0.17}{Figures~\ref{fig:ChebyshevScheMain},      
                    ~\ref{fig:ChebyshevSche},   
                ~\ref{fig:steps},          
                ~\ref{fig:weightscompared}}        
      & open
      & \cell{0.20}{uniform, $J = h = 1$}
      & \cell{0.35}{$H / \|H\|$, $\|H\| = 12.3815$} \\
    \addlinespace
    \cell{0.17}{Figures~\ref{fig:adaptive},~\ref{fig:order4},~\ref{fig:warmstart}}
      & open
      & \cell{0.20}{uniform, $J = h = 1$}
      & \cell{0.35}{$c\, H / \|H\|$, $c = 0.9474$ drawn from $U(0.8, 1)$, not known to the adaptive schedule} \\
    \addlinespace
    \cell{0.17}{Figures~\ref{fig:overlap}, ~\ref{fig:rate},~\ref{fig:saddle},~\ref{fig:far},~\ref{fig:mainNumerics}}
      & open
      & \cell{0.20}{random, $J_i, h_i$ i.i.d.\ $U(0.5, 1.5)$; 10 instances\textsuperscript{a}}
      & \cell{0.35}{each instance mapped onto $[-1, 1]$ by its exact extreme eigenvalues} \\
    \bottomrule
  \end{tabular}
  \caption{Hamiltonians of the figures: the open chain
  $H = -\sum_{i=1}^{9} J_i Z_i Z_{i+1} - \sum_{i=1}^{10} h_i X_i$ on the full $2^{10}$-dimensional space.
  \textsuperscript{a}The first ten seeds whose smallest gap among the five lowest levels is at least $0.3$ times
  their mean gap.}
  \label{tab:hamiltonians}
\end{table}

\paragraph*{Important notation distinction.} In the main text, we defined $\Delta E(\rho_k) =  E(\rho_k)- E^*$ as the excess energy at the $k$'th iteration and the theoretical conditions are framed in terms of this excess energy. In this appendix, we also track the energy excesses of individual orthogonal states $\ket{\phi_i^{(k)}}$, which we define as $\Delta E_i(\rho_k) := \bra{\phi_i^{(k)}} H\ket{\phi_i^{(k)}} - \lambda_i$. Most of the convergence curves are  plotted in terms of $\max_i \Delta E_i(\rho_k)$ over the number of iterations $k$. The only exception is Section~\ref{app:num_initial states} where we are verifying the theoretical conditions of Theorems~\ref{theorem:convergence},~\ref{theorem:convergence-rate} and we plot $\Delta E(\rho_k)$. These two quantities are related through $\Delta E (\rho_k) =\sum_{i} p_i \Delta E_i(\rho_k)$ and  $\max_i \Delta E_i(\rho_k) \geq \Delta E(\rho_k)$.  

Moreover, in most cases the algorithm either terminates after a fixed number of iterations or when the estimated energies stagnate and do not change more than a threshold ($\max_i \Delta E_i(\rho_k) \leq 10^{-10}$). This criteria for convergence is stricter than the main text in the sense that when  $\max_i \Delta E_i(\rho_k) \leq 10^{-10}$, $\Delta E(\rho_k) \leq 10^{-10}$ is guaranteed.

\subsection{The step size} \label{app:num_stepsize}
\begin{figure}
    \centering
    \includegraphics[width=\linewidth]{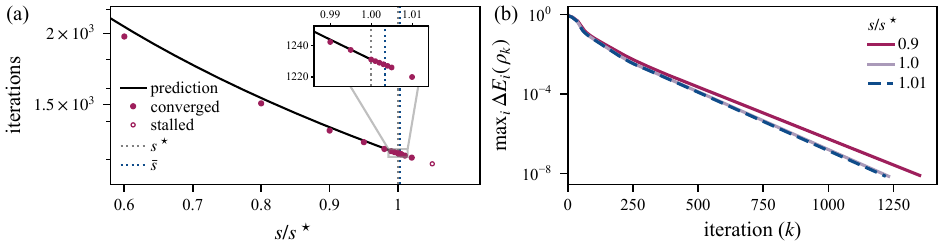}
    \caption{\textbf{Effect of the step size on the convergence of the algorithm.} Tracking the first 4 eigenstates of the 10-qubit TFIM with equal weights.  \textbf{(a)}~Iterations until the energy change satisfies $\max_i \Delta E_i < 10^{-10}$, against $s/s^\star$. 
Black line: the count at $s^\star$ rescaled by $\ln( |1 - s^\star\gamma_{\min}|)/\ln (|1 - s\gamma_{\min}|)$ according to predictions (Eq.~\eqref{eq:maxIters}). Red circle: the run at $s = 1.05\,s^\star$ meets the stopping criterion but stalls away from the
eigenstates. Dotted lines: $s^\star$ and $\bar{s} = 2/\gamma_{\max} = 1.0034\,s^\star$,
the stability edge of the linearized exact flow. Inset: the steps near $s^\star$.
\textbf{(b)}~$\max_i |\Delta E_i{(\rho_k)}|$, against the iteration for
$s/s^\star = 0.9$, $1$ and $1.01$.}
    \label{fig:steps}
\end{figure}

In this section, we perform numerical studies to establish the optimality of the proposed step schedules. First, we begin by considering fixed step sizes.  
 Figure~\ref{fig:steps} shows the number of required iterations  to reach an error threshold of $10^{-10}$ as a function of step size. Here, we track the first 4 eigenstates of the TFIM with equal couplings using an order 4 product formula. We verify the $1/\log(v_{\max})$ scaling of the iterations as predicted by Eq.~\eqref{eq:maxIters}. The numerically optimal step is $\hat{s}= 1.02s^\star$, which is slightly different from theory due to the use of product formulae instead of the exact implementation of the DBF. Moreover, as predicted by Theorem~\ref{theorem:convergence}, choosing steps higher than the stability threshold leads to the instability of the algorithm which is shown by the red circle marker. 

Next, we consider $m$-periodic Chebyshev schedules, which are expected to be optimal in the linear regime (Lemma~\ref{lemma:chebyshev}). Specifically, in the asymptotic limit of large $m$ the  optimal average contraction rate becomes 
\be \label{eq:ChebCRass}
v^\star \approx \frac{\sqrt{\kappa}-1}{\sqrt{\kappa}+1} \ .
\ee 
However, in general, the optimality of Chebyshev schedules does not hold when the algorithm is far away from the target and the problem cannot be linearized. In fact, this scheduling can be unstable when the state is far from target as it involves step sizes larger than the stability threshold $\bar{s} = 2/\gamma_{\max}$ required by Theorem~\ref{theorem:convergence}. This issue is more dominant for large $m$ because the largest step size grows closer to $\gamma_{\min}^{-1}$ as $m$ is increased (Eq.~\eqref{chebysch}).

\begin{figure}
        \centering
        \includegraphics[width=\linewidth]{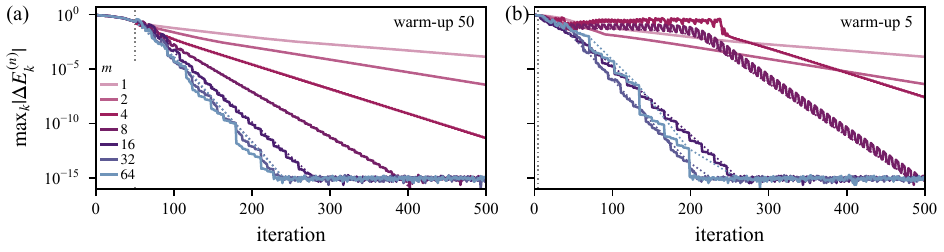}
        \caption{\textbf{Chebyshev Scheduling improves convergence rate.} The error of Chebyshev scheduling for various periods $m$ plotted over the number of iterations. The task is to find the first 4 lowest energy states of the TFIM Hamiltonian of Eq.~\eqref{eq:TFIM} with equal couplings. The initial states are uniform combinations of all the energy eigenstates and the weights are fixed to their optimal inverse-gap value. \textbf{(a)}All runs start with 50 rounds of warm up with $m=1$ to improve the initial state. \textbf{(b)} starts with only 5 iterations of warm up and two of the runs are slowed down. The product formula order is fixed to four for all cases.} 
        \label{fig:ChebyshevSche}
    \end{figure}

Moreover, for our product formula implementation, the instability issue is twofold. Namely, the large step sizes may affect the stability even in the linear regime because the approximation error of the product formula grows with the step size. Consequently, to maintain precision, higher order product formulae may be required, which increases the gate count of a single iteration. Therefore, although increasing the period $m$ decreases the total number of double-bracket iterations, it does not necessarily decrease the overall gate complexity of the algorithm. Nevertheless, for the specific task of  tracking the first 4 eigenstates of the TFIM with equal couplings, we are able to reach a period of $m=64$ without the need for higher order  product formulae.  Figure~\ref{fig:ChebyshevSche}(a) (same as Figure~\ref{fig:ChebyshevScheMain} in the main text) shows the error over the number of iterations.  The $m=64$ Chebyshev scheduling requires 150 iterations to reach the error of $10^{-8}$ compared to  the  1200 iterations required by constant step scheduling ($m=1$). The algorithm is initiated with a 50 iteration warm-up, where the constant step size is used across all periods to ensure stability. In panel~(b), this warm-up is reduced to 5 iterations and two of the runs converge slower and with more fluctuations. This highlights the role of the warm-up phase in the convergence of the algorithm. 

\subsection{The weights \texorpdfstring{$\mathbf{p}$}{}} \label{app:num_weights}
\begin{figure}
    \centering
    \includegraphics[width=\linewidth]{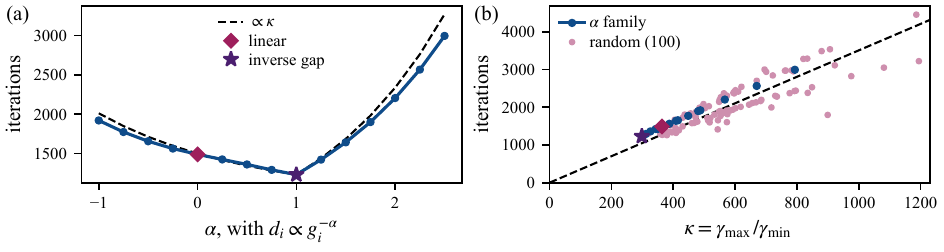}
    \caption{\textbf{Effect of the choice of weights on the convergence speed of the algorithm.} The number of iterations required for the algorithm to find the four lowest eigenstates of a 10 qubit TFIM model. \textbf{(a)} shows the number of iteration over a parameter of the probability space $\alpha$. $\alpha=1$ corresponds to the inverse gap rule and $\alpha=0$ corresponds to equally spaced weights. \textbf{(b)} shows the scaling of the number of iterations with the condition number $\kappa$ for 100 randomly drawn weights. The inverse-gap rule has both the lowest condition number and the lowest iteration count as predicted by the theory.}
    \label{fig:weightscompared}
\end{figure}

In Appendix~\ref{app:weights}, we showed that the fastest convergence is obtained by inverse-gap weights as they minimize the condition number of the problem.  
Figure~\ref{fig:weightscompared} shows the optimality of the proposed inverse gap weight selection. In panel~(a), the weights are parameterized in $\alpha$ as $\delta p_i \propto \delta \lambda_i^{-\alpha}$. The minimum number of iterations coincides with $\alpha=1$ implying the optimality of the inverse-gaps. In panel~(b), the number of iterations is plotted over  $\kappa$ for random and parametrized weights. The minimum happens when the weights are chosen according to the inverse-gap rule. Moreover, the scaling of required iterations with $\kappa$ is almost linear as predicted for the constant step size scheduling in Appendix~\ref{app:optimal_step_schedules}. 

\subsection{The reference states \texorpdfstring{$|\phi_i\rangle$}{}}
\label{app:num_initial states}
In this section, we numerically explore the role of the reference states $\ket{\phi_i}$ in the convergence of the algorithm. First, we numerically demonstrate that in order for the flow to converge to its stable fixed point of lowest energy, $\ket{\phi_i}$'s must have overlap with all the lowest $r$ energy eigenstates (Theorem~\ref{theorem:convergence}). Next, we demonstrate that for reference states that are close to one of the unstable saddle points the convergence is slowed down (consequence of Theorem~\ref{theorem:convergence-rate}). Therefore, ideally, a good reference state for the algorithm has to have two main properties: (i) it should have non-zero overlap with all the lowest $r$ eigenstates, and (ii) its resulting trajectory should stay away from all the fixed points before reaching the target at the end. 

\subsubsection{Necessary condition: overlap with the target eigenstates}
The flow can only reach eigenstates that the reference states $\ket{\phi_i}$'s overlap with. Let $P_r$ be the projector onto the span of the
$r$ reference states and $A_i = \bra{\lambda_i}P_r\ket{\lambda_i}$. If $A_i = 0$ for some tracked level $i \le r$, the span of the tracked states stays orthogonal to $\ket{\lambda_i}$ at every
iteration and the flow does not converge to the target fixed point $\rho^*$ (Theorem~\ref{theorem:convergence}).  More precisely, the final energy levels are $\ell_1 < \dots < \ell_r$ obtained by
removing $i$ from $\{1, \dots, r+1\}$, and the energy excess converges to
\begin{equation}
  \Delta E_{\mathrm{wrong}} = \sum_{j=1}^{r} p_j \left(\lambda_{\ell_j} - \lambda_j\right)  > 0.
\end{equation}

We test this on ten random transverse-field Ising chains with $N = 10$ qubits,
$H = -\sum_i J_i Z_i Z_{i+1} - \sum_i h_i X_i$, with $J_i, h_i$ drawn uniformly from $[0.5, 1.5]$. Each $H$ is normalized
to the spectrum $[-1, 1]$, and instances with a near-degenerate pair among the five lowest levels are discarded. We track
the $r = 4$ lowest eigenstates with inverse-gap weights and the optimal constant step $s^*$, using the fourth-order
product formula. For each Hamiltonian we draw four Haar-random orthonormal initial states. We then remove $\ket{\lambda_i}$
from all four, for $i = 1, \dots, 4$, and re-orthonormalize them. This gives five starts per Hamiltonian: one with all
$A_i > 0$ and four with a single $A_i = 0$. In floating point precision, round-off reintroduces a component of order $10^{-16}$
along $\ket{\lambda_i}$. This component would eventually
grow, because the confined fixed point is unstable in the full space. To isolate the exact-arithmetic behavior, we remove the component along $\ket{\lambda_i}$ after every iteration. This keeps
$A_i$ small throughout the algorithm. 

\begin{figure}[t]
  \centering
  \includegraphics[width=\textwidth]{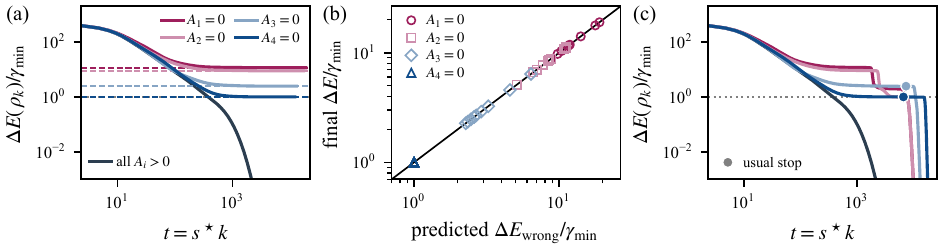}
  \caption{\textbf{Necessity of the overlap condition.} Ten random 10-qubit TFIM chains, $r = 4$ tracked states,
  inverse-gap weights, optimal step $s^\star$, fourth-order product formula. \textbf{(a)}~Excess energy, in units of
  $\gamma_{\min}$, against the flow time $t = s^\star k$, for one Hamiltonian. The initial states are Haar random (all
  $A_i > 0$), or the same states with the eigenstate $\ket{\lambda_i}$ removed ($A_i = 0$). The corresponding overlaps are removed after each iteration to stop the propagation of floating-point round-off error. Dashed lines show the predicted $\Delta E_{\mathrm{wrong}}/\gamma_{\min}$ of the fixed point without
  $\ket{\lambda_i}$, and the dotted line marks $\Delta E = \gamma_{\min}$. \textbf{(b)}~Final against predicted
  $\Delta E_{\mathrm{wrong}}/\gamma_{\min}$ for all ten Hamiltonians and $i = 1, \dots, 4$, with the same projection
  as in (a). \textbf{(c)}~The same setting as (a) but the round-off component along $\ket{\lambda_i}$ grows and every run escapes its plateau. For $A_1 = 0$
  and $A_2 = 0$ the flow then passes through a second plateau near another fixed point. Dots mark where the algorithm would have terminated without prior information as the largest energy change per iteration falls below
  $10^{-10}$.}
  \label{fig:overlap}
\end{figure}

Figure~\ref{fig:overlap} shows the results. In panel (a), with all overlaps nonzero, $\Delta E{(\rho_k)}$ decays exponentially to the target.
With $A_i = 0$, every run instead converges to a plateau at the predicted $\Delta E_{\mathrm{wrong}}$. All 40 runs
 end on the predicted levels and their final excess energy agrees with $\Delta E_{\mathrm{wrong}}$ (panel (b)). In panel (c), the corresponding $A_i$'s are not removed at every step and therefore the round-off introduces some overlap with eliminated states. In this case, all leave the plateaus and eventually reach the target. However, realistically, we would stop the iterations once the energies do not change more than a certain threshold. In two cases, this criterion terminates  the algorithm on a non-target plateau and  without reaching the target. 

In conclusion, a nonzero overlap with each of the $r$ lowest eigenstates is necessary. Convergence alone cannot 
detect the absence of the overlap, because the confined flow converges to a genuine fixed point made of exact eigenstates.

\subsubsection{Sufficient condition and fixed points}
 For an initial state with
$\Delta E{(\rho_{\mathrm{ref}})} < \gamma_{\min}$, Theorem~\ref{theorem:convergence} predicts that $\Delta E$ decays at the rate $2\gamma_{\min}$ per unit flow time
$t = s k$. 
We test this sufficient condition and its convergence rate for  three regimes: reference states satisfying the condition, reference states close to a non-target fixed point, and reference states that violate the condition but are away from any fixed point. Our results show that this condition is not necessary as reference states with higher initial excess energy can also converge with rate $2\gamma_{\min}$. All runs use the ten
Hamiltonians, weights, step and product formula of the previous subsection.

\paragraph{Below the lowest saddle.}
Here, reference states $\rho_{\mathrm{ref}}$ are prepared at a controlled value of $\Delta E{(\rho_{\mathrm{ref}})}$ below the lowest non-target fixed point.  We draw a random
anti-Hermitian generator $A$ acting on the span of the eight lowest eigenenergies, $A = \sum_{i,j \le 8} A_{ij} \ket{E_i}\!\bra{E_j}$ with
$A_{ij}$ independent complex Gaussian entries.  $\ket{\phi_i}$'s are obtained by applying $\exp(\theta A)$ to the four target states. The angle
$\theta$ is fixed by bisection so that $\Delta E{(\rho_{\mathrm{ref}})}/\gamma_{\min}= \{0.01, 0.1, 0.3, 0.6, 0.9\}$ exactly. We use
three random generators per Hamiltonian, giving 150 runs, each continued until $\Delta E(\rho_k) < 10^{-13}$.

We characterize each run by two decay rates.  
First, the late rate $\zeta_{\mathrm{late}}$ quantifies the decay at late times and close to the target. It is obtained by  the slope of a least-squares fit to $\ln\Delta E{(\rho_k)}$ against $t= s^\star k$, the flow time after $k$ iterations. The
fit uses the iterations with $k \ge 50$ and $10^{-12} < \Delta E{(\rho_k)} < 10^{-2}\Delta E{(\rho_\mathrm{ref})}$, which excludes the initial
transient stage. Second, the average rate
\begin{equation}
  \bar\zeta = \min_{k \ge 1} \left[ -\frac{1}{t} \ln\frac{\Delta E{(\rho_k)}}{\Delta E{(\rho_{\mathrm{ref}})}} \right] ,
\end{equation}
is the smallest average decay rate from the start to any iteration. A decay of the form
$\Delta E{(\rho_k)} \le \Delta E(\rho_{\mathrm{ref}}) e^{-\zeta t}$ holds at every iteration exactly  if $ \bar\zeta \ge \zeta$.

\begin{figure}[t]
  \centering
  \includegraphics[width=\textwidth]{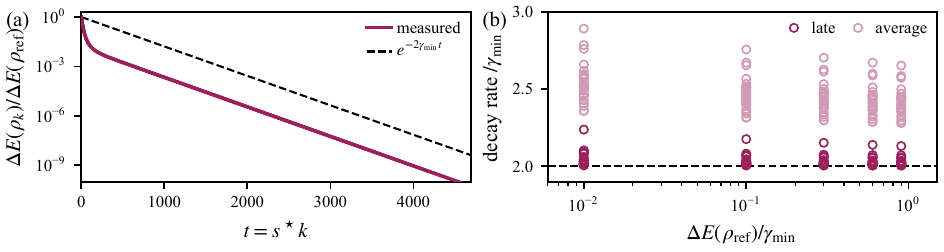}
  \caption{\textbf{Convergence rate below the lowest saddle.} Same Hamiltonians and parameters as
  Fig.~\ref{fig:overlap}. The $\ket{\phi_i}$'s are the four target eigenstates rotated by a random unitary acting on
  the eight lowest eigenstates, with the angle chosen so that
  $\Delta E(\rho_{\mathrm{ref}})/\gamma_{\min} \in \{0.01, 0.1, 0.3, 0.6, 0.9\}$. \textbf{(a)}~$\Delta E(\rho_k)/\Delta E(\rho_{\mathrm{ref}})$
  against the flow time $t = s^\star k$ for one Hamiltonian and one generator. The five measured curves coincide.
  The dashed line is the predicted decay $e^{-2\gamma_{\min} t}$. \textbf{(b)}~Decay rates, each in units of the
  $\gamma_{\min}$ of its own Hamiltonian, for all ten Hamiltonians and three generators each. Dark circles show the
  late rates $\zeta_{\mathrm{late}}$, and light circles show the average rates of the entire run. The dashed line is the predicted rate
  $2\gamma_{\min}$.}
  \label{fig:rate}
\end{figure}

Figure~\ref{fig:rate} shows the result. After a short transient stage, every run decays
exponentially.  The decay rates are independent of $\Delta E(\rho_{\mathrm{ref}})$: the curves for the five values of $\Delta E(\rho_{\mathrm{ref}})$ coincide once
divided by $\Delta E(\rho_{\mathrm{ref}})$. 

The late rates lie between $2.005\,\gamma_{\min}$ and $2.24\,\gamma_{\min}$, with mean
$2.03\,\gamma_{\min}$, all of which are above $2\gamma_{\min}$ as predicted by Theorem~\ref{theorem:convergence-rate}. 
 The decay also respects the prediction at every iteration. The average rate is $\bar\zeta \ge 2.28\,\gamma_{\min}$
for every run. This rate exceeds the late rate because it includes the fast initial transient stage. Hence,
$\Delta E(\rho_k) \le 0.89\,\Delta E(\rho_{\mathrm{ref}}) e^{-2\gamma_{\min} t_k}$ throughout. In addition, $\Delta E(\rho_k)$ decreases monotonically in
all 150 runs.

\paragraph{Close to a saddle.}
Above the initial excess energy of $\gamma_{\min}$ the sufficient condition of Theorem~\ref{theorem:convergence-rate} is violated, and the flow can pass close to a saddle. Then, the decay rate 
depends on the distance to the saddle.

We test this on the ten Hamiltonians. For each we take seven saddles. Four are the lowest energy saddles: energy levels 1 and 2, 2 and 3, or 3 and 4 exchanged, or level 4 moved to level 5. One is a higher saddle with levels 1
and 3 exchanged ($\Delta E\approx 4$--$6.7\,\gamma_{\min}$). Two are far from the target: all four states in reverse order
($\Delta E\approx  10$--$15\,\gamma_{\min}$), and all four states moved up one level ($\Delta E \approx 8.8$--$19\,\gamma_{\min}$). $\ket{\phi_i}$'s are obtained via rotating the saddle states by $e^{\theta A}$, with $A$ a random anti-Hermitian generator,
normalized so that the mean infidelity with the saddle state is $\theta^2$. We use
$\theta \in \{10^{-2}, \dots, 10^{-6}\}$, with the same $A$ for all $\theta$ of a given saddle and Hamiltonian, giving 350 runs. 

\begin{figure}[t]
  \centering
  \includegraphics[width=\textwidth]{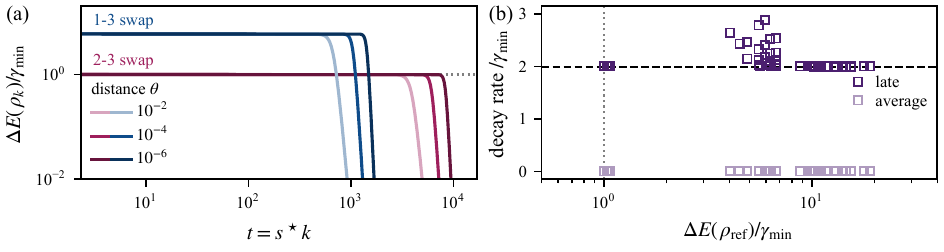}
    \caption{\textbf{Escape from saddles.} Same Hamiltonians and parameters as Figure~\ref{fig:overlap}. The initial states
  are a saddle configuration rotated by a random unitary $e^{\theta A}$, with mean infidelity $\theta^2$ to the
  saddle states. \textbf{(a)}~$\Delta E(\rho_k)/\gamma_{\min}$ against the flow time $t = s^\star k$ for one Hamiltonian,
  starting near the lowest saddle with states 2 and 3 exchanged ($\Delta E = \gamma_{\min}$, pink) and
  near the higher saddle with states 1 and 3 exchanged ($\Delta E = 5.9\,\gamma_{\min}$, azure), for $\theta = 10^{-2}, 10^{-4}, 10^{-6}$ (darker shades for smaller $\theta$). \textbf{(b)}~Late decay rate
  $\zeta_{\mathrm{late}}$ (dark) and  average decay rate $\bar\zeta$ (light), defined as in Fig.~\ref{fig:rate}, in units
  of $\gamma_{\min}$, against $\Delta E(\rho_{\mathrm{ref}})/\gamma_{\min}$. The points include seven saddles: the four state swaps (states 1 with 2, 2 with 3, 3
  with 4, and 4 with 5 exchanged; $\Delta E(\rho_{\mathrm{ref}}) \approx \gamma_{\min}$), the exchange of states 1 and 3
  ($\Delta E(\rho_{\mathrm{ref}}) \approx4$--$6.7\,\gamma_{\min}$), all states one level up ($\Delta E(\rho_{\mathrm{ref}}) \approx 8.8$--$19\,\gamma_{\min}$) and all states in reverse order
  ($\Delta E(\rho_{\mathrm{ref}}) \approx10$--$15\,\gamma_{\min}$).  The dashed line is the
  predicted rate $2\gamma_{\min}$, and the dotted line marks $\Delta E(\rho_{\mathrm{ref}}) = \gamma_{\min}$.}
  \label{fig:saddle}
\end{figure}

Figure~\ref{fig:saddle}(a) shows a typical run. The state first settles onto the saddle, and $\Delta E$ stays on a
plateau. Then, it leaves the saddle and decays to the target. The plateau is longer for initializations that are close to the saddle and higher energy saddles leave the plateau sooner than lower energy ones. During the plateau the energies hardly change, and the usual stopping rule,  $\max_i \Delta E_i \leq 10^{-10}$, fires on the plateau in most runs. Therefore, an energy-based test cannot tell a saddle from the target.

Figure~\ref{fig:saddle}(b) shows the two decay rates. The run average is close to zero for every saddle. Therefore, no bound of the form $\Delta E(\rho_k) \le \Delta E(\rho_{\mathrm{ref}}) e^{-\zeta t_k}$ with a rate of order
$\gamma_{\min}$ holds near a saddle. The guarantee of the sufficient condition is therefore lost as soon as
$\Delta E(\rho_{\mathrm{ref}}) \ge \gamma_{\min}$, even exactly at the threshold. 

Once the state has left the saddle, however, it converges at the same rate as below the lowest saddle. For the four
lowest saddles, the all-up configuration and the reversed order, the late rate lies between $1.99$ and
$2.01\,\gamma_{\min}$, independent of $\theta$ and of the height of the saddle. 

\paragraph{Above the lowest fixed point, away from saddles.}
Here we consider initial states with excess energy higher than $\gamma_{\min}$ but not close to any fixed point. We
use the construction of the low-energy rotations above, with $A$ restricted to the eight lowest eigenstates, and choose
$\theta$ so that $\Delta E(\rho_{\mathrm{ref}})/\gamma_{\min} \in \{2, 5, 20\}$. We use two random generators per value and Hamiltonian
(60 runs). Moreover, we add one set of Haar-random initial states per Hamiltonian, for which
$\Delta E(\rho_{\mathrm{ref}})/\gamma_{\min}\geq 290$. We measure the late rate $\zeta_{\mathrm{late}}$ and the average rate
$\bar\zeta$ defined above.  

\begin{figure}[t]
  \centering
  \includegraphics[width=\textwidth]{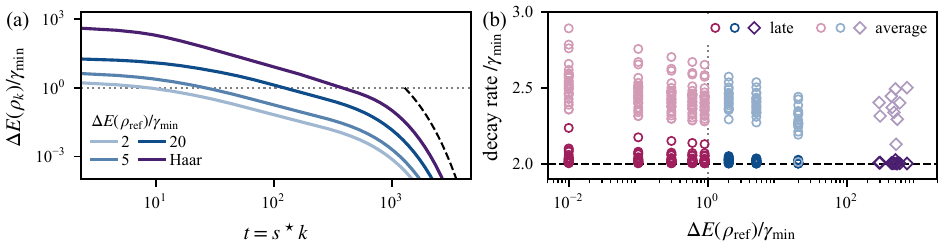}
  \caption{\textbf{Starts above the lowest saddle, away from any saddle.} Same Hamiltonians and parameters as
  Fig.~\ref{fig:overlap}. Cold starts are the four target eigenstates rotated by a random unitary acting on the eight
  lowest eigenstates, with the angle chosen so that
  $\Delta E(\rho_{\mathrm{ref}})/\gamma_{\min} \in \{2, 5, 20\}$. Haar-random starts have
  $\Delta E(\rho_{\mathrm{ref}}) = 290$--$730\,\gamma_{\min}$. \textbf{(a)}~$\Delta E(\rho_k)/\gamma_{\min}$ against the flow time
  $t = s^\star k$ for one Hamiltonian and one generator. Rotated starts are shown in blue, darker for larger
  $\Delta E(\rho_{\mathrm{ref}})$, and the Haar-random start in purple. The dashed line is a decay $e^{-2\gamma_{\min} t}$. The dotted line marks $\Delta E = \gamma_{\min}$. \textbf{(b)}~Late fit
  $\lambda_{\mathrm{late}}$ (dark) and run average $\bar\lambda$ (light) against $\Delta E(\rho_{\mathrm{ref}})/\gamma_{\min}$ for all ten Hamiltonians. Pink circles show the initial states generated by random rotations and purple diamonds show the Haar-random starts. The dashed line is the predicted rate $2\gamma_{\min}$, and
  the dotted line marks $\Delta E(\rho_{\mathrm{ref}}) = \gamma_{\min}$.}
  \label{fig:far}
\end{figure}

As shown in Figure~\ref{fig:far}, none of the runs have a plateau. Panel~(a) shows, $\Delta E$
decreases from the first iteration and crosses the height of the lowest saddle without slowing down (at roughly $t=700$ for Haar random initializations). For comparison, a start at distance $\theta = 10^{-6}$ from a lowest saddle stays on
the plateau for $t \approx 8000$ (Fig.~\ref{fig:saddle}(a)). Once below $\gamma_{\min}$, the runs follow the same
exponential decay as the starts below the lowest saddle. 

All 70 runs converge to the targets and the rates are shown in  panel~(b).  The late rate is always fitted below after energy excess has fallen below $\gamma_{\min}$ and is $\zeta_{\mathrm{late}} = 2.004$--$2.05\,\gamma_{\min}$ for all starts, which matches the predicted rate $2\gamma_{\min}$ for warm starts of Fig.~\ref{fig:rate}. The run average includes the whole part of the run above $\gamma_{\min}$, yet
it still satisfies $\bar\lambda \ge 2\,\gamma_{\min}$ for all runs. Hence, $\Delta E(\rho_k) \le \Delta E(\rho_{\mathrm{ref}}) e^{-2\gamma_{\min} t}$ holds at every iteration
for all of these runs, although none satisfies the sufficient condition. The condition
$\Delta E(\rho_{\mathrm{ref}}) < \gamma_{\min}$ is therefore sufficient but not necessary. Violating it slows the flow only when the
initial point lies close to a saddle (Fig.~\ref{fig:saddle}).

\subsection{Adaptive strategy}\label{app:num_adaptive}
In this section, we study an adaptive DBF algorithm that requires no knowledge of the spectrum. First, we introduce the algorithm and its hyperparameters. Next, we  numerically investigate the role of the hyperparameters and design choices in convergence. Finally, we present various numerical results regarding the performance and stability of the proposed algorithm on a TFIM Hamiltonian. 

\subsubsection{Method} 
\label{app:method}

The speed of the DBF is set by the weights and the step size, and the optimal choices of these parameters, i.e., the inverse-gap weights and Chebyshev scheduling, depend on the gaps of the spectrum, which are not known in advance. To circumvent this issue, we introduce an adaptive DBF algorithm (Algorithm~\ref{alg:adaptive}) that learns the spectrum dependent parameters while it runs. The main idea is to measure the energies and use them as estimates of the true spectrum to update the weights and steps throughout the algorithm. More precisely, this algorithm splits the iterations into
$B$ blocks of length $L$, and at the start of each block it measures the energies of the tracked states and uses their differences to estimate the gaps. 
The algorithm cannot directly measure the gap between the highest tracked level to the first untracked one and therefore estimates it with the  smallest measured gap. From these estimates the algorithm recomputes the inverse-gap weights. To avoid rapid changes, the newly obtained weights are blended with the old weights and the result is set as the weights for the upcoming block. The blending strength is controlled by a parameter $\alpha$. $\alpha=1$ completely discards the old weights and $\alpha=0$ means no weight is updated through the run.  Next, $\hat \gamma_{\min}, \hat \gamma_{\max}$ are computed from the new measured gaps and weights to update the step schedule accordingly. For the fastest rate estimate $\hat \gamma_{\max}$, the algorithm assumes the spectral width is 2, which is based on the guarantee  $\norm{H} \le 1$. This is an overestimation but helps the algorithm to maintain stability. Early in the flow, the estimated energies can be out of order as two tracked states can sit on each other's levels. Therefore, the estimated energies are then sorted. Moreover, to keep the flow stable, all the measured gaps are floored at a fixed threshold $g_{\min}$. After a warm-up of constant steps, which carries the flow through its nonlinear early phase, the steps can follow a Chebyshev cycle which speeds up the algorithm. 

\begin{algorithm}[t]
  \caption{Adaptive DBF}
  \label{alg:adaptive}
  \begin{algorithmic}[1]
    \REQUIRE Hamiltonian $H$ with $\norm{H} \le 1$; orthonormal initial states $\ket{\phi_1}, \dots, \ket{\phi_r}$;
      order $\ell$ of the product formula $\PC_\ell$; block length $L$; number of blocks $B$; Chebyshev period $m$, a power of two
      that divides $L$ ($m = 1$: constant step); warm-up $w$; mixing $\alpha \in (0, 1]$; gap floor
      $g_{\min} > 0$
    \ENSURE approximations $\ket{\phi_j^{(BL)}}$ of the $r$ lowest eigenstates of $H$, $j = 1, \dots, r$
    \STATE $p_j \gets 2(r + 1 - j) / [r(r + 1)]$ and $\hat g_j \gets 2/r$ for $j = 1, \dots, r$;
      $p_{r+1} \gets 0$
      \COMMENT{linear weights}
    \STATE $U_0 \gets I$
    \FOR{$b = 0, \dots, B - 1$}
      \STATE $k_0 \gets b\,L$
      \COMMENT{iterations done}
      \IF{$b > 0$}
        \STATE $\hat{E}_j \gets \bra{\phi_j} U_{k_0}^\dagger H U_{k_0} \ket{\phi_j}$ for $j = 1, \dots, r$
          \COMMENT{measure energies}
        \STATE $(\hat{E}_{(1)}, \dots, \hat{E}_{(r)}) \gets$ the measured energies $\hat{E}_j$ in ascending order
        \STATE $\hat g_j \gets \max(\hat E_{(j+1)} -\hat E_{ (j)}, g_{\min})$ for $j = 1, \dots, r - 1$
        \STATE $\hat g_r \gets \min_{j < r} \hat g_j$
          \COMMENT{gap to the first untracked level is set to the minimum of other gaps}
        \STATE $p'$: $p'_j - p'_{j+1} \propto 1 / \hat g_j$, $p'_{r+1} = 0$, $\sum_j p'_j = 1$
          \COMMENT{new inverse-gap weights computed from estimated gaps}
        \STATE $p \gets (1 - \alpha)\, p + \alpha\, p'$ \COMMENT{convex combination of new and old  weights}
      \ENDIF
      \STATE $\hat \gamma_{\min} \gets \min_j (p_j - p_{j+1})\, \hat g_j$, \quad $\hat \gamma_{\max} \gets 2 p_1$
      \IF{$m > 1$ and $k_0 \ge w$}
        \STATE $\mu_i \gets \frac{\hat \gamma_{\max} + \hat \gamma_{\min}}{2}
          + \frac{\hat\gamma_{\max} - \hat\gamma_{\min}}{2} \cos\frac{(2i - 1)\pi}{2m}$ for $i = 1, \dots, m$ \COMMENT{roots of the Chebyshev polynomial}
        \STATE $s_i \gets 1 / \mu_{(i \bmod m) + 1}$ for $i = 0, \dots, L-1$
          \COMMENT{$L/m$ Chebyshev cycles}
      \ELSE
        \STATE $s_i \gets 2 / (\hat \gamma_{\min} + \hat \gamma_{\max})$ for $i = 0, \dots, L-1$
          \COMMENT{constant step}
      \ENDIF
      \STATE $\rho_{k_0} \gets \sum_{j=1}^r p_j\, U_{k_0} \ket{\phi_j}\!\bra{\phi_j} U_{k_0}^\dagger$ \COMMENT{reference state with the new weights}

      \FOR{$i = 0, \dots, L-1$}
        \STATE $k \gets k_0 + i$
        \STATE $V_k \gets \PC_\ell(\rho_k, H, s_{i})$
        \STATE $\rho_{k+1} \gets V_k\, \rho_{k}\, V_{k}^\dagger$
        \STATE $U_{k+1} \gets V_k U_{k}$
      \ENDFOR
    \ENDFOR
    \STATE \textbf{return} $\ket{\phi_j^{(BL)}} = U_{BL} \ket{\phi_j}$, $j = 1, \dots, r$
  \end{algorithmic}
\end{algorithm}

\subsubsection{Setup of Experiments}
\label{app:setup}

All four experiments of this section target the $r = 10$ lowest eigenstates of the open TFIM with $N = 10$
qubits, and the fourth-order product formula unless stated otherwise. We count the cost in
Hamiltonian evolutions per tracked state, which are 14 per iteration at fourth order and 139 at sixth order. The cost of energy
measurements at the block starts is not included in this analysis but can taken into account through the block length $L$. 

The first two experiments (Figs.~\ref{fig:last-gap}
and~\ref{fig:grid}) study the effect of hyperparameters and design choices on the performance of the algorithm. The spectrum is normalized
to exactly $[-1, 1]$. On this spectrum the smallest gap among the targets is $0.0042$ (levels 8 and 9), and the gap from
the highest target to the next level is $0.0126$. The initial states are a single fixed set of states, each
of which overlaps every eigenstate equally. All runs are compared at the first iteration at which they fall below
$2 \times 10^{-7}$.  

The last two experiments (Figs.~\ref{fig:order4}
and~\ref{fig:warmstart}) study the stability and query complexity of the algorithm for the TFIM model using the setup of the main text. Namely, the spectrum is scaled by an unknown factor $c \approx 0.947$, and
21 sets of Haar-random initial states are considered. The plots show the normalized error as median and interquartile range. Moreover, we run the algorithm with the exact inverse-gap weights and Chebyshev steps as reference for comparison with the adaptive runs.

\subsubsection{Gap to the first untracked level}
\label{app:last-gap}

\begin{figure}[t]
  \centering
  \includegraphics[width=\textwidth]{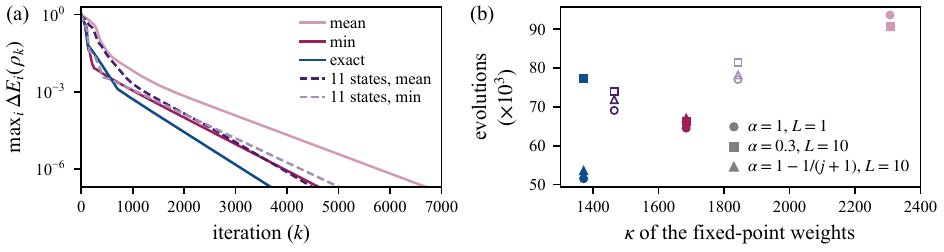}
  \caption{\textbf{Rule for the untracked gap.} On the chain normalized to
  $[-1, 1]$ with fixed initial states. \textbf{(a)} $\max_i {\Delta E_i{(\rho_k)}}$ over the 10 targets at $\alpha = 1$,
  $L = 1$ for the mean and the min rule of the measured gaps, the exact gap, and an eleventh tracked state with the mean
  or min rule above it (dashed). \textbf{(b)} Hamiltonian evolutions per target state until the error falls below
  $2 \times 10^{-7}$, against $\kappa$ of the weights each rule converges
  to. Marker shapes show the update setting, colors indicate the rule as in {(a)} and open markers show 11 tracked states.}
  \label{fig:last-gap}
\end{figure}

The proposed algorithm measures the energies of the $r$ evolved states throughout the run. The energy gaps are then estimated by the difference between the sorted energy estimates. However, the last gap, i.e. the difference between the energy of the last tracked state and the first untracked one, cannot be measured directly. Therefore, we  use rules to indirectly estimate the last energy gap from the other gaps. Figure~\ref{fig:last-gap} compares four rules for the estimation of the last gap: 
\begin{enumerate}
    \item  The last gap is set to the mean of all the measured gaps. 
    \item  The last gap is set to the smallest (min) of all the measured gaps.
    \item  The exact gap is used as a reference.
    \item  An extra eleventh state whose energy measures the gap is tracked. Its own last gap again is set by the mean or min rule.
\end{enumerate}
In this experiment an update is skipped when the energies
were out of order. Panel (a) shows the error over iterations for $\alpha = 1$, $L = 1$. The exact gap needs 3680 iterations, the min rule 4610 and the mean
rule 6692.

Panel (b) shows the number of required Hamiltonian evolutions over the $\kappa$ that each rule converges to.  The min rule costs almost 25\,\% more than the exact gap  and less than the two
eleven-state variants, which pay for the extra state. The order of the adaptive rules
is maintained at all three update settings (choices of $\alpha$, $L$). Therefore, we use the min rule without an extra state for the rest of the experiments. 

While the min rule is the most cost effective and accurate for the TFIM, this is not necessarily true for every Hamiltonian. For Hamiltonians whose gaps shrink for higher energy eigenstates, the min rule works well, but  if the energy levels are equidistant mean rule can be more effective. This is an important design choice in the algorithm that can be guided by partial information about the distribution of energy levels. 

\subsubsection{Mixing and block length}
\label{app:grid}

\begin{figure}[t]
  \centering
  \includegraphics[width=\textwidth]{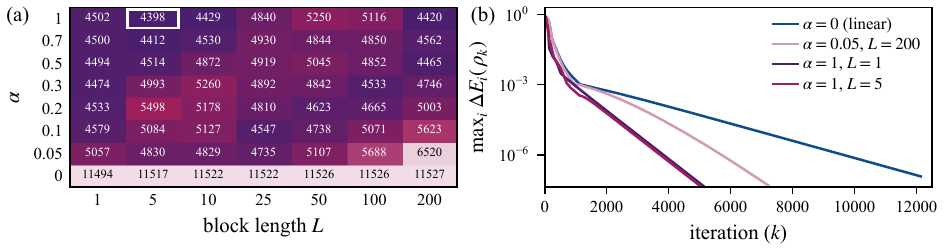}
  \caption{\textbf{Mixing $\alpha$ and block length $L$ of the adaptive DBF with the min rule and sorted energies}, on the
  chain normalized to $[-1, 1]$ with fixed initial states. \textbf{(a)} Iterations until
  $\max_i |{\Delta E_i{(\rho_k)}}| < 2 \times 10^{-7}$; $\alpha = 0$ keeps the linear weights and adapts only the
  step. The color scale is clipped at 6600. White box: the setting of the main text. \textbf{(b)} The error for four
  settings.}
  \label{fig:grid}
\end{figure}

Figure~\ref{fig:grid} varies how strongly ($\alpha$) and how often ($L$) the weights are updated, with the min rule
and sorted energies. With $\alpha = 0$ the weights stay linear and only the step adapts. This takes about 11\,500
iterations. Every $\alpha > 0$ moves the weights towards the optimal weights of the min-rule fixed point. However, for small $\alpha$ and long blocks, the weights adapt slowly and convergence is slowed down. Therefore, the fastest setting, $\alpha = 1$ and $L = 5$, is the setting used in the main text. In this case, the weights can simply be replaced at every block start without the need to be blended with the old weights. 

\subsubsection{Chebyshev schedules with the fourth-order product formula}
\label{app:order4}

\begin{figure}[t]
  \centering
  \includegraphics[width=\textwidth]{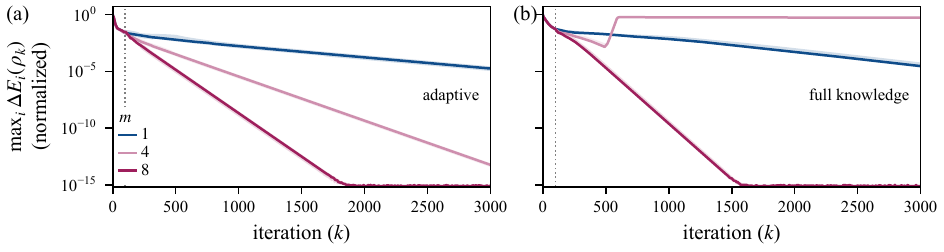}
  \caption{\textbf{DBF algorithm with Chebyshev schedules with periods $m = 1,4,8$} at product-formula order four, on
  the scaled TFIM chain of Fig.~\ref{fig:adaptive} using \textbf{(a)} Adaptive and \textbf{(b)} full-knowledge strategies.  Normalized error
  $\max_i {\Delta E_i(\rho_k)}/ \max_i \Delta E_i(\rho_\mathrm{ref})$ is plotted for 21 sets of Haar random initial states. Dark curves show the median and the shaded areas show the interquartiles. All runs have a $m=1$ warm-up phase for the first 100 iterations (dotted line).}
  \label{fig:order4}
\end{figure}

Figure~\ref{fig:order4} adds the Chebyshev period $m = 4$ to Fig.~\ref{fig:adaptive}. Both adaptive Chebyshev
schedules converge to the precision floor, after around  3000 iterations.  With full knowledge, $m = 4$ does not converge:  the error falls until iteration 500, then jumps to about $0.5$ and plateaus. This issue is due to product-formula error, which grows with large step sizes of the Chebyshev schedule and the energy gaps of the pairs. Consequently, at the long steps of a cycle, the modes no longer follow the Chebyshev polynomial optimality, and the schedule amplifies them. Apart from the period of the cycle, this instability also depends on the spectrum and weights, which is why the adaptive runs do not suffer from it. Finally, this instability can be mitigated by 
a higher-order product formula which motivates the next experiment. 

\subsubsection{Warm-started Chebyshev at sixth order}
\label{app:warmstart}

\begin{figure}[t]
  \centering
  \includegraphics[width=\textwidth]{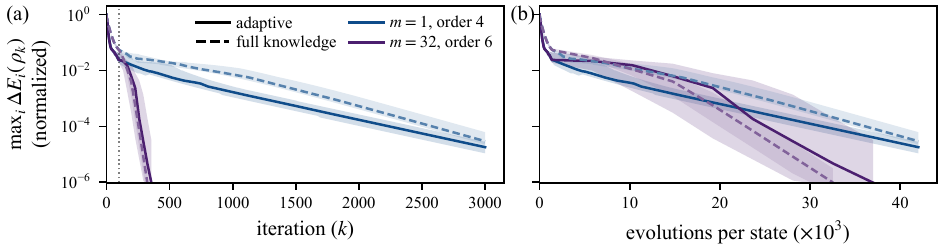}
  \caption{\textbf{Constant step ($m = 1$) at product-formula order four and Chebyshev schedule with $m = 32$ at order six},
  both started by a 100-iteration warm-up phase (dotted line). Normalized error is plotted against
  \textbf{(a)} the iteration and \textbf{(b)} the Hamiltonian evolutions per state. Solid and dashed lines show adaptive and full knowledge runs respectively. Dark lines and shaded areas represent the median
  and interquartile range over 21 sets of Haar-random reference states.}
  \label{fig:warmstart}
\end{figure}

Figure~\ref{fig:warmstart} demonstrates the convergence of a Chebyshev schedule with $m=32$ and a 100-iteration warm-up with $m=1$. The large Chebyshev period requires a sixth-order product formula, which costs about ten times more than
the fourth-order one per iteration. This additional Hamiltonian query cost is only justified if it is compensated by the speed-up achieved by the Chebyshev schedule. Therefore,   

the  initial warm-up phase runs at fourth order with the
$m=1$ schedule, which is cheap and stable. After 100 iterations the run switches to Chebyshev cycles of period
$m = 32$ at sixth order, and stops at the first cycle end at which the normalized error is
below $10^{-6}$. For comparison, we also run an $m=1$ Chebyshev cycle with the fourth order product formula. 

Panel (a) shows the number of iterations for this $m=32$ adaptive run versus the reference $m=1$ schedule at order 4. When compared at the level of required iterations, the $m=32$ schedule converges almost 10 times faster than the $m=1$ schedule. However, the true cost of the schedules is revealed in panel (b) where the errors are plotted over the number of Hamiltonian evolutions. Here, the $m=32$ costs 10 times more due to the higher order product-formula implementation and is only advantageous when the error drops below $10^{-3}$. Therefore, longer Chebyshev cycles do not automatically outperform shorter cycles and the true optimal period depends on the problem and the choice of the product formula order.

In both panels, adaptive runs follow the full-knowledge runs closely and all runs are stable. This further demonstrates the power of the adaptive DBF algorithm that is completely agnostic to the spectral gaps. 

\end{document}